# Advances in the Theory of III-V Nanowire Growth Dynamics

*Peter Krogstrup[†,1], Henrik I. Jørgensen[2], Erik Johnson[1], Morten Hannibal Madsen[1], Claus B. Sørensen[1], Anna Fontcuberta i Morral[3], Martin Aagesen[2], Jesper Nygård[1], Frank Glas[4]*

[1]Center for Quantum Devices, Niels Bohr Institute, University of Copenhagen, 2100 Copenhagen, Denmark

[2]SunFlake A/S, Universitetsparken 5, 2100 Copenhagen, Denmark

[3]Laboratoire des Materiaux Semiconducteurs, Ecole Polytechnique Fédérale de Lausanne, 1015 Lausanne, Switzerland

[4]CNRS, Laboratoire de Photonique et de Nanostructures, Route de Nozay, 91460 Marcoussis, France

**Abstract:** *Nanowire crystal growth via the Vapor-Liquid-Solid (VLS) mechanism is a complex dynamic process involving interactions between many atoms of various thermodynamic states. With increasing speed over the last few decades many works have reported on various aspects of the growth mechanisms, both experimentally and theoretically. We will here propose a general continuum formalism for growth kinetics based on thermodynamic parameters and transition state kinetics. We use the formalism together with key elements of recent research to present a more overall treatment of III-V NW growth, which can serve as a basis to model and understand the dynamical mechanisms in terms of the basic control parameters, temperature and pressures/beam fluxes. Self-catalyzed GaAs nanowire growth on Si substrates by Molecular Beam Epitaxy is used as a model system.*

† email: krogstrup@nbi.dk

# 1. Introduction

Nanowire (NW) crystals are wire-like single crystal structures with diameters typically constrained to tens of nanometers and with lengths of micrometers. The finite lateral size gives rise to many new physical properties which are not seen in bulk materials. In particular, there has been an enormous interest in controlling and understanding the crystal growth of semiconductor nanowires over the recent years, as this is key for the control of the opto-electronic properties and nanowire morphology[1,2,3,4,5,6,7,8]. The vapor-liquid-solid (VLS) mechanism was first proposed in 1964 by Wagner and Ellis[9] as an explanation for unidirectional Si crystal growth in the presence of a liquid Au droplet. They concluded on the basis of a set of observations that the liquid phase acts as a sorption center for growth material arriving from the vapor phase, and that the NW formation takes place by precipitation of growth material from the droplet. Today the VLS method is the most common way of achieving NW formation, and NWs are now being grown using various growth methods and with a wide range of materials such as oxides, group IV, III-V and II-VI semiconductors and metals. Here we focus on III-V materials, however, the general theoretical approach can be extended to other types of materials. The most typical methods for III-V NW formation are Metal Organic Vapor Phase Epitaxy (MOVPE) and Molecular Beam Epitaxy (MBE). In all cases there is a supersaturated liquid droplet which initiates and maintains NW growth. Typically, the growth direction is [111]B in the case of the cubic zinc blende (ZB) structure (ABC-ABC, 3C stacking) and [0001]B for hexagonal wurtzite (WZ) structure (AB-AB, 2H stacking), see Figure 1-2. Higher order stacking sequences such as 4H (ABCB-ABCB) and others are possible but are occurring very rarely and only in small segments, see Johansson et al.[10] for a detailed discussion on higher order polytypes.

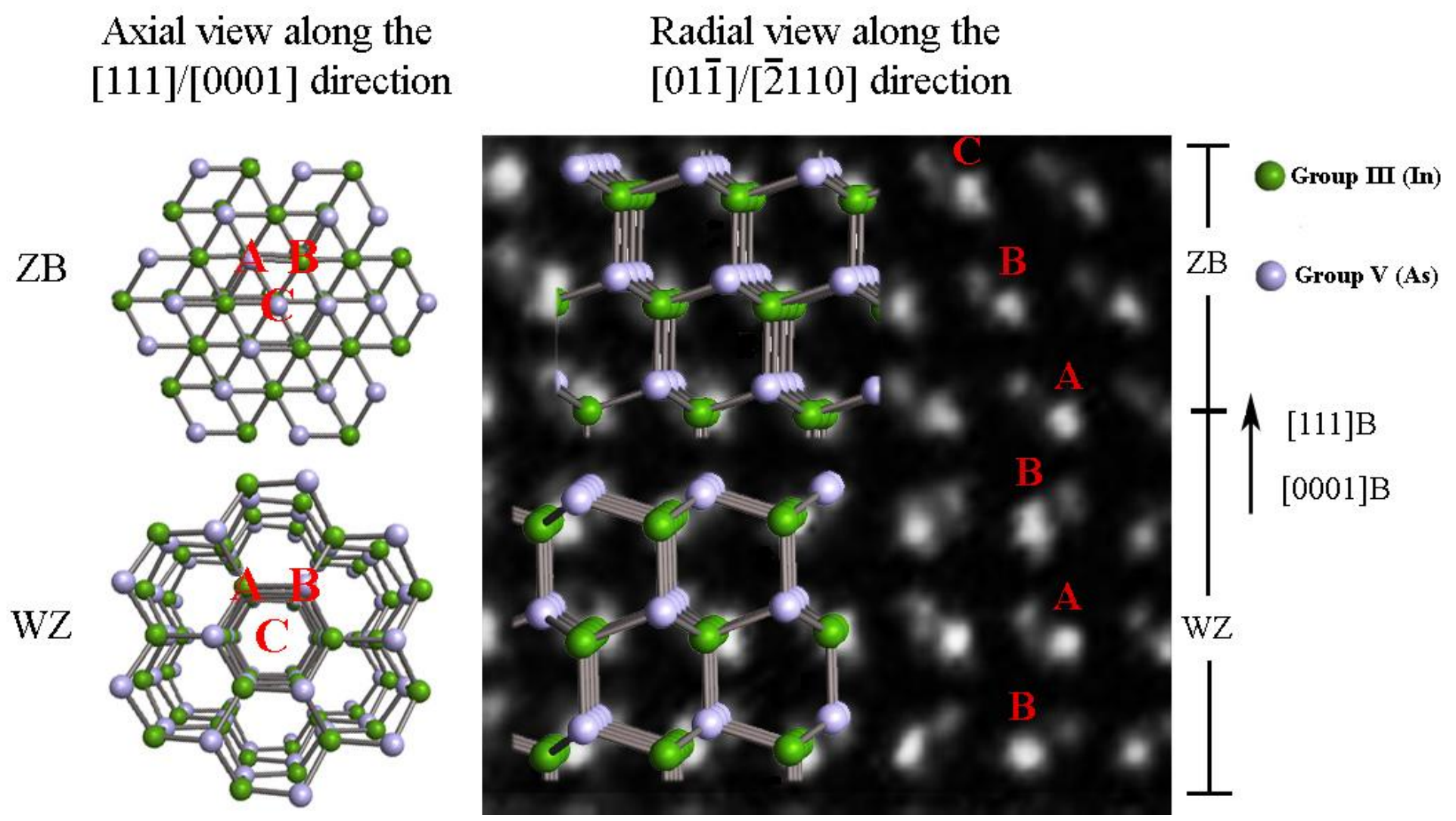

**Figure 1. The two most common crystal structures in III-V NWs, ZB (ABC-ABC stacking) and WZ (AB-AB stacking), viewed along the axial [111]/[0001] NW crystal growth directions and radial [01-1]/[-2110] crystal directions. The background of the high resolution radial view is a High Angle Annular Dark Field (HAADF) Scanning Transmission Electron Microscopy (STEM) image of a InAs NW, see ref.[11].**

In the 1970s, theoreticians proposed the first advanced growth models, where fundamental aspects of VLS growth, such as axial and radial growth rates, size effects, nucleation and diffusion phenomenon were discussed (see for example Givargizov and references therein[12]). Even though groups started more detailed analyses of III-V NW growth in the 1990s,[13] the VLS models from the 1970's where not significantly refined until Dubrovskii et al.[14,15,16] in 2004 and Johansson et al.[17] in 2005 proposed detailed VLS growth models of III-V NWs. Similar mechanisms such as the vapor-solid-solid mechanism (VSS) were also discovered as a variation of VLS.[18] Since then, the understanding of the complex growth mechanisms and the experimental control of the crystal phases, morphology and many different kind of heterostructure growth has undergone a huge progress. Today it is well accepted that group III species is adsorbed at the NW sidefacets and substrate surfaces and effectively diffuse to the growth region as adatoms[19,20,21], while group V species such as As and P are contributing to the axial growth primarily via either direct impingement from the beam (MBE) or as secondary absorbed species[22,23,24]. Today it is a fact that the shape of the NWs, and hence their potential applicability, is strongly dependent on the shape and morphology of the liquid-solid growth interface during growth[7,25,26,27,28,29]. Thus, understanding and controlling the dynamics of NW growth is of great practical importance, and elucidating the effects of growth kinetics on especially the NW crystal shape, composition and on its crystalline quality have become major research topics[30,31,32,33,34]. Figure 2 shows post-growth transmission electron microscope (TEM) images of the two most common types of GaAs NW growth today, Au-catalyzed and self-catalyzed growth. Since the work of Wagner and Ellis[9], Au has always been the preferred material to promote axial nanowire growth via the VLS mechanism. However, since 2008 research on self-catalyzed growth of GaAs NWs has received renewed interest[35,36]. It is today a highly appreciated growth mode for GaAs NWs by MBE and the control of the morphology and crystal phases has quickly reached a high level (see for example the recent growth experiments by Yu et al.[37] and Munchi et al.[38]).

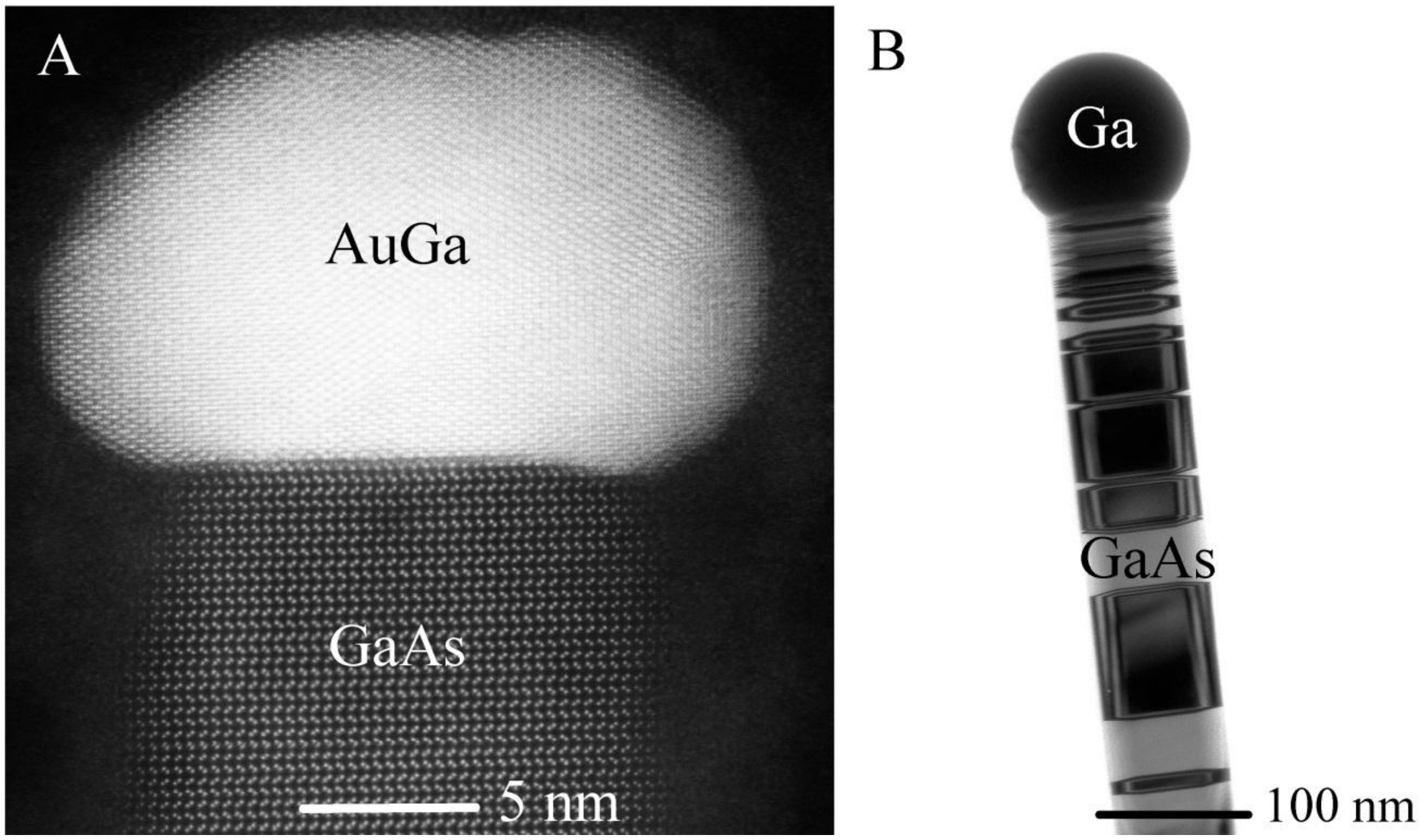

**Figure 2. Two post growth images of the most common types of growth of GaAs NWs via the VLS mechanisms; (A) Au catalyzed and (B) self-catalyzed growth. (A) shows a thin GaAs NW with a solidified AuGa crystal cap. The image was acquired with the high angle annular dark field (STEM) technique, using the probe corrected TEAM0.5 microscope. This technique makes it possible to resolve the atomic columns of the dumbbells revealing a perfect As-terminated WZ structure. In (B) a relatively thick multiply twinned ZB structured GaAs NW with a liquid Ga droplet on top, is shown. This image is acquired with a 200keV CM20 microscope on film which is ideal for low magnification images with a large field of view. Both the AuGa and the Ga cap have been emptied of As upon cooling down to room temperature after growth termination.**

While most analyses of NW growth kinetics are based on post-growth characterization and static analyses of complete NWs, recent progress has been made experimentally by in-situ growth characterization[39,40,41,42,43] and ex-situ study of NWs with markers inserted during growth[23,44], and dynamic modeling[45,46,47]. For a more complete understanding of growth one should understand in detail the dependence of the basic control parameters (i.e. temperature and pressures/beam fluxes) on the growth mechanisms. Moreover, as local conditions on the growth front change during NW growth, it is necessary to include the time dependence in the analysis. However, to do this in a general manner, all essential features need to be incorporated into one coherent description of the growth dynamics, including a detailed treatment of all the main types of transitions involved in the process. Schwarz and Tersoff[45] presented in 2009 a pioneering continuum model for NW growth dynamics via the VLS process, where they could follow the evolution from a eutectic droplet at the substrate surface into a nanowire. Even though the kinetic equations governing this two dimensional modeling is simplified to barrier-free kinetics without any explicit temperature dependence, it is able to describe some basic properties of the dynamical evolution. However, as will be explained here, transition barriers and temperature dependence play a very important role on the crystal structure and morphology. As an example, another pioneering work was presented two years earlier by Glas et al. (2007)[25], who proposed that the liquid to solid phase transition at the (111) topfacet of III-V NWs was nucleation limited, and that the structure of each monolayer was determined by the structure of the two-dimensional nucleus which is needed to overcome the transition barrier. Thus, to understand the structural details of the III-V nanowire growth, the temperature dependence cannot be neglected. In general, the temperature dependence on a given barrier limited transition rate is described with an Arrhenius dependence, or more specifically transition state theory[48]. Thus, here we will combine various theoretical models into more general dynamical and quantitative approach where the formalism, which will be explained in detail, is based on transition state kinetics driven by a Gibbs free energy minimization process. The modeling is based on the quantitative description of all the relevant dynamic processes, such as mass transfer, nucleation and dynamical reshaping of interfaces, and consists of many time-dependent and coupled equations involving the material parameters and growth conditions. We give various examples of modeling the self-catalyzed GaAs NW growth and match the theoretical predictions directly with growth experiments, while stressing that the theoretical formalism will be useful for other NW material systems. The aim of this review is to give a detailed theoretical insight into the III-V NW growth dynamics in an as pedagogical manner as possible. The focus will be on combining the knowledge which has been gained about III-V nanowire growth so far within the general framework of chemical kinetics, and to present a general theoretical formalism for III-V NW growth kinetics, which can serve as a tool to analyze and predict the evolution of NW growth, in terms of temperature and pressures/beam fluxes.

We will here give a brief outline of the content: Section 2 presents the general theoretical formalism and is divided into three subsections. Subsection 2.1 formulates the kinetics of the atomic movements, i.e. the probabilities of atomic state transitions in terms of rates, based on transition state theory. Here the effective transition rates between the various types of states are derived as a function of intrinsic parameters describing the ‘local’ environment. We then turn to the actual crystal formation at the liquid-solid interface in subsection 2.2. There, we discuss the framework needed to analyze the liquid-solid phase transition to a facetted nanowire crystal where transitions on certain facets can be nucleation limited. A specific topic which has attracted huge attention, is the mechanisms controlling the relative formation rates of ZB, WZ or other types of crystal structures in III-V NWs[25,26,27,28]. This is treated and discussed in detail in the framework of the present theory in sections 2.3 and 3.5. Section 3 show examples of self-catalyzed GaAs NW growth experiments and how to use the theory to analyze and understand NW growth dynamics. First, growth simulations of the overall NW morphologies are presented in sections 3.1-3.4, and section 3.5 present detailed simulations of the anisotropic liquid-solid NW growth dynamics and discuss the results.

*A list of the symbols and abbreviations used in the following is given at the end of the paper (section 6).*

## 2. Theoretical formalism

NW growth is a process far from thermodynamic equilibrium and in order to quantify the growth in terms of thermodynamic parameters, it is convenient to refer to an *equilibrium reference state* (*ERS*). Because the solid III/V stoichiometry is assumed to be fixed at 1:1 (which is verified to a very good accuracy), the chemical potential of the infinite solid phase is a function of temperature only and therefore serves as a natural reference state for the *ERS*. The *ERS* chemical potential of group III (or V) is equal to the liquid chemical potential when the liquid and solid are in equilibrium

$$\mu_{III(V)}^{ERS} \equiv \mu_{l,III(V)}^{\infty}(x_{III}^{ERS}, x_{V}^{ERS}) = \mu_{s,III}^{\infty} + \mu_{s,V}^{\infty} - \mu_{l,V(III)}^{\infty}(x_{III}^{ERS}, x_{V}^{ERS}) \qquad (1)$$

Here $x_i^{ERS}(T)$ is the *ERS* mole fraction of group $i$ in the liquid, and '$\infty$' refers to large phases (i.e. without size effects, such as the Gibbs-Thomson effect). For the growth in a MBE chamber, we distinguish between five main types of states for each element $i$; beam flux (*b,i*), vapor (*v,i*), adatom/admolecule (*a,i*), liquid (*l,i*) and solid (*s,i*). Here the *v* states are all other states in the gas phase which are not a part of the direct beam flux, i.e. mainly what is reemitted from the neighboring surfaces and evaporated form the droplets (and possibly reabsorbed). 6 intrinsic parameters are needed to describe the *ERS* in the case of self-catalyzed growth; temperature $T$, liquid concentration $x_V^{ERS}$ (group III concentration follows from $x_{III} + x_V = 1$), the partial vapor pressures $p_{III}^{ERS}$, $p_V^{ERS}$ and the *ERS* adatom densities $\rho_{III}^{ERS}, \rho_{V}^{ERS}$ (note that the beam flux cannot be a part of an equilibrium system). The *ERS* for self-catalyzed growth has one degree of freedom, which means that the *ERS* is determined by the choice of one parameter, e.g. the temperature. For an example of calculating the *ERS* parameters for self-catalyzed growth of GaAs or InAs we refer to section 3.1. For growth catalyzed by a foreign element (such as gold) which is only present in the liquid phase, the *ERS* has one additional degree of freedom. This means that we can choose for instance both the temperature and the group III concentration in the liquid to determine fully the *ERS* state. For liquids including Au, see ref.[49].

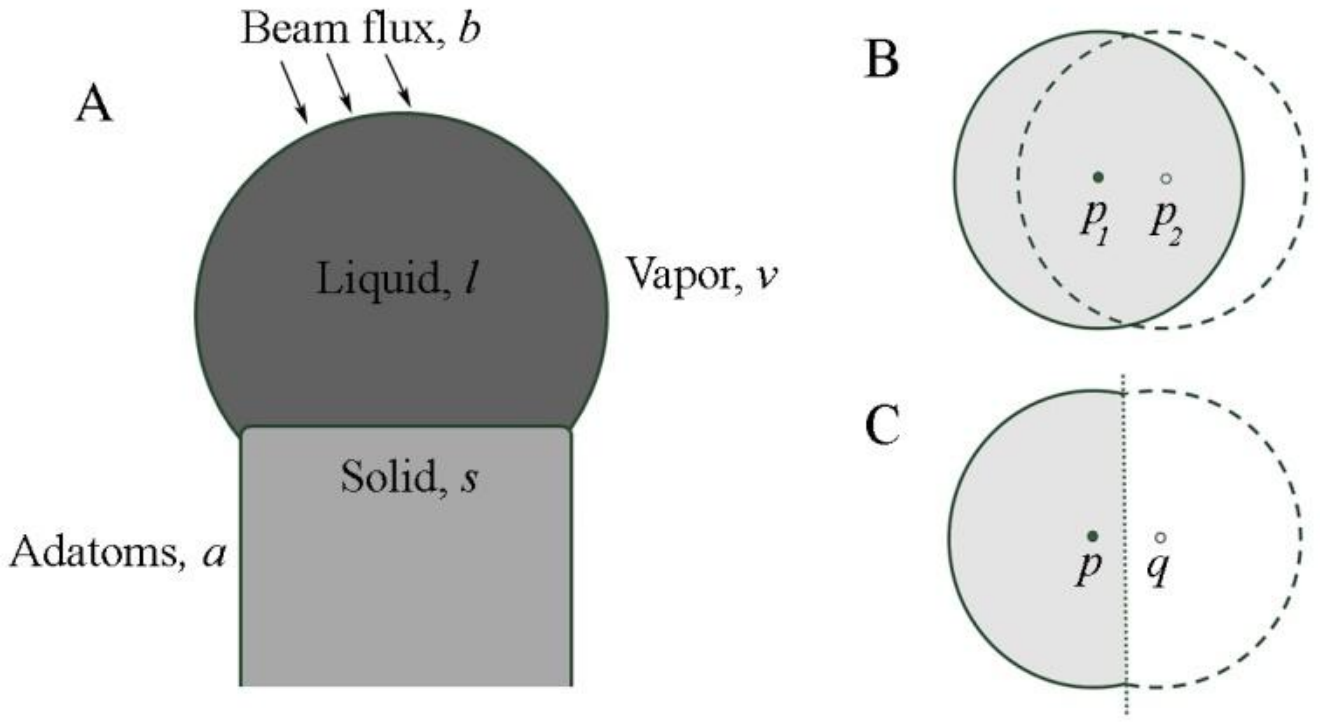

**Figure 3. A) The five types of states considered during the NW growth process. B) The principle of describing atomic transition rates in a continuum language relies on the choice of small volume segments in the vicinity of the atomic state in which every property of the microstate *p* takes on average values of such ensemble. Within one of the main states shown in A) two adjacent local states (here $p_1$ and $p_2$) are described with almost the same parameters. C) Between two distinct types of states we choose a dividing interface where local states on each side of the phase boundary are described with mean parameters from a small volume segment within each respective main state. Thus in this formalism a discontinuous jump in the chemical potentials between two adjacent main states is possible during growth.**

## 2.1 Growth kinetics

Within each of the main types of states (Figure 3A), a 'local state' *p* are characterized by the mean intrinsic properties of some local surrounding (the 'local ensemble'), see Figure 3B, which is large enough to represent the thermodynamic characteristics and small enough to represent the local environment when the global system is out of equilibrium. At interfaces between two main types of states, a single interface is typically chosen, which means that one distinguish between particles on each side of the interface with local state properties depending only on the local environment of the main state to which they belong (Figure 3C). A 'single Gibbs interface' is usually introduced to attach interface excess quantities to an assumed infinite sharp interface between two phases, i.e. no atoms belong to the interface, only excesses. To describe the growth dynamics we need to treat the *b* and *a* states as separate states, but only consider interface excesses between the classical *v*, *l* and *s* phases. To insure a consistent treatment of the kinetics in terms of the intrinsic thermodynamic parameters, it is convenient to measure the chemical potentials of the all various states with respect to the chemical potential in the *ERS,*

$$\delta\mu_{p-ERS,i} = \mu_{p,i} - \mu_i^{ERS} \qquad (2)$$

where $\mu_{p,i}$ are the chemical potential of the state $p$. The chemical potentials with respect to the four *ERS* states are:

$$\delta\mu_{v-ERS,i}\left(p_i, T\right) = k_B T \ln\left(\frac{p_i}{p_i^{ERS}}\right) \qquad (3)$$

$$\delta\mu_{a_j-ERS,III(V)}\left(\rho_{j,III}, \rho_{j,V}, T\right) = k_B T \ln\left(\frac{\bar{\rho}_{j,III(V)}\left(1-\bar{\rho}_{j,III}^{ERS} - \bar{\rho}_{j,V}^{ERS}\right)}{\bar{\rho}_{j,III(V)}^{ERS}\left(1-\bar{\rho}_{j,III} - \bar{\rho}_{j,V}\right)}\right) \qquad (4)$$

$$\delta\mu_{l-ERS,i}(x_{III}, x_V, T) = \mu_{l,i}^{\infty}(x_{III}, x_V, T) + \gamma_{vl}\frac{\partial A_{vl}}{\partial N_{l,i}} - \mu_i^{ERS} \qquad (5)$$

$$\delta\mu_{s-ERS,III-V}^{X} = \sum_j \gamma_j \frac{\partial A_j}{\partial X}\frac{\partial X}{\partial N_{s,III-V}} + \Delta\varepsilon_s \qquad (6)$$

where $\rho_{j,i}$ is the adatom density on the $j$'th facet, and $A_j$ and $\gamma_j$ are the area and interface energy of the $j$'th interface respectively. The form of eq.(4) is a simplified version which stems from a detailed calculation of the partition function, see ref.[50]. For a full expression, the following two terms should be added to Eq.(4): $-\bar{Z}_{j,aa}\left(B_{j,III(V)}\left(\bar{\rho}_{j,III(V)} - \bar{\rho}_{j,III(V)}^{ERS}\right) + B_{j,III-V}\left(\bar{\rho}_{j,V(III)} - \bar{\rho}_{j,V(III)}^{ERS}\right)\right)$. $\bar{Z}_{aa}$ is a reaction constant (including coordination number) for facet $j$, and $B_{j,III(V)}$ and $B_{j,III-V}$ are the binding free energies for III-III(V-V) and III-V bonds on the $j$ surface, respectively[51]. If the adatom concentrations and binding energies are low, eq.(4) can be approximated by an ideal behavior, $\delta\mu_{a_j-ERS,i}(\rho_{j,i}, T) \cong k_B T \ln\left(\frac{\rho_{j,i}}{\rho_{j,i}^{ERS}}\right)$, which strongly reduces computation time. The relative chemical potential of the solid $\delta\mu_{s-ERS,III-V}^{X}$ (eq.(6)) in terms of a given parameter, $X$ is the change in Gibbs free energy per pair due to a corresponding change in $X$, such as a length or an angle. In this continuum approach it describes the mean thermodynamic properties for the chosen parameter $X$. For a full description of the NW crystal an complete set of independent parameters, $\{X\}$, is needed. That is, adding matter to a nanosize crystal will change not only its volume, but also its shape, and therefore the interface excesses. In addition to its volume, the crystal must thus be defined by a set of parameters {X} such as facet areas, projected facet heights, facet angles, edge lengths or local interface curvature. It is important to notice that {X} is a chosen set of independent parameters that fully define the choice of crystal geometry (several choices are possible; an example is given in section 3.5). Then, the change of energy of the crystal when matter is added to it comprise a first term, associated to its change of shape, and a second term associated to its change of volume (which is simply related to the chemical potential of the reference infinite solid, as introduced in eq.(1)). In the first term of eq.(6), the independence of the X parameters and the effect of the changes of these parameters on the areas of the interfaces to which excess energies are associated, are taken into account. The second term is the difference in bulk cohesive energy between the standard reference of the *ERS* (typically ZB) and the actual formation structure *s*. The liquid-solid system will tend towards the equilibrium shape which is the one where the sum of all chemical potentials of the set are equal (See for example Carter et al.[52] for a treatment of a fully facetted solid in two dimensions using the concept of *weighted mean curvature*[53]). Note that if the crystal structure *s* is the same as the *ERS*, $\delta\mu^X_{s-ERS,III-V}$ is only a size effect as the bulk chemical potential is the same as the *ERS* (i.e. $\Delta\varepsilon_s = 0$). See section 2.2 for more details. In addition to these interface size effects, it was suggested by Schmidt et al.[54] and Schwartz and Tersoff[45] that an excess TL energy, which may arise from an in-balance of capillary forces at the TL, plays an important role on the dynamics of NW growth. See appendix 5.6 for a discussion.

As mentioned in the introduction, the general average rate at which a given $p \rightarrow q$ transition takes place depends exponentially on the Gibbs free energy of activation for reaching the transition state (*TS*), as $P_{pq,i} \propto \exp\left(-\frac{\delta g^{TS}_{pq,i}}{k_B T}\right)$. $\delta g^{TS}_{pq,i}$ is taken as the difference in free energy per atom between the state *p* (calculated from the thermodynamic parameters describing this state) and the transition state of the particle between *p* and *q*. If a given transition requires a bond-dissociation of molecules into single atoms (e.g. $As_2 \rightarrow 2As$), the dissociation enthalpy and entropy should be added to $\delta g^{TS}_{pq,i}$.

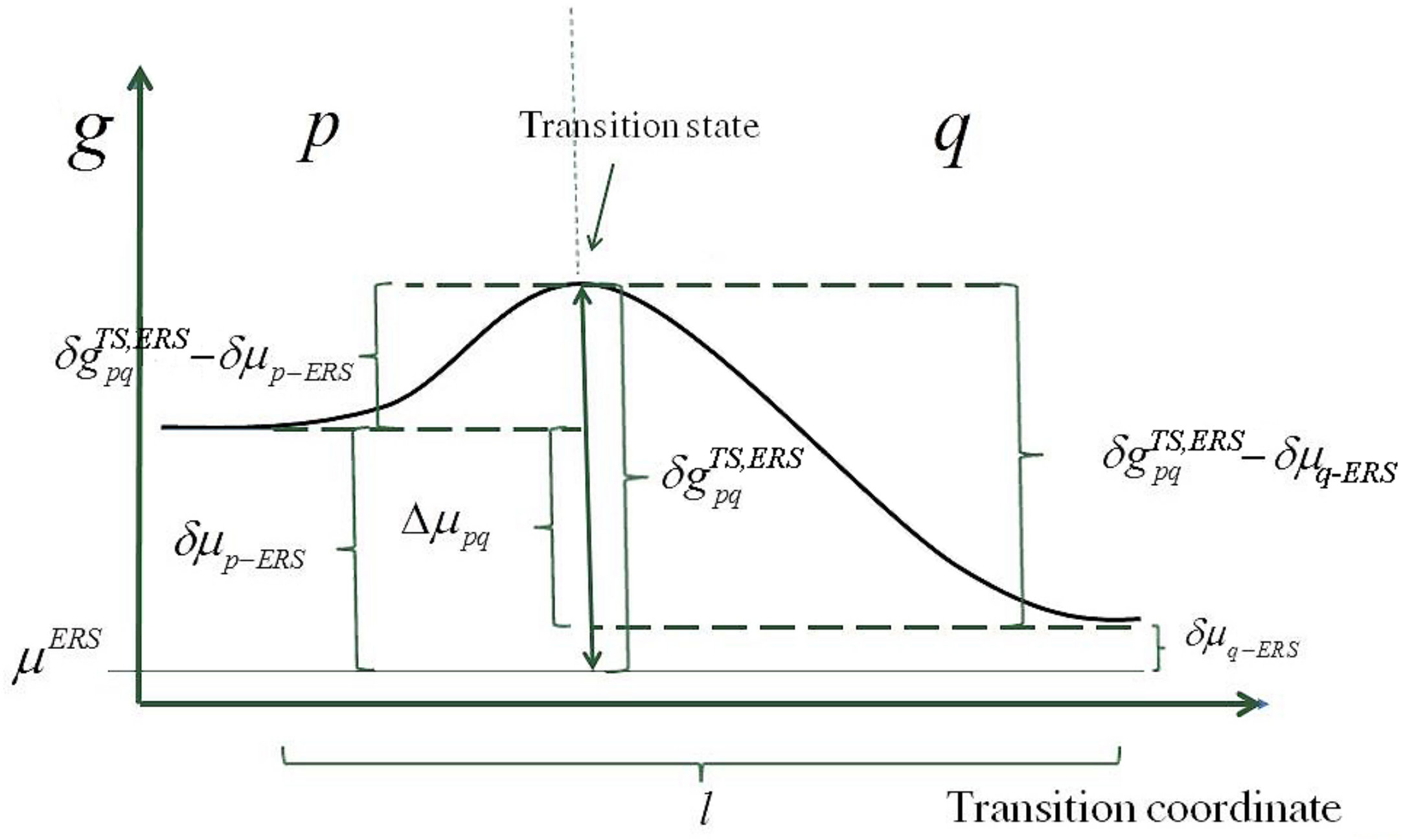

**Figure 4. One dimensional illustration of the free energy barrier associated with a *pq* state transition. Here the equilibrium transition state barrier is symmetric (i.e. $\delta g_{pq,i}^{TS,ERS} = \delta g_{qp,i}^{TS,ERS}$), as would be the case for a reversible state transition without requirements for dissociation/formation of bonds only one way. Note that even though the illustration is a typical sketch of a single particle barrier it is treated in a continuum approach as the free energies are based on mean parameter values of the local ensemble.**

The activation energy for reaching the *TS* can be written as, $\delta g_{pq,i}^{TS} = \delta g_{pq,i}^{TS,ERS} - \delta\mu_{p-ERS,i}$, where $\delta g_{pq,i}^{TS,ERS}$ is the activation energy for a *p* to *q* transition, and $\delta\mu_{p-ERS,i}$ is the chemical potential with respect to the *ERS*. The mean flux of atoms in the state *p* crossing the *pq* boundary per unit area (or length) is given by:

$$\Gamma_{pq,i} = \begin{cases} \Xi_{pq,i}\bar{c}_{p,i} \exp\left(-\dfrac{\delta g_{pq,i}^{TS,ERS} - \delta\mu_{p-ERS,i}}{k_B T}\right) & if \quad \delta g_{pq,i}^{TS,ERS} \geq \delta\mu_{p-ERS,i} \\ \Xi_{pq,i}\bar{c}_{p,i} & if \quad \delta g_{pq,i}^{TS,ERS} \leq \delta\mu_{p-ERS,i} \end{cases} \tag{7}$$

where $\Xi_{pq,i}$ is a 'single atom flux' prefactor accounting for the number of attempts per atom to pass from the *p* state to the *TS* between *p* and *q* per unit time and unit area. $\bar{c}_{p,i}$ is the normalized density of group *i* atoms in state *p*, i.e. the probability of having an atom in the state. When $\delta g_{pq,i}^{TS,ERS} < \delta\mu_{p-ERS,i}$, the transition is considered to be barrier-free. The form of $\Xi_{pq,i}$ can be very different depending on the type of transition. If the *p* state is part of a condensed state (*a, l or s*), the prefactor can be written as, $\Xi_{pq,i} = Z_{pq,i}\nu_{p,i}$, where $Z_{pq,i}$ is the steric factor[55] of the *p* to *q* transition per unit area and $\nu_{p,i}$ is a vibration frequency. For the gas states (*b or v*), we are only interested in the transitions to condensed states, and the prefactor can be written as, $\Xi_{b(v),i} = \frac{S_{b(v)q,i} f^{\perp}_{b(v),i}}{\bar{c}_{b(v),i}}$. Here $f^{\perp}_{b(v),i}$ is the effective flux of atoms/molecules from the *b*(or *v*) states impinging normal to the interface of the *q*=*l* (or *s*) states. In order to calculate the effective flux across a *pq* boundary, the backward *q* to *p* flux needs to be subtracted from the forward *p* to *q* flux, $\Delta\Gamma_{pq,i} = \Gamma_{pq,i} - \Gamma_{qp,i}$. Under *ERS* conditions, we can apply an equation of *detailed balance* (i.e. the net fluxes of material across a boundary equal zero, $\Delta\Gamma^{ERS}_{pq,i} = 0$), which implies that

$$\Xi_{qp,i} = \Xi_{pq,i} \frac{\bar{c}^{ERS}_{p,i}}{\bar{c}^{ERS}_{q,i}} \exp\left( -\frac{\Delta g^{TS,ERS}_{pq,i}}{k_B T} \right) \qquad (8)$$

with $\Delta g^{TS,ERS}_{pq,i} = \delta g^{TS,ERS}_{pq,i} - \delta g^{TS,ERS}_{qp,i}$ (Note that if the *ERS* transition state barrier is symmetric the exponential simply vanishes, as would be the case for a reversible state transition without requirements for dissociation/formation of bonds only one way). This is a general consequence of the detailed balance assumption when merging thermodynamics and transition state kinetics. The detailed balance provides an equilibrium relation between the ratios of coordination factors, attempt frequencies, possibly asymmetries for transition barriers at fixed *ERS* compositions. Finally, using eq.(7) and eq.(8), the net transition flux across the *pq* boundary is given as

$$\Delta\Gamma_{pq,i} = \Xi_{pq,i} \exp\left( -\frac{\delta g^{TS,ERS}_{pq,i}}{k_B T} \right) \left( \bar{c}_{p,i} \exp\left( \frac{\delta\mu_{p-ERS,i}}{k_B T} \right) - \frac{\bar{c}^{ERS}_{p,i}}{\bar{c}^{ERS}_{q,i}} \bar{c}_{q,i} \exp\left( \frac{\delta\mu_{q-ERS,i}}{k_B T} \right) \right) \qquad (9)$$

As in eq.(7), if $\delta g^{T,ERS}_{pq,i} \le \delta\mu_{p-ERS,i}$, i.e. $\exp\left( -\frac{\delta g^{TS,ERS}_{pq,i} - \delta\mu_{p-ERS,i}}{k_B T_p} \right)$ is set to one. The entropy in the first exponential can be put into a new prefactor, $\Xi'_{pq,i} = \Xi_{pq,i} \exp\left( \frac{\Delta s^{TS,ERS}_{pq,i}}{k_B} \right)$, that can be used as a temperature independent fitting parameter.[56]

To keep track of the atomic movements involved in the axial NW growth, a mass transfer equation are used to describe the atomic flow to and from the liquid phase,[22]

$$\frac{d}{dt} N_l = I_{III} + I_V - I_{inc} \qquad (10)$$

Here the liquid sorption currents $I_i$ of group *i* atoms,

$$I_i = \int \Delta\Gamma_{al,i} dl_{TL} + \int \Delta\Gamma_{(vb)l,i} dA_{vl} \qquad (11)$$

describe the effective 'adatom to liquid' and 'gas to liquid' currents. $I_{inc}$ is the effective atomic incorporation current from the liquid into the solid, $N_l$ is the number of atoms in the liquid, $l_{TL}$ is the triple line (TL) length and $A_{vl}$ is the projected liquid-vapor surface area[57]. If the equilibrium vapor pressure, $p_i^{eq}$, of a large liquid phase with a given composition is known (see section 3.1), the liquid to vapor transition rate from a liquid in such a state must fulfill the criteria, $\Gamma_{lv} \le \frac{p_i^{eq}}{\sqrt{2\pi m_i k_B T}}$, simply due to mass conservation. However, this criteria may be violated when size effects play an important role. Following the transition state approach, a simple version (sufficient in most cases) would be to assume no transition state barrier for sorption and a single vapor species for each element:

$$\Delta\Gamma_{(vb)l,i} \cong f_{i,\perp} - \frac{x_i}{x_i^{ERS}} \frac{p_{i_n}^{ERS}}{\sqrt{2\pi m_{i_n} k_B T}} \exp\left(\frac{\delta\mu_{l-ERS,i}}{k_B T}\right) \qquad (12)$$

where $f_{i,\perp} = f_{b,i,\perp} + f_{v,i,\perp}$ is the effective impinging flux of group *i*. For typical growth conditions where $f_V > f_{III}$, the vapor pressure of group V can be assumed to be proportional to the incoming flux, $f_{v,V} \propto f_{b,V}$. This is because a huge contribution of the excess As species must come from secondary adsorption, see Ramdani et al.[23]. Secondary adsorption of group III can typically be neglected, although for growth on substrates covered with a thermally grown oxide layer it can play a significant role, as shown by Rieger et al.[58]. The *va* and *al* transition flux can be written respectively as,

$$\Delta\Gamma_{(vb)a,i} \cong f_{i,\perp} - \frac{\rho_i}{\rho_i^{ERS}} \frac{p_{i_n}^{ERS}}{\sqrt{2\pi m_{i_n} k_B T}} \exp\left(\frac{\delta\mu_{a-ERS,i}}{k_B T}\right) \qquad (13)$$

$$\Delta\Gamma_{al,i} = \Xi'_{la,i} \exp\left(-\frac{\delta h_{la,i}^{TS,ERS}}{k_B T}\right)\left(\frac{\bar{\rho}_i}{\bar{\rho}_i^{ERS}} x_i^{ERS} \exp\left(\frac{\delta\mu_{a-ERS,i}}{k_B T}\right) - x_i \exp\left(\frac{\delta\mu_{l-ERS,i}}{k_B T}\right)\right) \qquad (14)$$

Finally, the net sorption currents (eq.(11)) are given as,

$$
\begin{aligned}
I_{(vb)l,i} &= \int \Delta\Gamma_{(vb)l,i} dA_{vl} \cong A_{vl}\left( f_{v,\perp} - \frac{x_i}{x_i^{ERS}} \frac{p_{i_n}^{ERS}}{\sqrt{2\pi m_{i_n} k_B T}} \exp\left( \frac{\delta\mu_{l-ERS,i}}{k_B T} \right) \right) + A_{vl}^{'} f_{b,i} \\
I_{al,i} &= \int \Delta\Gamma_{al,i} dA_{vl} \cong L_{TL}\Xi_{al,i}\left( \bar{\rho}_i - \bar{\rho}_i^{ERS} \frac{x_i}{x_i^{ERS}} \exp\left( \frac{\delta\mu_{l-ERS,i}}{k_B T} \right) \right)
\end{aligned} \quad (15)
$$

In eq.(15) all information about the transition state barriers from the $l$ to the $v$ or $a$ states is stored in the *ERS* parameters, due to the detailed balance assumption at equilibrium. Only a given projection of the liquid surface $A_{vl}^{'}$ is exposed to the incident beam flux, depending on the beam direction and droplet geometry[57]. $L_{TL}$ is the length of the *TL*. Note that if $\delta g_{lq,i}^{TS,ERS} < \delta\mu_{l-ERS,i}$, the exponentials vanish in eq.(15) according to eq.(7).

To get a more intuitive feeling of the effect of growth conditions on the adatom kinetics in terms of an effective diffusion length, adatom migration on a large homogenous planar interface serves as a good example.[59] Even though this approach is not accurate for modeling the growth dynamics, it is instructive and intuitive, and sufficient to understand many overall growth phenomena as function of growth conditions. There are three main transition paths for an adatom, namely surface diffusion (*aa*), desorption (*av*) and incorporation (*as*). Using the *TS* approach, see appendix 5.1, the diffusion length of an adatom on a surface $j$ can be written as

$$
\lambda_{j,i} \cong \sqrt{\bar{Z}_{aa,i}^{'} l_{a,i}^{2} \exp\left( -\frac{\delta h_{aa,i}^{TS,ERS}}{k_B T} \right) \left( \bar{Z}_{as,i}^{'} \exp\left( \frac{\delta\mu_{a-ERS,i}}{k_B T} \right) + \bar{Z}_{av,i}^{'} \exp\left( -\frac{\delta h_{av,i}^{TS,ERS}}{k_B T} \right) \right)^{-1}} \quad (16)
$$

where we assume that $1-\bar{\rho}_{j,i} \approx 1$, and that the density of incorporation sites is given as $\bar{c}_{inc,III(V)} \propto \exp\left( \frac{\delta\mu_{a-ERS,i}}{k_B T} \right)$. $l_{a,i}$ is the lattice site spacing and the entropy change is included in the prefactors, $\bar{Z}_{pq,i}^{'} = \bar{Z}_{pq,i} \exp\left( \frac{\delta s_{pq.i}}{k_B} \right)$. In Figure 5 we show estimations of diffusion lengths as a function of growth conditions, using parameters given in Appendix 5.3.

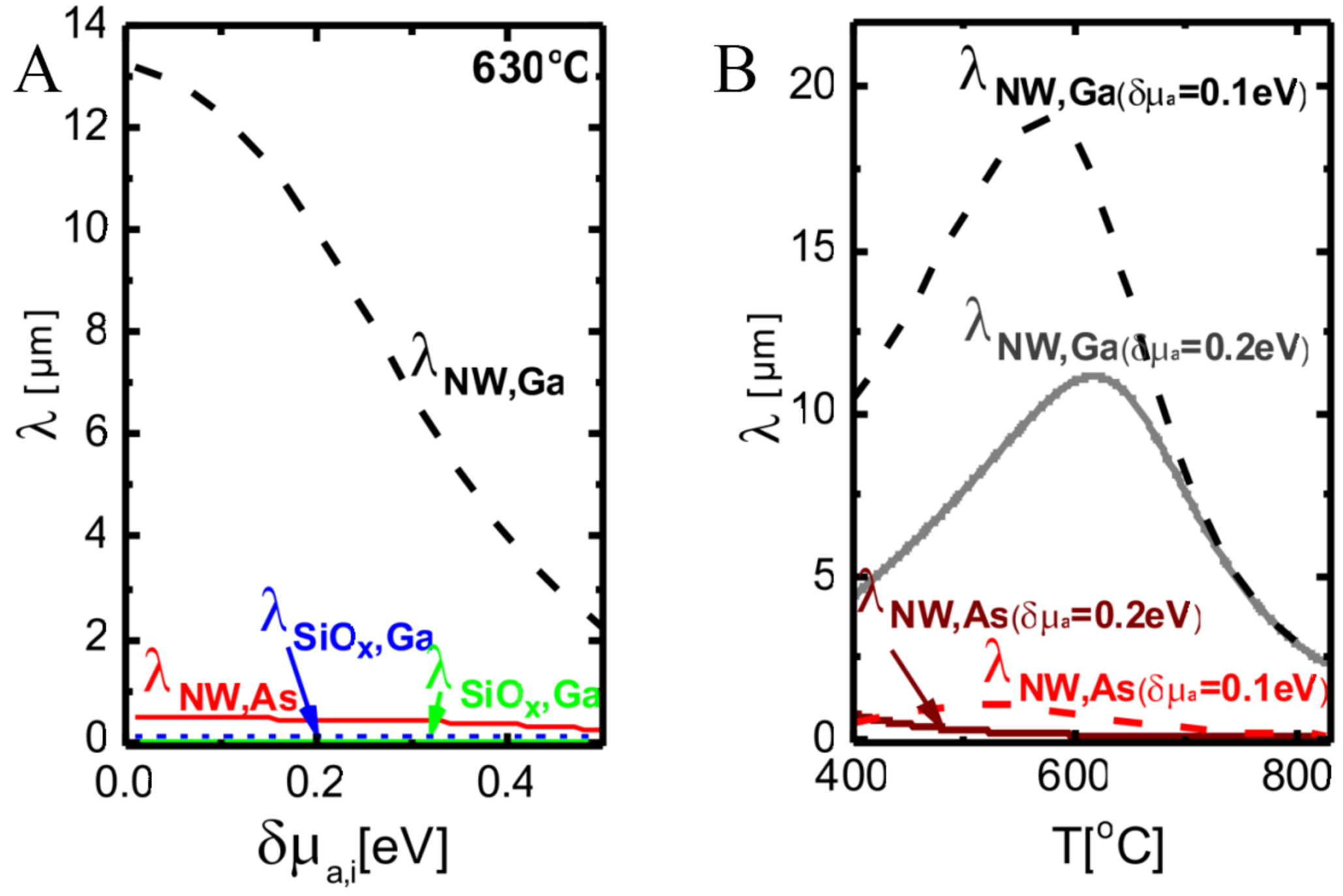

**Figure 5 Diffusion length estimations for uniform diffusion of Ga and As on the NW sidefacets and thermal oxide at $T = 630°C$, using activation enthalpies listed in Appendix 5.3. It is seen that on the oxide surface the diffusion length is independent of the chemical potential because it is in the desorption limited regime where the chemical potential does not play a role according to eq.(16). But the diffusion length for Ga adatoms on the crystalline facets (here (110) sidefacets) depends very strongly both temperature and chemical potential, see appendix 5.1.**

Above we have treated the static case of adatom diffusion. An approach to treat the dynamics of adatom diffusion and the adatom collection to the liquid phase is discussed in detail in appendix 5.2, where we show how to merge a ‘Dubrovskii/Johansson’ static diffusion scheme into the dynamic formalism using the *TS* kinetics with a uniform diffusivity along NW and substrate, a method which is used for the modeling in section 3.

## 2.2 The liquid-solid phase transition

We now turn to the actual crystal formation at the *ls* interface. Here, we will for simplicity assume that the liquid diffusion is fast on the time scale of NW growth, and that the liquid phase is homogenous. The dynamic treatment including non-homogenous liquids can be carried out if a reference composition and liquid diffusivity are known. See for example Jong et al.[60] and Connell et al.[61] for treatment of non-homogenous liquids. The possibility of fast diffusion along the growth interface during VLS is assumed negligible, as indicated by a study by Dick et al.[62]. As shown by Schwarz and Tersoff[45], if the solid were isotropic, the equilibrium shape would be with a curved $ls$ interface. But as the authors also pointed out in a later publication, in the anisotropic case (which is relevant for III-V NW growth), the morphology is strongly faceted[46]. It is complicated to treat the dynamical evolution if the solid is partially wetted by a droplet which at the same time is changing in size during growth. For such a system the preferential orientations of the facets depends on the liquid phase size and it is necessary to describe the crystal growth in terms of both facet sizes and facet orientations (and therefore an independent parameter set $\{X\}$ of both 'areas' and 'orientations', as explained in section 2.1). An additional complication affects the evolution of the crystal shape if the facets are limited in their growth rate by the formation of a small nucleus[63], see section 2.3 for a treatment of the nucleation limited axial growth at the topfacet. For VLS growth one only considers growth at the $ls$ interface and distinguishes between two types of $ls$ transitions:

**Nucleation free growth**: Facets which are limited in their growth rate or in their change of orientation by the transfer of single pairs to the growth front, as described by eq.(9).

**Nucleation limited growth**: Facets which are limited in their growth rate by the formation of a small nucleus, or more generally limited in their change of $X$ due to an energy barrier which is larger than the single pair transition state barrier.

As in ref.[26], the $ls$ growth system will be divided into two main regimes (mainly due to traditional reasons as explained below).

**Regime *I***: The TL stays in contact with the topfacet.

**Regime *II***: The TL is not in contact with the topfacet, or possibly only for a short time during a nucleation event at the topfacet.

The vast majority of literature on the nucleation at the topfacet has assumed an ideal regime *I*, where the $ls$ interface is perfectly flat, see for example ref.[25]. However, it is very uncertain under which material systems and growth conditions ideal regime *I* conditions applies. It is likely that it is only relevant under non steady state conditions where the liquid decreases significantly in size, such as immediately after closing the shutter of the group III source or upon during cool down where the nucleation barrier is lowered.[27] But as recent in-situ TEM experiments[40,41,42] strongly suggests and as shown in the modeling examples in section 3.5, regime *II* may be a dominant *VLS* steady state growth mode.

Many studies suggest that the dominating type of growth at the topfacet is strongly nucleation limited (see for example ref.[67]) while small truncation facets at the edges of the growth interface might be nearly nucleation free[41,42]. The chemical potential of the solid depends on the stacking type of the crystal structure (e.g. $s$: WZ(2H), ZB(3C), 4H, 6H, ect.), with ZB and WZ being the most common sequences, where the ZB structure has the lowest cohesive energy for most III-V's and are therefore favored in bulk materials.[64,65] The liquid needs to reach a critical level of supersaturation (typically of the order of a few hundreds of meV per III-V pair)[25] before the nucleation barrier at the topfacet can be overcome. Under this constraint other facets which are not nucleation limited will reshape in respond to the elevated liquid chemical potential at a rate determined by eq.(9), and the whole growth system is therefore in a configuration far from equilibrium. For *VLS* growth, group V is typically the less abundant specie in the liquid i.e. $x_{III} > x_V$. For a fixed solid stoichiometry, the activation energy for the nucleation free single pair *ls* transitions, eq.(9) can be written as,

$$\Delta\Gamma^X_{ls,III-V} = \Xi_{ls,III-V} \exp\left(-\frac{\delta g^{TS,ERS}_{ls,III-V}}{k_B T}\right)\left(x_V \exp\left(\frac{\delta\mu_{l-ERS,III-V}}{k_B T}\right) - x_V^{ERS} \exp\left(\frac{\delta\mu^X_{s-ERS,III-V}}{k_B T}\right)\right) \quad (17)$$

As the liquid chemical potential $\delta\mu_{l-ERS,III-V}$ is an oscillating function due to the nucleation limited growth at the topfacet[67,] the parameter $X$ (describing nucleation free facet size or angle) will therefore oscillate accordingly. Because the chemical potentials depend on location and system morphology, so do the transition fluxes, and the free energy minimization needs to be described with respect to an appropriate set of independent parameters, $\{X(\omega)\}$. Generally speaking, the larger the parameter set the more accurately the modeling, but also the more computations are needed. In three dimensions, the chosen set of parameters $\{X(\omega)\}$ will depend on $\omega$ which is defined to be the angle between the middle of the sidefacet and position as measured from the center of the top facet, see ref.(26) for clarification.

As shown in the stereographic projection in Figure 6(A), if only considering ZB and WZ stacking, it is sufficient to divide the $\omega$-dependence of the crystal into 3 sections because the ZB crystal structure has 3-fold symmetry. The WZ crystal structure has 6-fold symmetry around the growth axis and is therefore also described completely within this region. In Table 1 in Appendix 5.3, we give the interfaces with lowest energies for the ZB and WZ structure (we restrict ourselves to the upper half hemisphere with polar (111) ZB or [0001] WZ directions). To describe the NW diameter as a function of $\omega$ in terms of the cross sectional Wulff shape[66], we need to look at the energies of the facets in the $\theta = 90°$ plane (the outer ring) in the stereographic projection in Figure 6 (A). For a cross sectional six-fold symmetric NW it is enough to describe the NW diameter at the growth interface in the range $\omega = [-30° : 30°]$ as

$$d_{NW}(\omega) = \frac{d_{NW}(\omega = 0^\circ)}{\left(1 + \omega\eta(\omega)\right)\cos(\omega)} \quad (18)$$

where the function $\eta(\omega)$ determines the cross sectional shape of the growth interface. $\eta(\omega)$ is a complicated function that depends on many factors. We can simplify it as $\eta(\omega)=\eta_0\left(\cos^{-1}(\omega)-1\right)\omega^{-1}$, where $\eta_0=0$ for complete hexagonal facetting and $\eta_0=1$ in the isotropic case (complete axi-symmetric cross section). In the case of the ZB structure which has a three-fold symmetric crystal structure, it is very likely that the NW cross section does not have a perfect six-fold geometrical symmetry. In this case we need to take account of the possibility of a three-fold symmetric cross section where the diameter is given by $d_{NW}(\omega)=r_+(\omega)+r_-(\omega)$ with $r_+(\omega)$ and $r_-(\omega)=r_+(\omega+180°)$ being the radius as measured from the center of the NW crystal. For complete facetting ($\eta_0=0$) and a constant NW volume, the relation between $r_-$ and $r_+$ is:

$$r_+=2r_- -\sqrt{3r_-^{\,2}-2\sqrt{3}\cdot A_c} \tag{19}$$

where $A_c$ is the cross sectional area. According to Wulff, in the absence of a liquid phase, the cross sectional equilibrium shape of the NW crystal would be given by $\frac{\gamma_A}{\gamma_B}=\frac{r_+}{r_-}$, where $\gamma_A(\gamma_B)$ is the effective vertical surface energy of the facet normal to the $r_-(r_+)$ vector.

For a more complete description of the dynamics we need the values of the anisotropic surface and interface energies for the different crystal structures. To carry out the iterative minimization of the free energy are desirable. To this end, we need a $\gamma$ plot with rounded cusps that can approach arbitrarily close to the sharp cusps of faceted orientations. This can be realized by summing a set of 2D Lorentzian functions centered on the facets of high symmetry, which have the lowest interface energies. The angular dependence of the interface energy is then described, in angular coordinates $(\theta,\omega)$, by:

$$\gamma_{vs,j}(\theta,\omega)=\gamma_{vs0}-\sum_{hkl}c_{hkl}\frac{\mathrm{I}_{hkl}}{1+\left(\frac{\phi_{hkl}(\theta,\omega)}{w_{hkl}}\right)^2} \tag{20}$$

where $\phi_{hkl}(\theta,\omega) = \arccos\left(\cos(\theta-\theta_{hkl}) + \sin(\theta)\sin(\theta_{hkl})\left(\cos(\omega-\omega_{hkl})-1\right)\right)$ is the angle between the facet $hkl$ (see table 1) and direction $(\theta,\omega)$, where $\theta = 0$ corresponds to the growth direction (see Figure 6 (B) and (C) for ZB and WZ structure).

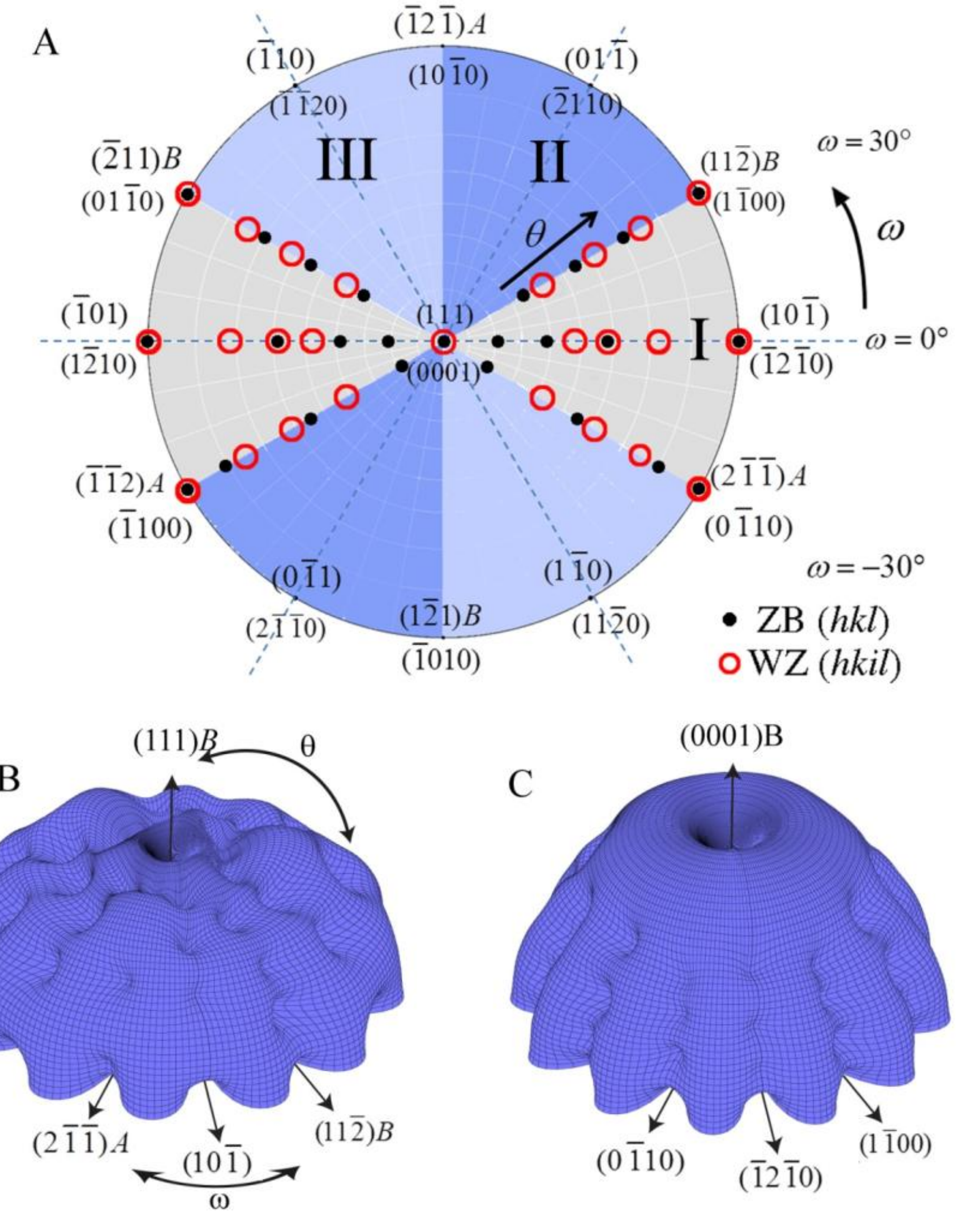

**Figure 6. III-V NW crystal anisotropy for ZB and WZ structures. (A) A stereographic projection of the upper hemisphere along the [111] ([0001]) zone axis of a ZB (WZ) crystal. Due to the 3-fold symmetry of the ZB structure along (111), we only need to consider the grey areas, which are described in the range $\omega = [-30°; 30°]$. The black dots represent the facets $(hkl)$ with the lowest predicted energy of the ZB structure and the red rings represent the corresponding facets $(hkil)$ of the WZ structure. The edge of the projection represents the plane normal's perpendicular to the growth axis, $\theta = 90°$. Lower hemisphere orientations are found by mirroring the upper hemisphere orientations in the zone axis and change sign of the miller indices. The specific angles shown are given in Appendix 5.3. 3D gamma plot in spherical coordinates $(\theta,\omega)$ of the anisotropic *ls* interface energy for (B) ZB and (C) WZ structure using eq.(20) with the lowest miller index facets in the 12 directions between the (111) growth direction and {1-10} or {11-2} families (see Appendix 5.3). The distance between origo and the surface is proportional to the interface energy of the given orientation.**

In eq.(20), the maximum interface energy is noted $\gamma_{vs0}$, and the decrease in interface energy at each high symmetry facet is given by the 'intensity' $I_{hkl} = \gamma_{vs0} - \gamma_{vs,j}$. The values of $\gamma_{vs,j}$ for the main orientations can either be found in the literature or obtained from density functional theory calculations. $w_{hkl}$ is a scale parameter which specifies the half-width at half maximum of the energy increase around the $(hkl)$ facet. $c_{hkl}$ is a constant close to unity, but if $w_{hkl}$ is large the interface energy may have to be adjusted to a value slightly lower than unity because the contributions from adjacent facets may overlap. We will simply assume that the *ls* interface energy is given by $\gamma_{ls}(\theta,\omega) = \sigma\gamma_{vs}(\theta,\omega)$, where $\sigma$ is typically assumed to be a constant of the order $0.3-0.5$.

## 2.3 Nucleation limited axial growth in the (111)/(0001) direction

We will here treat the nucleation limited growth which takes place at the *ls* top facet separately because this is where the axial growth and where the final crystal structure of the NW is formed. Many recent experimental studies have indicated that growth on the dominating *ls* (111)/(0001) top facet is limited by the formation of a nucleus, which means that the liquid supersaturation needs to exceed a certain critical value before a new monolayer can be formed, see for example ref.[39] and ref.[67]. This implies that the topfacet is stabilized as long as the difference in chemical potentials between the liquid and topfacet is smaller than a critical value, due to large activation energies both ways. Because the mother phase (the liquid) is small, the liquid supersaturation drops far below the critical level after a ML formation and probability of having a subsequent second nucleation is unlikely. We are therefore only interested in single nucleation events. To describe the probability of forming a critical nucleus we need to take account of the stochastic nature of the phase fluctuations which causes nucleation. But first, we need an expression for the mean nucleation rate.

If the movement of atoms in and out of clusters of various sizes (smaller than the critical nucleus) at the growth interface, takes place on a timescale much smaller than the time between nucleation events, the nucleation probability can be derived assuming steady state nucleation rate conditions[68,69,70], which is the typical assumption in NW growth theory[10,19,20,25]. It is reasonable to assume that the attachment/detachment frequency of III-V pairs to and from the clusters on the (111)B topfacet is limited by the group V elements. This is not only because the concentration of group V is low in the liquid but also because the group III elements are attached with only one covalent bond on average in the 'B' terminated surface when group V is absent. Once group V is present, the pair is stabilized leaving only one free covalent bond per pair on average (re-construction is not considered in the continuum formalism). With this, the mean nucleation rate at given site with coordinates $(r,\omega)$ at the top facet ($r$ measured from the center) can then be written as

$$j_{s(r,\omega)} = A_{n^*} Z_{s(r,\omega)} c_1 \Xi_{ls,III-V} x_V \exp\left(-\frac{\delta g_{ls,\text{int}}^{TS,ERS}}{k_B T}\right) \exp\left(-\frac{\Delta G_{n^*,s(r,\omega)}}{k_B T}\right) \qquad (21)$$

where $A_{n^*}$ is the step area of the critical nucleus of $n^*$ pairs, $Z_{s(r,\omega)} = \frac{1}{n^*}\sqrt{\frac{\Delta G_{n^*,s(r,\omega)}}{4\pi kT}}$ is the 2D Zeldovich factor and $\Delta G_{n^*,s(r,\omega)} = -\sum_{i=2}^{n^*}\left(\delta\mu_{l-ERS,III-V} - \delta\mu^i_{s(r,\omega)-ERS,III-V}\right)$ is the formation free energy of the nucleus, with $\delta\mu^i_{s(r,\omega)-ERS,III-V}$ being the chemical potential of a cluster of $i$ pairs at $(r,\omega)$. $s$ denotes solid structure described by its stacking type (*ZB (3C), WZ (2H), 4H ...*). Consistent with transition state approach described above, the forward flux from the liquid to the cluster, $\Gamma_{ls,III-V}$, is assumed independent of the size of the cluster, (the backward flux from the clusters depends on the cluster size but cancels out in the derivation). $\delta g_{ls,\text{int}}^{TS,ERS}$ is the transition state barrier for attachment of a single pair to the clusters at the interface. The detailed kinetics at the interface is unknown; we thus simply assume that the concentration of single III-V pairs attached to the interface $c_1$ (single pair clusters) is equal to the concentration of the group V in the liquid, $c_1 \approx x_V$. Once the nucleation event has occurred, the ML is completed in a non-nucleation limited manner, at a rate given by eq.(17), and the liquid supersaturation builds up slowly again until the next nucleation event takes place.

The nucleus formation free energy can be written as in a more familiar form,

$$\Delta G_{n^*,s(r,\omega)} = -\Delta\mu^{\infty}_{ls,III-V} n^* + h\sum_{k=1}^{m} l_k \gamma_{step(r,\omega),k} \qquad (22)$$

where the first term is the formation free energy required to form the volume part of the nucleus. The second term is the excess free energy due to the formation of a dividing step. $\gamma_{step(r,\omega),k}$ and $l_k$ are the free energy and length of the $k^{\text{th}}$ step facet, respectively. As the nucleation takes place when the number of pairs in the cluster exceeds the critical value, $n \geq n^*$, which is associated with the maximum free energy increase given by the condition,

$$\frac{d\Delta G_{n^*,s(r,\omega)}}{dn} = -\Delta\mu^{\infty}_{ls,III-V} + h\sum_{k=1..m}\left(\frac{dl_k}{dn}\gamma_{step(r,\omega),k} + l_k\frac{d\gamma_{step(r,\omega),k}}{dn}\right) = 0 \qquad (23)$$

we can derive an explicit expression for the nucleation barrier $\Delta G_n^*$ by extracting $n^*$ from eq.(23) ($l_k$ depends on $n^*$) and insert it into eq.(22). The last term in the summation of eq.(23) is typically neglected in continuum models, as the interface energies are assumed constant as a function of interface area.

For regime *II* we will divide all the possible nucleation sites into three main classes (see Figure 7).

**A)** At the edge between the topfacet and truncated facet, see ref.[42]. Here the nucleus forms an extension to the truncated facet a crystal structure different from the equilibrium bulk structure can be dictated by the orientation of this facet (similarly to what was proposed in the case of a nucleus in contact with a vapor by Glas et al.[25]).

**B)** Nucleation at the center of the top facet, see for example ref.[19]. The preferential crystal structure here is the structure with the lowest cohesive energy which is typically ZB.[71]

**C)** It is possible that the truncation size becomes positive at a given $\omega$, before one to the other types of nucleation events takes place. Then a TL nucleation event will be induced at the topfacet and the necessary step for step flow is formed. A fast completion of the monolayer will lower the supersaturation and move the truncation back to negative values (provided that the barrier of forming the truncation facet is small enough). For a six-fold crystal geometry it is likely that such an event will take place at the corners, i.e. $\omega = 30°$. If the liquid size is decreasing TL nucleation becomes more and more dominant and the system will eventually move into regime *I*.

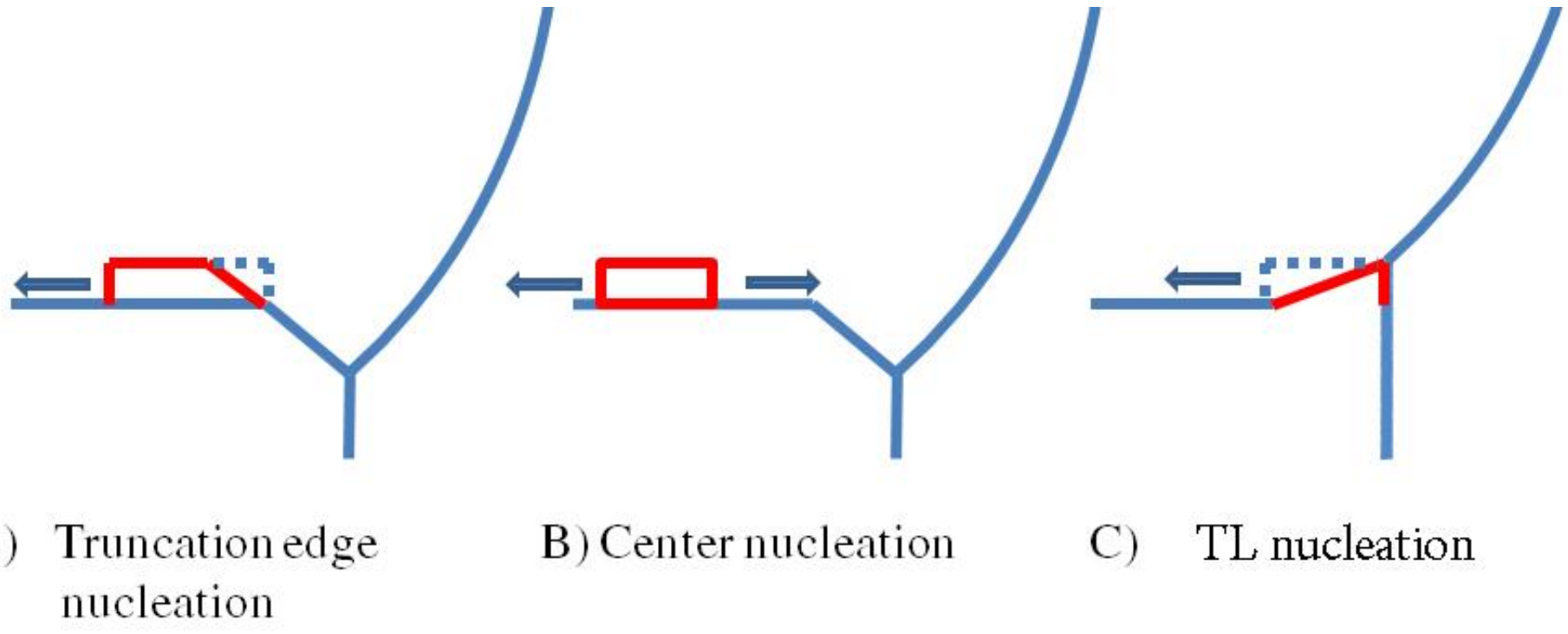

**Figure 7. Cross section view on the triple line region at a given $\omega$, showing three different ways to form an energetically favorable step on the topfacet. A) A step formed due to a nucleation event at the corner between the topfacet and a truncated facet, a regime II type nucleation. B) A step formed due to a nucleation event at the center of the top facet. C) If the relative droplet size is sufficiently small and/or the liquid supersaturation is sufficiently high at nucleation, it is possible that the truncation size becomes positive which will induce a TL nucleation event at the topfacet and the necessary step for step flow is formed.**

Under conditions where the time needed to reach steady state composition in the liquid is smaller than the time between the formation of two consecutive MLs, the total center-nucleation rate can be written as $J_c \cong 2\pi j_c \int_{Center} \left( \frac{d_{NW}(\omega)}{2} - \Delta z(\omega) \tan\left(\theta_T(\omega)\right) - l_m^* \right) d\omega$, where $\Delta z(\omega) \tan\left(\theta_T(\omega)\right)$ is the decrease in topfacet length at $\omega$ due to the truncation. For truncation edge nucleation we will integrate over the part with a negative truncation, $J_T \cong \pi \int_{l_T} l_m^*(\omega) \left( d_{NW}(\omega) - \Delta z(\omega) \tan\left(\theta_T(\omega)\right) \right) j_T(\omega) d\omega$. In order to carry out a more realistic modeling of the $s$-stacking probabilities we can account for the stochastic nature of nucleation by multiplying eq.(21) by a random number between 0 and 1, $\Re(0,1)$, at each time step, and define a normalized value $\delta$ above which nucleation will take place

$$\int_{Topfacet} j_{s(r,\omega)} dA_{topfacet} \cdot \Re(0,1) \geq \delta \tag{24}$$

Finally, whenever one or more sites fulfill eq.(24) the rate of the subsequent step flow and completion of a ML are determined by eq.(17). However it is possible that the truncation ($\Delta z(\omega) \tan\left(\theta_T(\omega)\right)$) under certain conditions goes to zero and at certain positions becomes positive before eq.(24) is fulfilled (Figure 7C). In this case a step is naturally provided at the triple line (TL) and completion of a monolayer will take place at the same time as the truncation most likely goes to negative again due to a lowering of the liquid supersaturation.

# 3. Dynamical modeling examples of self-catalyzed GaAs NW growth

In this section we will show examples of how to use the theoretical formalism (presented in section 2) to analyze and understand the dynamics of self-catalyzed GaAs NW growth. We start with analyses of the evolution of the overall morphology of self-catalyzed GaAs NW growth on Si substrates (section 3.1-3.4).

## 3.1 Calculating the *ERS* and size effects for self-catalyzed GaAs NW growth simulations in the axi-symmetric approximation

To simulate a specific process such as self-catalyzed GaAs NW growth on Si (111) substrates, requires the relevant *ERS* parameters and size effects based on the assumptions made for the simulation. Thus, before giving detailed simulation examples of the overall NW growth, we will first go through the specific calculations needed for this system. As mentioned, modeling the overall morphology does not require detailed information on the shape of the *ls* interface, and in this section we will therefore assume an axi-symmetric cross section ( $\omega$ dependence can be neglected) and an ideal regime *I* with a single flat *ls* interface. In this case we do not need to define an independent parameter set, but only use the liquid size evolution and nucleation at the topfacet to determine the evolution of the crystal morphology in terms of the diameter $d_{NW}$ at the growth interface and the *vl* contact angle $\theta$ with respect to the topfacet. The size effect terms in eq.(5) and (6) for the chemical potential can be found using the trigonometric relations, $A_{ls} = \frac{\pi d_{NW}{}^2}{4}$, $A_{vl}(\theta) = \frac{\pi d_{NW}{}^2 \left(1-\cos(\theta)\right)}{2\sin^2(\theta)}$ and $N_l(d_{NW},\theta) = \frac{\pi d_{NW}{}^3}{24\Omega_l} \frac{\left(1-\cos(\theta)\right)^2 \left(2+\cos(\theta)\right)}{\sin^3(\theta)}$, where $\Omega_l$ is the atomic volume in the liquid. A change in $d_{NW}$ implies not only a change in the *ls* and *vl* areas but also the formation of a new *vs* area corresponding to the absolute change in the *ls* area. We have not taken account here of the possibility of wetting the sidefacets for a cylindrical shaped cross section, as described in ref.[72] and [73]. For a detailed analysis of the wetting in regime *I* (i.e. on a flat hexagonal top facet) see ref.[26]. To calculate the chemical potentials of the *ERS* (eq.(1)), we need to calculate the liquid chemical potentials when the liquid phase is in equilibrium with the solid. For liquid binaries (self-assisted growth), the chemical potential is given by the tangent method, or correspondingly;

$$\mu_{l,i}^{\infty}\left(x_V,T\right) = g_l^{\infty}\left(x_V,T\right) + (1-x_i)\frac{\partial g_l^{\infty}\left(x_V,T\right)}{\partial x_i} \qquad (25)$$

Here the liquid free energy per atom of an infinitely large binary alloy is given by, $g_l^{\infty}\left(x_V,T\right) = \left(1-x_V\right) g_{l,III}\left(T\right) + x_V g_{l,V}\left(T\right) + g_{l,mix}\left(x_V,T\right)$ where $g_{l,mix}(x_V,T) = \left(1-x_V\right)x_V\left[L_0(T) - L_1(T)\left(1-2x_V\right)\right] + RT\left[\left(1-x_V\right)\ln\left(1-x_V\right) + x_V\ln\left(x_V\right)\right]$ accounts for the asymmetry in the compositional effect on the free energy by using the Redlich-Kister formalism[74] as in ref.[75] with two liquid interaction parameters $L_0$ and $L_1$. These parameters together with the free energy values of the pure components $g_{l,i}$ are given for $g_{Ga_{1-x_V}As_{x_V}}$ and $g_{In_{1-x_V}As_{x_V}}$ in Table 2 of Appendix 5.3, where the equilibrium concentrations are estimated from fitting the liquidus values reported in ref. [95]. All Gibbs free energies and chemical potentials are relative to the enthalpy of the standard element reference ($HSER_i$),[75] and denoted $g_l{}'(T)$ and $\mu_{l,i}{}'(T)$, respectively. Using these data, the *ERS* chemical potential $\mu_i^{ERS}\,' = \mu_{l,i}^{\infty}\,'(x_V^{ERS}, T)$ is calculated using eq.(25) and the relative chemical potential is simply $\delta\mu_{l,i}(x_V, T) = \mu_{l,i}\,'(x_V, T) - \mu_i^{ERS}\,'$.

To calculate the partial vapor pressures over a liquid of given composition, we note that $\mu_{v,i_n}\,'^{n} = n\mu_{l,i}\,'$, where $n$ is the number of atoms in the molecule considered and $'^{n}$ denotes that the value is given with respect to $n$ times the standard reference. Using the thermodynamic data from appendix 2 in Ansara et al.[75], we find an expression for the Gibbs free energy of a pure $i_n$ species, $g_{v,i_n}^{pure}\,'^{n}(T) = \Psi_i\,'^{n}(T) + RT\ln(P)$, where $P$ is the total pressure and $\Psi_i\,'^{n}(T)$ is a function of temperature only (see Table 3 in Appendix 5.3 for the thermodynamic data). Now, since $\mu_{v,i_n}\,'^{n} - \mu_{v,i_n}^{pure}\,'^{n} = RT\ln\left(\frac{p_{i_n}}{P}\right)$, where $\mu_{v,i_n}^{pure}\,'^{n} = g_{v,i_n}^{pure}\,'^{n}$, we can write the following expression for the vapor pressure of element $i_n$,

$$p_{i_n}(x_V, T) = \exp\left(\frac{n\mu_{l,i}\,'(x_V, T) - \Psi_i\,'^{n}(T)}{RT}\right) \tag{26}$$

The corresponding ERS pressures are then found by setting $\mu_{l,i}\,'(x_V, T) = \mu_i^{ERS}\,'$.

From Figure 8A we see that the only species that may have significant partial pressures are the Ga and $As_2$ species. As the liquid supersaturation increases (increasing $x_{As}$), the vapor pressure of Ga remains almost constant. This means that the *vl* and *al* transition fluxes for the Ga species are roughly independent of the supersaturation. On the other hand, as the supersaturation increases the desorption of the As species increases very strongly (note the log scale).

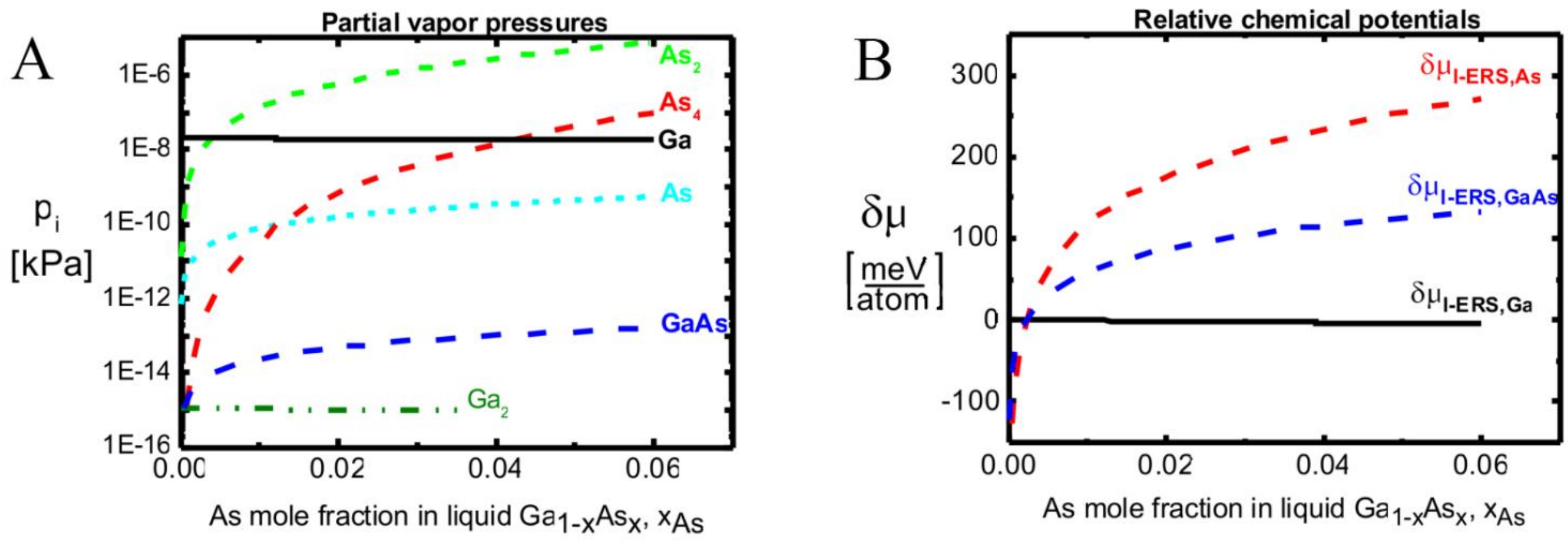

**Figure 8. Partial vapor pressures (A) and relative liquid chemical potentials (B) of the relevant species in the liquid Ga-assisted case for GaAs NW growth, as a function of the As mole fraction at $T = 630°C$. The critical value $\delta\mu^{c}_{l-ERS,III-V}$ is typically of the order 100meV per atom which corresponds to few percent of As in the liquid as shown in (B). The As concentration is kept low in the liquid due to the fast increasing vapor pressure of $As_2$, and there exists a certain threshold value of beam flux/vapor pressure where the steady state concentration of As in the liquid exceeds the critical value for nucleation at the topfacet.**

To complete the *ERS* description, we need to calculate the adatom densities, $\rho^{ERS}_{NW,i}$ and $\rho^{ERS}_{sub,i}$, which we do by using kinetics. For the adatom collection we follow the approach outlined in Appendix 5.2, and the ERS adatom densities are found using eq.(36) under ERS conditions, where $\rho^{ERS}_{NW,i}$ is calculated by setting $L_{NW} \to \infty$ and $\Delta\Gamma_{al,i} \to 0$, and $\rho^{ERS}_{sub,i}$ is found by setting $r \to \infty$, both under conditions of the calculated *ERS* beam fluxes found above. Using the parameters listed in appendix 5.2, the *ERS* adatoms densities are $\rho^{ERS}_{NW,Ga}\left(T = 630°C\right) = 5.3 \cdot 10^{17} m^{-2}$ and $\rho^{ERS}_{NW,As}\left(T = 630°C\right) = 0.16 \cdot 10^{17} m^{-2}$.

Tuning the fitting parameters can be time consuming. The fitting values of the relevant prefactors and activation free energies for adatom desorption and incorporation used in the simulations presented below are given in appendix 5.3. In order to use the diffusion lengths given by eq.(16), we need estimates of the activation energies $\Delta g^{TS,ERS}_{pq,i} = \Delta h^{TS,ERS}_{pq,i} - T\Delta s^{TS,ERS}_{pq,i}$. As the entropy change as a function of temperature is negligible compared to the enthalpy change, we include the entropy contribution into the temperature independent prefactors as $\bar{Z}'_{as,i} = \bar{Z}_{as,i} \exp\left(\frac{\Delta S^{ERS}_{pq,i}}{k_B}\right)$. This leaves us with enthalpy barriers which can be estimated from zero temperature ab initio calculations such as Density Functional Theory methods.[91] After having built up the simulation framework, it can be used to analyze a variety of features and systems. Here, we will only give a few examples.

## 3.2 Dynamics of self-catalyzed GaAs NW growth on Si(111) at low V/III ratios

For typical *MBE* growth of self-assisted GaAs NWs on a Si (111) covered with a thin native $SiO_x$ layer, Ga beam fluxes corresponding to planar growth rates of $0.1-0.3\frac{\mu m}{hr}$ are commonly used, with a V/III flux ratio in the range $5-100$ and a substrate temperature around $T = 630^{\circ}C$ [76,77]. There exists a certain "growth parameter window", namely ranges of values for the basic growth parameters (temperature and beam fluxes), where it is possible to obtain NW growth (as a rule of thumb, the higher the temperature the higher the V/III ratio[22]). A general feature of the simulations is that there are sharp and well defined boundaries for the growth parameter window. As the critical liquid supersaturation needed for nucleation at the topfacet is almost independent of the applied pressures (beam fluxes)[26], the axial growth rate is simply dictated by the time it takes for the liquid to reach the critical concentration of As, $\delta\mu_{l-ERS,As}^{crit}$, after being lowered upon a nucleation event and subsequent ML formation. If we neglect for simplicity the surface diffusion of As species and account for the impinging $v$ states by simply using that the beam flux hits the total vl interface, the minimum As flux needed to obtain growth is roughly given as:

$$f_{(bv)l,As,\perp}^{crit} \approx \frac{x_{As}^{crit}}{x_{As}^{ERS}} f_{lv,As}^{ERS} \exp\left(\frac{\delta\mu_{l-ERS,As}^{crit}}{k_B T}\right) \tag{27}$$

Here $x_{As}^{crit}$ is the critical concentration of As needed for a nucleation event and $f_{lv,As}^{ERS} \approx \sum_n \frac{p_{As_n}^{ERS}}{\sqrt{2\pi m_{As_n} k_B T}}$ is the flux of material evaporating from the liquid under *ERS* conditions. This means that the critical As flux is strongly dependent on the nucleation barrier and is only very little dependent on the Ga flux as long as there is a large liquid Ga phase. For the simulation shown in Figure 9 (A), the critical impinging As flux needed to overcome the nucleation barrier is roughly $f_{(bv)l,As,\perp}^{crit} \approx 100 \cdot f_{lv,As}^{ERS}$. To examine how the axial growth rate depends on the incoming fluxes, we need to look at the time it takes to refill the liquid phase after ML formation in order to recover the critical level. The outgoing $lv$ flux of As depends on the liquid chemical potential roughly as $\Gamma_{lv} \propto \left(\frac{x_{As}}{x_{As}^{ERS}}\right)^2$ (because $\delta\mu_{l-ERS,i}$ depends on the As concentration roughly as $\ln\left(\frac{x_{As}}{x_{As}^{ERS}}\right)$). Now, because a small (large) droplet size will lead to a large (small) decrease in the As concentration immediately after a ML formation, the time needed to refill the liquid to the critical concentration depends on the droplet size. Thus, especially in the regions of the growth parameter window where the droplet size changes during growth, the incoming flux of Ga may also play an important role on the growth rate. In Figure 9 B, it is seen that the droplet size increase at low V/III ratios, but as the V/III is increased the expansion of the droplet slows down as growth accelerates and Ga is incorporated faster into the NW. For moderate V/III ratios, where the droplet stays in a steady state regime, the growth rate becomes more or less linear with the As flux until it reaches a limit where the droplet gets small and eventually gets consumed.[78] The apparent linear relation between NW length and As flux at moderate V/III ratios is consistent with previous reports.[79] At very high incoming As fluxes, As just consumes the droplet and NW growth becomes impossible.

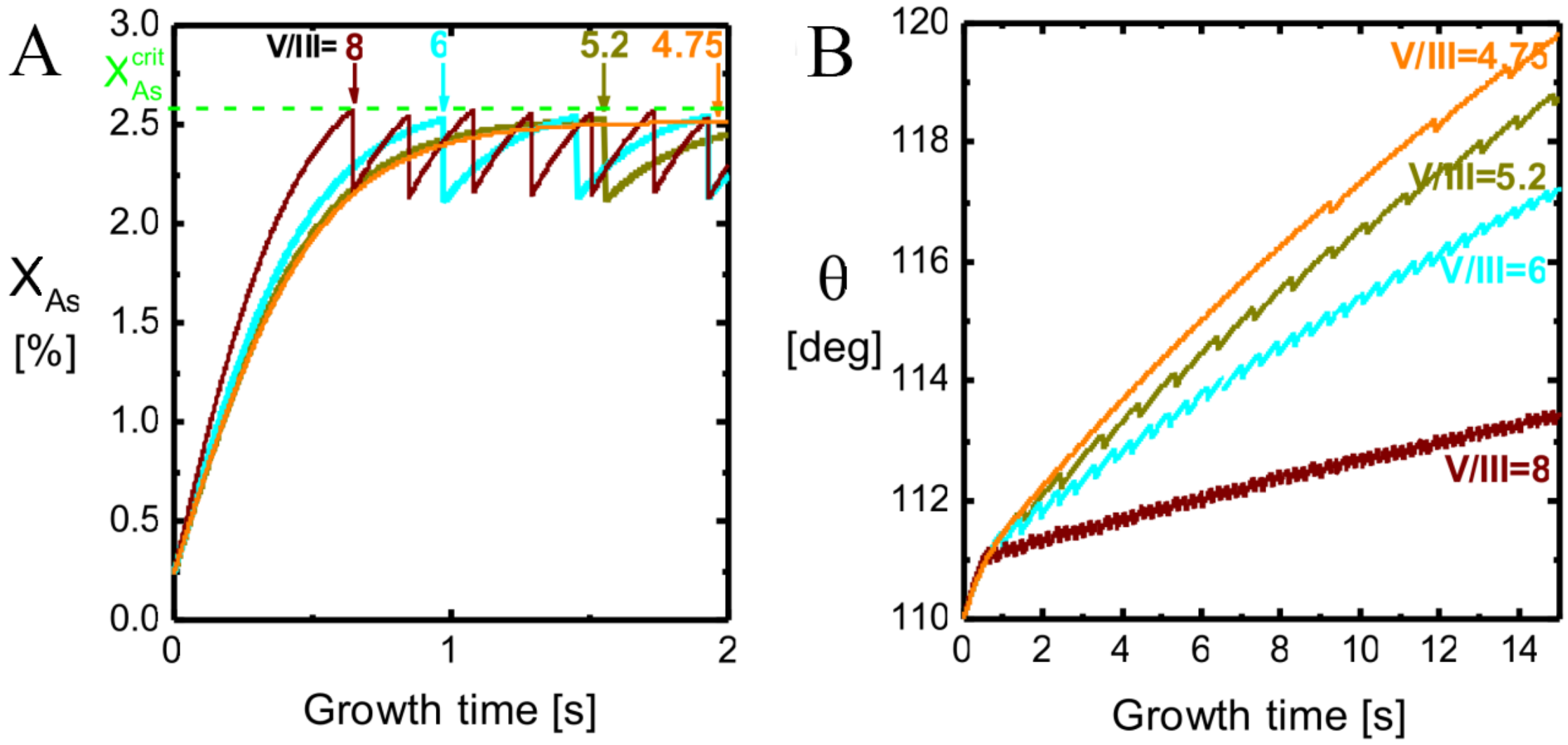

**Figure 9. Initial transitory stage for the self-catalyzed growth of GaAs NWs on Si(111) at $T = 630°C$ using a Ga flux equivalent to a planar growth rate of $GR_{planar} = 0.3\frac{\mu m}{hr}$. The initial contact angle and NW diameter were set to $\theta_{initial} = 110°$ and $d_{NW,0} = 50\,nm$, and the time step was set to 0.001 sec. (A) The As molar fraction in the $Ga_{1-x}As_x$ liquid phase and (B) contact angle just after opening the As shutter, are shown for four different V/III ratios close to the lower limit of the growth window. A fast drop in the curve corresponds to a nucleation event and the formation of one monolayer at the topfacet (for V/III=4.75 it takes about 10 sec before the first nucleation event takes place and for lower V/III ratios it becomes impossible overcome the nucleation barrier). This event lowers the liquid chemical potential $\delta\mu_{l-ERS,As}$ and $\Delta\Gamma_{vl,As}$ and $\Delta\Gamma_{al,As}$ immediately increase and forces the As molar fraction back to a level sufficient to overcome the nucleation barrier again.**

In Figure 10B, a series of 6 min simulations shows an example of the huge change in morphology when changing the $As_2$ flux around the lower limit of the growth window. The NW diameter increase when the droplet reaches a size where the contact angle exceeds the wetting angle on the side walls. A higher V/III ratio implies less tapered NWs, because the droplet does not increase in size at the same speed as for lower V/III ratios (Figure 9).

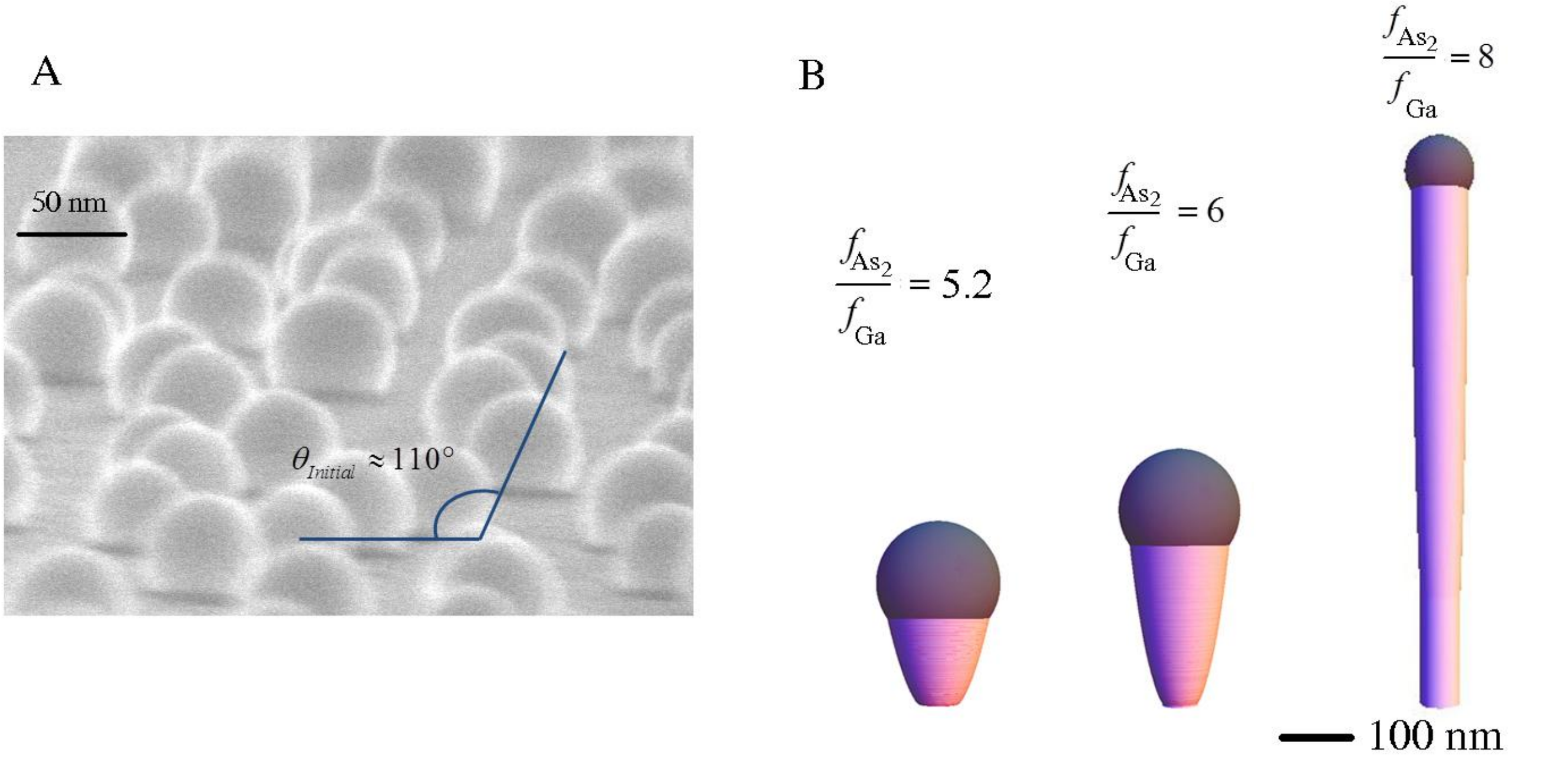

**Figure 10. Around the lower limit of the V/III growth parameter window, a small change in the incoming $As_2$ beam flux may cause a big change in the NW morphology. (A) To estimate the initial contact angle and liquid size in the case of self-catalyzed GaAs on Si(111) covered with a native oxide layer, Ga was deposited at the same initial conditions as before a typical NW growth (here 1 min of Ga pre-deposition) but without opening the valve to the As cell. These initial conditions were used for the simulations shown in Figure 9 (A) and (B). In (B) the same growth conditions as for the simulations shown in Figure 9 have been used.**

## 3.3 Relating the structure along the NW length to the relative size of the droplet

It is well known that it is generally possible to affect the crystal structure adopted by the NWs by tuning the growth conditions (for a review see K. A. Dick et al.[80]). In the case of self-catalyzed GaAs nanowires the preferential structure under quasi steady state growth conditions is typically ZB[81]. However, as shown by Jabeen et al.[82] and Spirkoska et al.[83] and many others, the density of twin planes (TPs) is generally observed to be highest at the beginning and at the end of the growth. This is another indication that changes in the growth conditions change the probabilities of forming ZB and WZ. However, there can be a wide variety in the distribution of crystal phases and defects along the NW length, since these depend on the complicated interplay between the various growth parameters. In particular, it is difficult to obtain a perfect crystal structure throughout the whole NW because the effective V/III ratio, $\frac{I_V}{I_{III}}$, changes as the NW grows. This is seen in a typical TEM image of a self-catalyzed GaAs NW (Figure 11), where the temperature and beam fluxes are kept constant during growth. To explain this, we have to use dynamics.

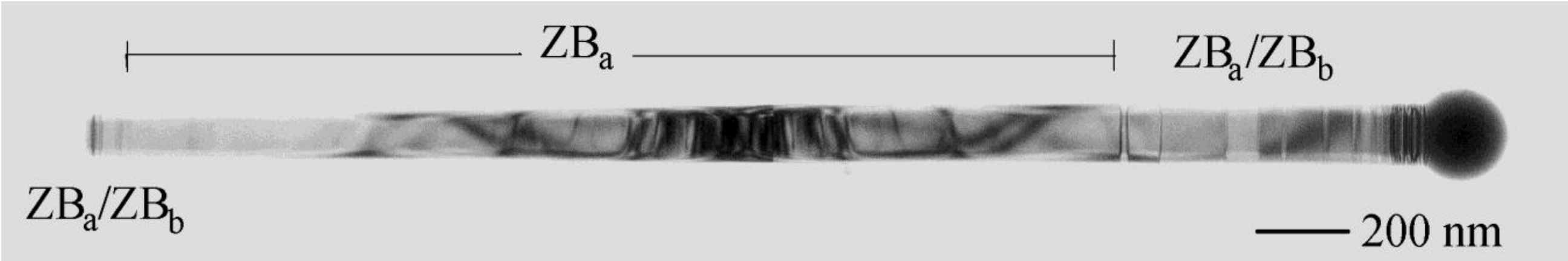

**Figure 11. A TEM image of a GaAs NW grown for 40 min with a V/III ratio of 8 and a pyrometer temperature of 635°C at $GR_{planar} = 0.3\mu m/hr$. The distribution of TPs appearing at the bottom and at the tip is typical of Ga-catalyzed GaAs NWs. The structural distribution depends on the relative size of the liquid phase (which changes during growth, see Figure 12) because the latter has a huge influence on the nucleation statistics (see sections 2 and 4).**

As proposed by Ramdani et al.[23], secondary adsorption is to a good approximation proportional to the beam flux of the material in excess (i.e. As), and such contributions are simply taken account of by assuming that the beam impinges on the total liquid surface. This gives effectively a higher collection from the gas states than if we only had considered direct impingement from the beam states. In these simulations, the nanowire diameter typically stays constant because the contact angle stays between the wetting angles on the topfacet and sidefacet, but the evolution of the liquid size is not monotonous. Relating the typical structural distribution seen in Figure 11 to the typical evolution of relative size of the liquid predicted from the simulations (shown in Figure 12), shows good agreement with theoretical predictions by Krogstrup et al.[26] using the flat topfacet assumption (regime *I*).

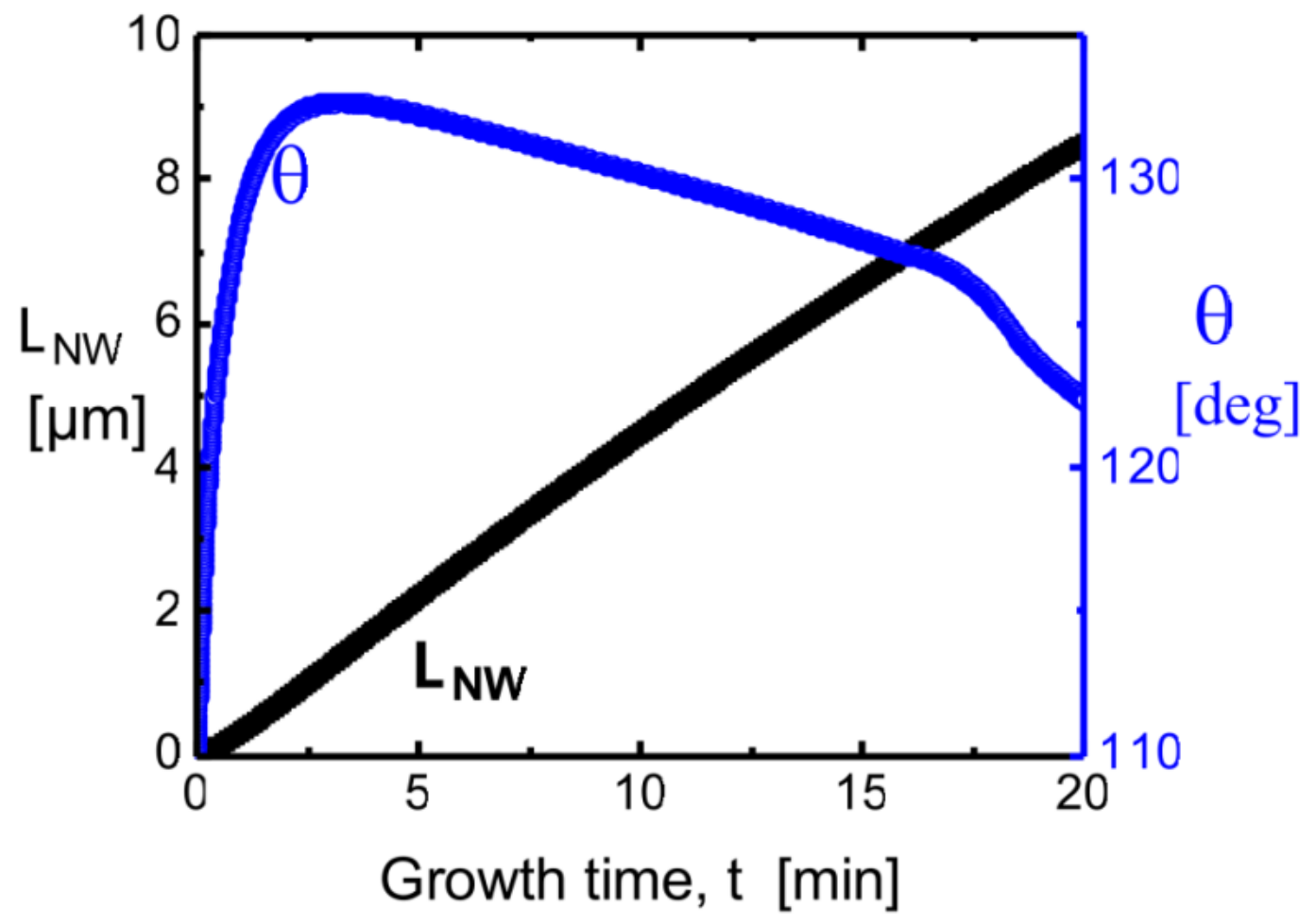

**Figure 12 A typical evolution of NW growth rate and contact angle during a complete self-catalyzed GaAs NW growth simulation. The initial contact angle is** $\theta_{initial} = 110°$ **and the nanowire diameter 100 nm. Main growth parameters are:** $\frac{f_{As_2}}{f_{Ga}} = 20$, $GR_{planar} = 0.3\frac{\mu m}{hr}$ **and** $T = 630°C$.

The crystal structure with the highest formation probability depends on the size of the liquid phase relative to the growth interface area. This match apparently well with the present simulations, which are done in regime *I*. However, it should be noted that whether the overall modeling it is done in regime *I* or *II*, the evolution of the droplet size seems to be qualitatively the same. As will be seen for regime *II* modeling in the next sections, truncation edge nucleation might also favor WZ at relative small droplets.

## 3.4 Growth of self-catalyzed GaAs NWs on patterned Si(111)/$SiO_x$ substrates

As shown by Bauer et al.[84] and Plissard et al.[85] it is possible grow positioned self-catalyzed GaAs NW arrays using a Si/$SiO_x$ template, and when growing the wires using a hole array in a $SiO_x$ layer thermally grown on the Si substrate, approximately the same growth temperatures as above is used, but the Ga flux needs to be equivalent to a planar growth rate of $0.8-1.2\frac{\mu m}{hr}$ and the V/III flux ratios need to be in the range 1-5 (see ref.[84] and ref.[85] for details). This is a much higher Ga flux than for growth on untreated substrates with native oxide and is an indication that the *av* transition rate from the thick thermally grown oxide layer is dominant for the adatom state, as also seen in. Thus, for growth on a patterned oxide layer of approx. 20-30 nm of $SiO_x$, the diffusion length is strongly limited by desorption for both Ga and As species. As it has not been possible to find activation enthalpies for *av* transitions on oxide surfaces in the literature, we simply take it to be half the value on a corresponding crystalline surface. The density of incorporation sites at the oxide surface is set to zero. For the growths on Si (111) wafers with both the native oxide layer and the thermal oxide layer, it is assumed that the low energy pathway of diffusion is one dimensional on the NW sidefacets (along the NW growth axis) and isotropic on the substrate.

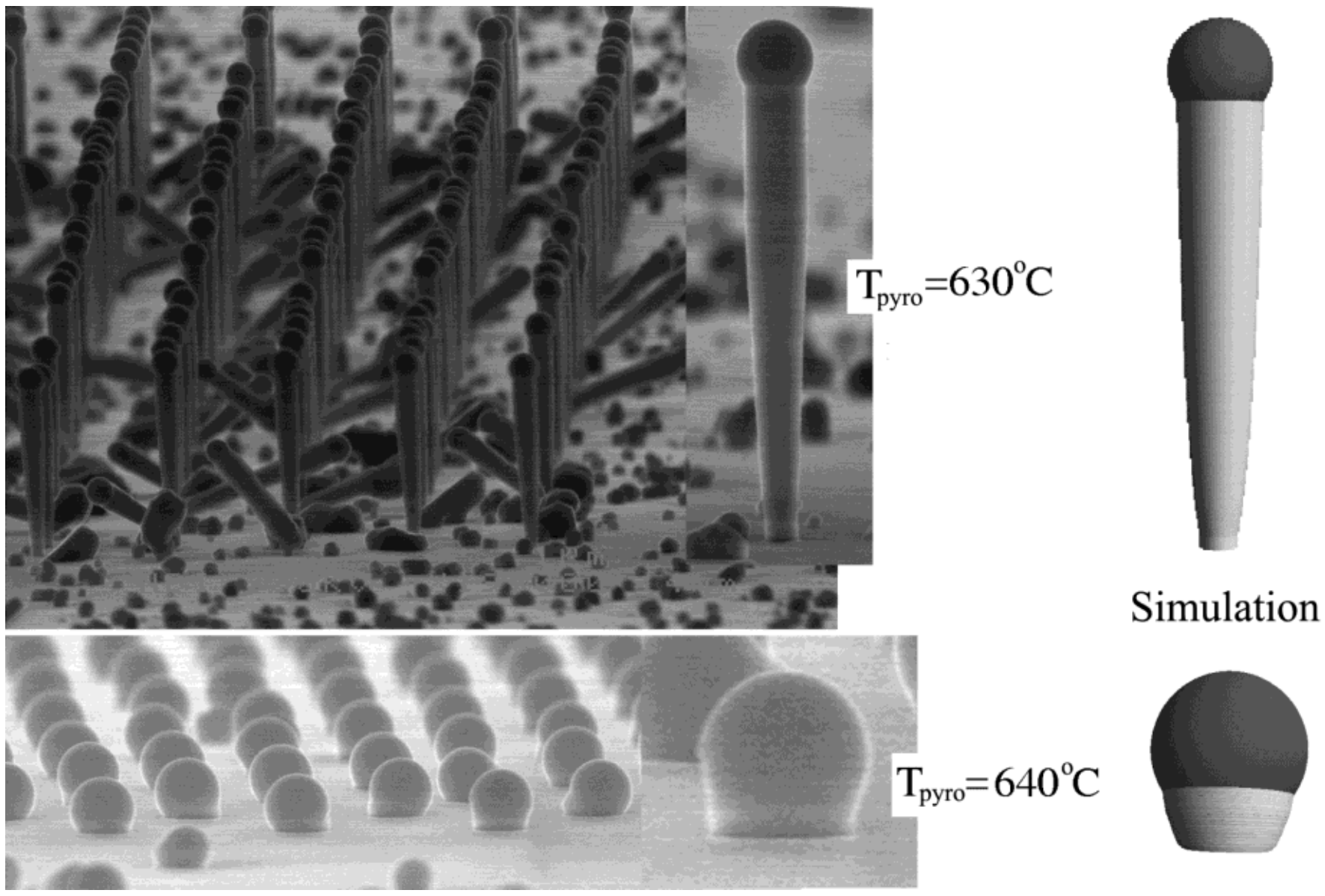

**Figure 13. Investigation of the upper growth temperature limit at a given V/III ratio for positioned self-catalyzed GaAs NW growth on a Si 111 substrate with a 30 nm thick $SiO_x$ layer. The preparation of the holes in the oxide layer is done with e-beam lithography on the same 2" wafer and all processing was carried out before the wafer was cut into four ¼" wafers just before loading into the buffer chamber. This ensures that differences due to preprocessing steps have a minimal effect on the final results when comparing the growths. The two growths are grown under exactly same conditions for 20 min, a Ga flux corresponding to a planar GaAs growth rate of $GR_{planar} = 0.9\frac{\mu m}{hr}$ and a measured flux ratio of $\frac{f_{As_2}}{f_{Ga}} = 3$ (measured with an ion-gauge filament), but with two different temperatures that were measured just before initiation of the growth with a pyrometer as $T_{pyro} = 630°C$ and $T_{pyro} = 640°C$. The activation enthalpy for the *av* transition of Ga adatoms on the oxide is set to half the value of the modeling on native oxide and the As species was set to desorb immediately from the oxide (i.e. $\Gamma_{al,As} = 0$). Using the same basic conditions in the simulations (shown on the right) the sharp temperature transition occurred at $T_{simulation} = 661°C$ and $T_{simulation} = 667°C$, which just means that there is still some fine tuning of parameters left to be done. The NW crystal formation completely stopped at $T_{simulation} = 669°C$.**

## 3.5 Liquid-solid growth dynamics - The single slice construction

As mentioned in section 2.2 and 2.3, and as indicated by many recent experiments[40,41,42], the assumption of a perfect flat liquid-solid interface is in most cases not a good assumption when analyzing the details of the liquid-solid dynamics and structural formation probabilities. To analyze the liquid-solid dynamics at the growth region in more detail we need to define a more complete parameter set $\{X\}$. To do this, we first write the Gibbs free energy of the total *ls* NW growth system in the form,

$$G_{sys,j} = \int_{0}^{180^\circ} \Delta G_{sys,j}(\omega)\, d\omega \tag{28}$$

where $\Delta G_{sys,j}(\omega)$ is the free energy of a representative thin 'double cake piece' throughout the whole *ls* system, as shown in the top view illustration of a NW with a typical six-fold axial symmetry in Figure 14(A).

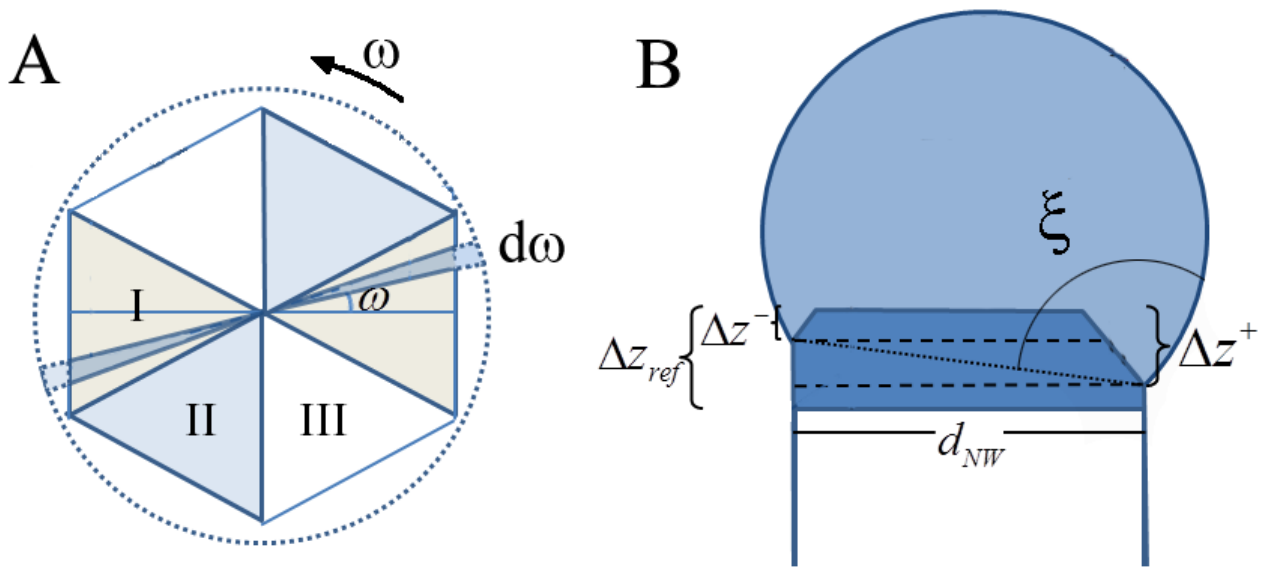

**Figure 14. (A) Top view illustration of the liquid-solid growth region where three sections indicated with a color and roman numerals are identical but rotated $60^\circ$ degrees, both in case of ZB and WZ structure. $\Delta G_{sys,j}(\omega)$ is the Gibbs free energy of a single slice throughout the growth region. (B) Side view illustration of a suggested growth system at a given $\omega$. The colored region indicates the growth system (dark blue: solid, light blue: liquid), where $\Delta z_{ref}$ is a reference length to a position from where the solid is considered to be fixed as measured from the topfacet. $\Delta z_+$ and $\Delta z_-$ are the truncation heights at $\omega = 0^\circ$ and $\omega = 30^\circ$, respectively. $\xi$ is the contact angle of the constant curvature construction and is a function of $\omega$.**

The construction of the *ls* growth system at a given $\omega$ considered in this section is sketched in (B). For a single faceted solid crystal the equilibrium shape (called the 'Wulff shape') and can be calculated exactly if the surface energy function in eq.(20) is known.[86] But the equilibrium shape of a liquid-solid system is extremely complex to derive and we will make simplifying assumptions in order to make qualitative predictions of a corresponding liquid-solid 'Wulff shape'. In appendix 5.7 we discuss the complete three dimensional *ls* system of constant liquid curvature and complete facetting.

In eq.(28), the integration over $\omega$ using analytical equations is difficult to carry out, thus we will start by looking at a single slice construction for which the liquid curvature stays constant. The choice of the parameter set $\{X(\omega)\}$ used with eq.(6) when describing the dynamics of the NW growth system is obviously crucial for the overall evolution of the structure and morphology in the simulation. Recent in-situ growth experiments have suggested that a truncated morphology at the growth interface edge is a general phenomenon, and it will therefore be taken into account here. Thus, we will choose, $\{X\} \in \{d_{NW}, \Delta z_+, \Delta z_-, \theta_{T+}, \theta_{T-}, \xi\}$ as our parameter set (See Figure 15), where $\theta_{T+}$ and $\theta_{T-}$ are the truncation angles at $\omega = 0°$ and $\omega = 180°$ respectively. Note that all these parameters are functions of $\omega$, but only considering a single cut of a finite thickness (of say $d\omega = 1°$) can give us an idea of the mean properties of the total three dimensional growth region. The Gibbs free energy for a single slice, $\Delta G_{sys,ZB(WZ)}(\omega = 0°)d\omega$, can be found using basic trigonometric relations (see appendix 5.4).

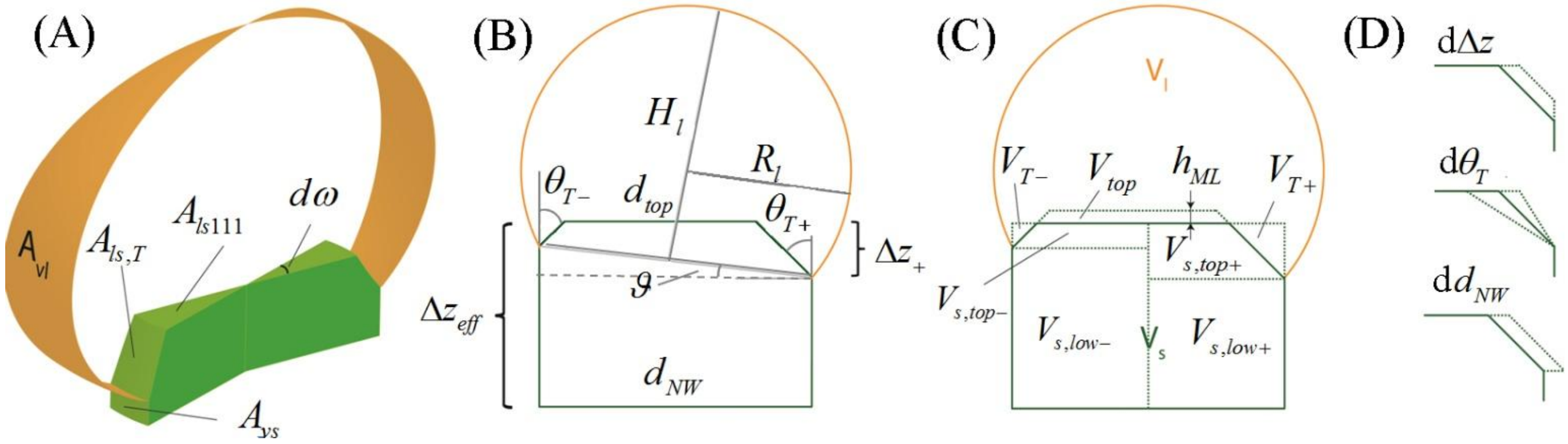

**Figure 15: Single slice construction in regime II. (A) 3D figure showing the nanowire growth system used in the single slice model. (B) 2D illustration of the involved trigonometric quantities and (C) the volume elements. (D) 2D illustration of the three different ways the truncation can change during growth. The Gibbs free energy of this construction can be calculated using basic trigonometric relations, see Appendix 5.4.**

The *ls* system is continuously adjusting towards $\Delta\mu_{ls}^{X} = \delta\mu_{l-ERS} - \delta\mu_{s-ERS}^{X} = 0$ conditions, but the input of free energy from the beam fluxes, vapor and adatoms and the interplay with anisotropic solid and the nucleation limited growth on the top facet keep the system out of equilibrium. Under certain conditions, the solid can enter a regime where undesired facets are locked in because a free energy barrier has to be overcome in order to form a facet lowering the total free energy of the system. The liquid-solid driving forces of liquid $Ga_{1-x}As_x$ assisted GaAs NW growth is plotted as a function of truncation height $\Delta z$ for a certain set of fixed parameters in using eq.(6) and the single slice construction around $\omega = 0°$. In Figure 16 (A) it is seen that the equilibrium value of $\Delta z_{\pm}$ is larger for smaller systems. For the single slice construction the equilibrium morphology will always have a negative truncation. However a non-steady state evolution of the growth system can force the system into regime *I* and to get back to regime *II* will require nucleation of a truncated facet which requires a certain formation free energy. Figure 16 (B) shows that, in the single slice construction, varying the truncation on one side have a small effect on the truncation on the opposite side. Figure 16 (C) shows an important general trend, namely that truncation heights are generally smallest at smallest droplet sizes. This means that relatively small droplets has higher tendency of going into regime I than larger droplets, in accordance with ref.[26]. In Figure 16 (D), it is seen that a strong dependence of the liquid concentration on the truncation size indicates that it is the composition which plays a dominating role on the oscillating morphology.

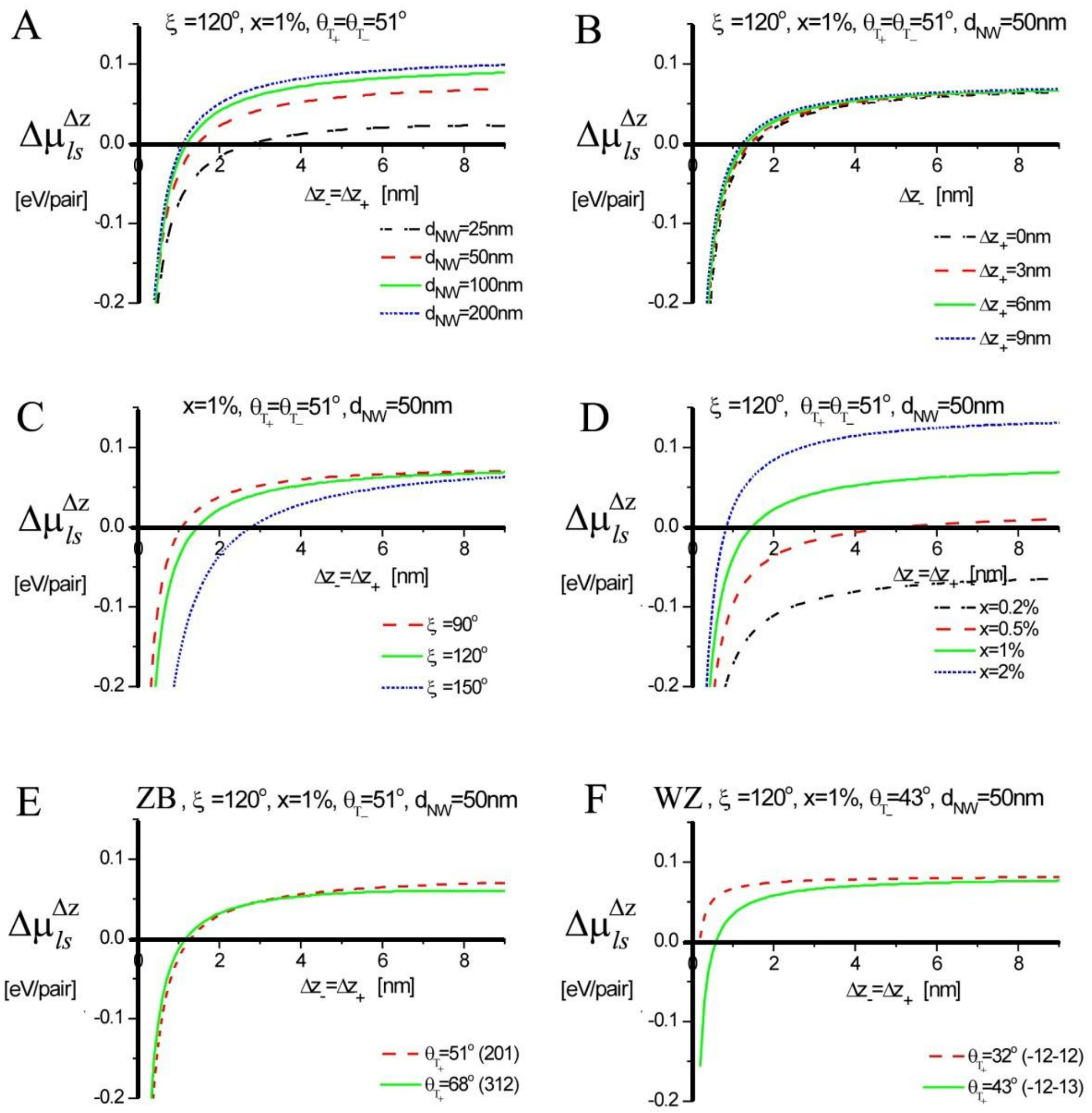

**Figure 16. Plots of driving forces $\Delta\mu^{\Delta z}_{ls,III-V} = \delta\mu_{l-ERS.III-V} - \delta\mu^{\Delta z}_{s-ERS.III-V}$ of Ga Assisted GaAs NW growth as a function of the truncation size for the single slice construction at $\omega = 0°$ with $\{1\bar{1}0\}$ type sidefacets at $T = 630°C$. The equilibrium value of the parameters under the chosen growth conditions are where $\Delta\mu_{ls} = 0$. (A) The equilibrium value of $\Delta z_{\pm}$ is larger for smaller systems. For the single slice construction the equilibrium morphology will always have a negative truncation. (B) By varying the truncation on one side has a small effect on the truncation on the opposite side in the single slice construction. (C) Truncation heights are generally smallest at smallest droplet sizes. (D) A strong dependence of the liquid concentration on the truncation size indicates that it is the composition which plays a dominating role on the oscillating morphology. (E) and (F) shows the driving forces around certain facets in the case of ZB and WZ structures, respectively.**

Figure 16 (E) and (F) shows the driving forces around certain facets in the case of ZB and WZ structure, respectively. It is seen that for certain sets of orientations and parameters, the system needs to form another facet orientation to reach a quasi-equilibrium state. If the potential barrier to form such a facet is large, the system can enter an unstable growth mode.

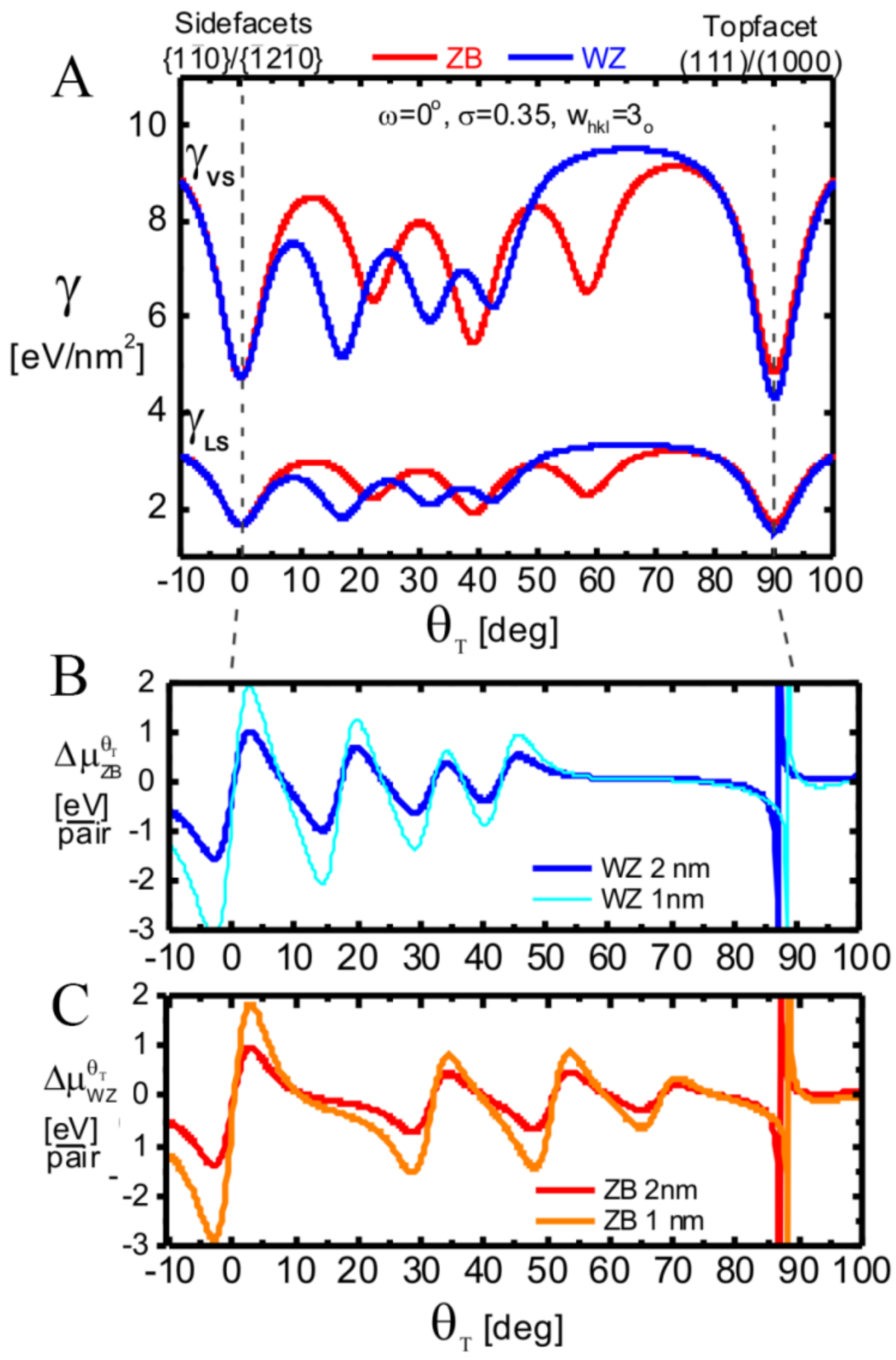

**Figure 17. (A) The 2D $\gamma$-plots using eq.(20) for the *vs* and *ls* interfaces around $\omega = 0°$ and $\gamma_{pq0} = 10\frac{eV}{nm^2}$. It is seen that WZ is dominant at small truncation angles $\theta_T$ and ZB is dominant at large $\theta_T$. (B),(C) Driving forces $\Delta\mu^{\theta_T}_{ls,i} = \delta\mu_{l-ERS.i} - \delta\mu^{\theta_T}_{s-ERS.i}$ of ZB and WZ is plotted as a function truncation facet angle for ZB and WZ, respectively. The parameter set are $d_{NW} = 50\,nm$, $\xi = 120°$, $x_V = 1\%$ and $\theta_{T+} = 51°$ for ZB and $\theta_{T+} = 43°$ for WZ. The stable points are the ones where the driving force is zero and the gradient of the driving force is positive. The plot tells us that it is not possible to switch freely between facet orientations. The singularities close to 90° is due to the definition of the truncation angle shown in Figure 15 (D), because the system cannot approach a single topfacet for a fixed $\Delta z$ value (see Appendix 5.3 for the low energy orientations used).**

In Figure 17 (B) and (C) it is shown that the truncation angles affect the driving forces in a more complicated way than the other parameters which are considered here. This implies that the liquid-solid growth region can stay in a dynamical metastable and still steady state regime (see appendix 5.5).

The surface energies and the interface energy function given by eq.(20) play a crucial role on the NW growth simulations in general and on the truncation dynamics in particular. The interface energy function is plotted for the single slice construction in Figure 17 (A). $w_{hkl}$, which specifies the half-width half maximum of the energy decreases around the $(hkl)$ facet is an important parameter for the dynamical system. If $w_{hkl}$ is small, the corresponding truncation facet orientation is locked to a low energy facet orientation and it is unlikely that the facet can overcome the energy barrier $\Delta G_{ls,III-V}^{\theta_{T1}-\theta_{T2}}$ needed to form another facet and a more preferable configuration. In Figure 17 (B) and (C) shows how the *ls* driving force for truncation angle (i.e. the free energy change per pair due to a change in $\theta_T$) depends on the orientation for a given parameter set, where $\Delta z = 2nm$ is closer to equilibrium than $\Delta z = 1nm$. We emphasize that in this continuum approach with single truncation facets it has not been taken into account that facet orientations becomes discrete when $\Delta z$ becomes small. The formation of a new facet orientation can be nucleation limited if the barrier is larger than the single transition state barrier $\Delta G_{ls,III-V}^{\theta_{T1}-\theta_{T2}} > \delta g_{ls,III-V}^{TS,ERS}$ and such a transition has to be treated in a framework similar to that of section 2.3. However these transitions will not be treated in detail here; instead, in the simulations the probability of forming another truncation facet orientation simply depends on the evolution of the system morphology. For large values of $w_{hkl}$ the angle of the truncation facet can change more or less freely and it will oscillate in accordance with the oscillations of the growth system. However, to make qualitative predictions about a given growth process and the structural formation probabilities it is necessary to have reasonably good estimates of the parameters describing the surface energy functions. This is indicated in Figure 18 where each set of simulation parameters gives different results.

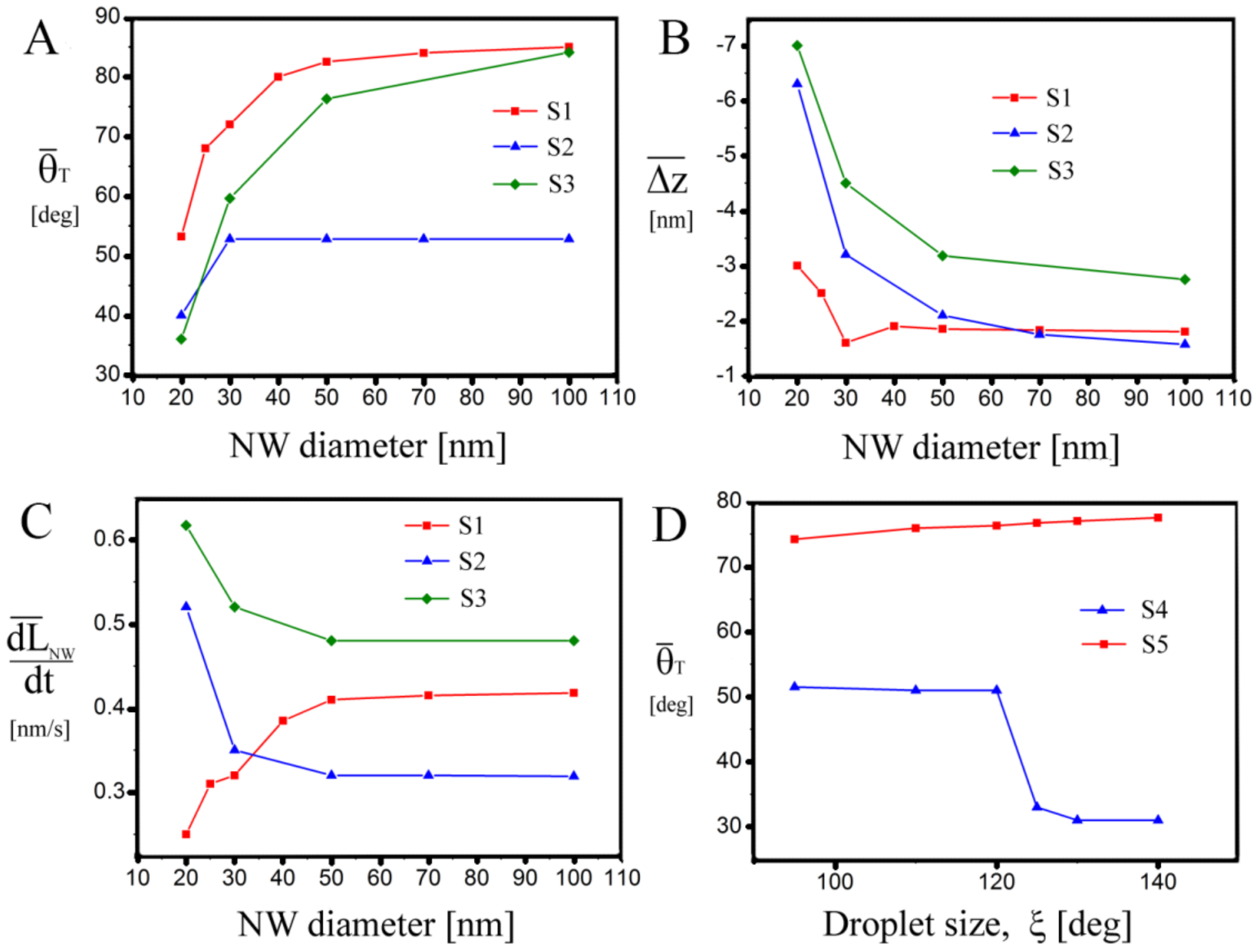

**Figure 18 Average values taken from growth simulations of self-catalyzed GaAs growth in the single slice construction (at $\omega = 0°$ where the system choose symmetry, i.e. $\Delta z = \Delta z_- = \Delta z_+$ and $\theta_T = \theta_{T-} = \theta_{T+}$ ). (A,B,C) Three different simulations with parameters, S1, S2, S3 (listed in Appendix 5.3) of the truncation dynamics show some general trends as a function of the system size. The quantities which are plotted here are average values after reaching a quasi-steady state, see Appendix 5.5 for examples of the truncation dynamics. See text for discussion.**

For the single slice construction, it is possible to predict the impact of the growth interface size and morphology on the relative formation rates of the ZB and WZ stacking's, given a set of simulation parameters. In Figure 18 (A,B,C) three different simulations S1, S2, S3 of the truncation dynamics show some general trends as a function of the growth interface diameter, even though huge quantitative differences are seen due to changes in the parameters determining the shape of the interface energy functions. The quantities which are plotted here are average values after reaching a quasi-steady state; see Appendix 5.5 for examples of the truncation dynamics. Parameters used for the simulations are listed in Appendix 5.3. If truncation edge nucleation dominates at the top facet, WZ would be favored at small diameters and ZB at larger diameters (Figure 18 A) if considering the interface energy function as plotted in . In Figure 18 (B) it is seen that the truncation facet seems to be smaller with increasing size of the liquid-solid growth region, even though it does not necessarily have a monotonic dependence due to the anisotropic interface energy. In Figure 18(C) it is shown that the axial growth rate is strongly dependent on the size of the growth region. However, the actual dependence is not simple depends on the simulation parameters used, where the relative liquid size does also play an important role on the truncation angle as seen in Figure 18 (D). Thus, because values for the surface energy function, the various energy barriers and kinetic reaction constants are not well known, the modeling of NW growth is still far from being a supplement to NW growth experiments.

## 4. Summary

We have presented a detailed review on and overall treatment of the theoretical formalism of III-V NW growth dynamics, using the current understanding of NW growth. The overall treatment can be used analyze and model the dynamics of axial III-V nanowire growth via the vapor-liquid-solid mechanism as a function of the basic growth parameters, partial pressures/beam fluxes and substrate temperature. The formalism relies on transition state kinetics driven by minimization of free energy of the total system. All chemical potentials are measured with respect to a common equilibrium reference state where the total system is in a thermodynamical equilibrium. The formalism makes it possible to understand the complex mechanisms of nanowire growth dynamics in greater detail and can serve as strong analyzing tool when optimizing VLS growth of III-V nanowires. We have implemented the theoretical framework into a computer simulation model, and even though the program is in a preliminary stage, the modeling examples show growth good agreement with experiments and that the theory can be used to model NW growth dynamics in a new level of detail.

**Acknowledgement**

This work was supported by the Danish National Advanced Technology Foundation through project 022-2009-1, the Danish Strategic Research Council through project 09-065736 and the ERC starting grant ‘UpCon’. Yun Yamasaki from

# 5. Appendices

## 5.1 ADATOM COLLECTION

For the condensed adatom regime, it can be shown (using mass conservation) that the general equation for steady state adatom collection can be written in a relatively compact form,

$$\Delta\Gamma_{al,i} = \frac{2\pi}{l_j} \sum_{r'=\frac{d_{NW}}{2}}^{\infty} r'\left(\Delta\Gamma_{(vb)a_{r'},i} - \Delta\Gamma_{a_{r'}s,i}\right) + \sum_{l=0}^{\frac{L_{NW}}{h_{ML}}} \left(\Delta\Gamma_{(vb)a_l,i} - \Delta\Gamma_{a_l s,i}\right) \tag{29}$$

where the first summation accounts for the net transition flux from the substrate to the sidefacets and the second summation for the net generation of adatoms from the beam and the vapor along the NW sidewalls. $a_{r'(l)}$ is the adatom site at *r'* (or *l*) along the substrate (or NW) surface as measured from the NW foot. $l_j$ is the distance between the two adjacent adatom sites. The substrate diffusion is assumed to be isotropic, which is a reasonable for growths carried out on substrates with a (111) orientation or substrates covered with an amorphous oxide layer. A simple approach is to assume that there exist certain effective collection areas characterized by corresponding effective diffusion lengths.[87] Even though this approach is not very accurate for modeling the growth dynamics, it is instructive and intuitive. To get a more intuitive feeling of the adatom kinetics in terms of a diffusion length in the transition state approach, adatom migration on a large homogenous planar interface serves as a good example. Then all parameters are translation invariant and there is no net diffusion. Since in this case we do not have to distinguish between the adatoms as all states are independent of position in the continuum approach we will just label all adatoms with an '*a*'. There are three main transition paths for an adatom; surface diffusion (*aa*), desorption (*av*) and incorporation (*as*). The *as* mechanism can be further divided into two types of incorporation mechanisms: incorporation at a favorable site (such as a kink) leading to radial growth, or by interdiffusion which can take place at all sites. Incorporation by interdiffusion is only relevant for impurities such as dopants since exchange of group III and V element will not have a net effect on the adatom state, and will therefore be neglected here. In such conditions, an adatom diffusion length is a well-defined quantity. The mean length displacement (i.e. the mean distance between the location of 'birth' and 'death' events, where 'death' is determined by either an '*as*' or '*av*' transition) is $\lambda_{j,i} = \sqrt{D_{j,i}\tau_{j,i}}$, where $D_{j,i} = Z_{aa,i}\upsilon_{a,i} l_j^2 (1-\bar{\rho}_{j,i}) \exp\left(-\frac{\delta g_{aa,i}^{TS,ERS} - \delta\mu_{a-ERS,i}}{k_B T}\right)$, is the mean adatom diffusivity, and $\tau_{j,i} = \left(\tau_{j,i,as}^{-1} + \tau_{j,i,av}^{-1}\right)^{-1}$ is the average adatom lifetime. Here $l_j$ is the distance between the two adjacent adatom sites along the lowest energy direction(s), with activation free energy $\delta g_{aa,i}^{TS}$. For simplicity, higher energy directions are ignored. If an adatom occupies a given site, it is impossible for another adatom to jump into the same site. Thus, the concentration of free sites, $1-\bar{\rho}_{j,i}$, is included in the diffusivity, $\bar{\rho}_{j,i}$ being the normalized adatom density. The lifetimes ended by an '*as*' or '*av*' state transition are inversely proportional to the respective transition rates, $\tau_{j,i,as} = \frac{1}{\bar{Z}_{as,i}\bar{c}_{inc,i}\nu_{a,i}} \exp\left(\frac{\delta g_{as,i}^{TS,ERS} - \delta\mu_{a-ERS,i}}{k_B T}\right)$ and $\tau_{j,i,av} = \frac{1}{\bar{Z}_{av,i}\nu_{a,i}} \exp\left(\frac{\delta g_{av,i}^{TS,ERS} - \delta\mu_{a-ERS,i}}{k_B T}\right)$, respectively. $\bar{c}_{inc,i}$ is the normalized density of probable incorporation sites (kinks or possibly steps at high adatom densities and/or low temperatures). This important factor illustrates a major difference between the *av* and *as* transitions, namely that desorption can take place everywhere, which is not the case for incorporation. $\tau_{j,i,as}$ is conditioned by the incorporation of both a group III and a group V element because of the fixed 1:1 stoichiometry of the III-V solid. For the modeling, $\tau_{j,i,as}$ is limited by incorporation of an adatom at a kink site, which means that the solid chemical potential equals the *ERS* potential, and possible sidefacet nucleation events will not be considered. For desorption of adatoms, the intrinsic activation barrier[88] $\delta g_{av,i}^{TS,ERS}$ is independent of the other components.

The general equation for the adatom diffusion length at a given point at a homogenous interface is therefore:

$$\lambda_{j,i} = \sqrt{\bar{Z}_{aa,i} l_{a,i}^{2} \left(1-\bar{\rho}_{j,i}\right) \exp\left(-\frac{\delta g_{aa,i}^{TS,ERS}}{k_B T}\right) \left(\bar{Z}_{as,i}\bar{c}_{inc,i} \exp\left(-\frac{\delta g_{as,i}^{TS,ERS}}{k_B T}\right) + \bar{Z}_{av,i} \exp\left(-\frac{\delta g_{av,i}^{TS,ERS}}{k_B T}\right)\right)^{-1}}$$

(30)

which is apparently independent of chemical potential and vibration frequency of the adatoms. However the number of incorporation sites $\bar{c}_{inc,i}$ depends on the local adatom densities of both components and on the orientation of the local facet, and is therefore also dependent on the chemical potential of the local adatom state, see eq.(4).

## 5.2 ADATOM TRANSITION STATE DIFFUSION CALCULATIONS USING AN UNIFORM 'DUBROVSKII/JOHANSSON' DIFFUSION SCHEME

Calculating the adatom density distribution using the general adatom diffusion equation, eq.(29), in terms of growth conditions has proved to be difficult. Here we will show a simplified approach by using the transition state fluxes in a classical Fickian diffusion scheme to find the adatom density distribution in terms of the basic growth parameters. As in previous studies[89,90] we only distinguish between two types of facets, the NW sidefacets (*NW*) and a planar substrate facet (*sub*). If we for simplicity assume that; $1-\bar{\rho}_{j,i} \approx 1$, and that $\bar{c}_{inc,i}$ is a constant along the length, meaning that the diffusion length only varies with time and does not vary along a given facet. Thus two coupled diffusion equations, $D_{NW,i}\frac{d^2}{dz^2}\rho_{NW,i}(z) = D_{NW\,i}\frac{\rho_{NW,i}(z)}{\lambda_{NW,i}^2} - f_{NW\,i,\perp} - \Gamma_{sa\,(NW\,i)} - \Gamma_{va\,(NW\,i)}$ and $D_{sub,i}\frac{1}{r}\frac{d}{dr}\left(r\frac{d}{dr}\rho_{sub,i}(r)\right) = D_{sub,i}\frac{\rho_{sub,i}(r)}{\lambda_{sub,i}^2} - f_{sub,i,\perp} - \Gamma_{sa(sub),i} - \Gamma_{va(NW),i}$ need to be solved. Here the diffusivity will be assumed uniform, $D_{j,i} = Z'_{aa,i}\upsilon_{a,i}l_j^2\exp\left(-\frac{\delta h_{aa,i}^{TS,ERS}}{k_B T}\right)$. If shadowing effects and influence from other NWs on the substrate are ignored we can assume that $\left.\frac{d\rho_{sub,i}}{dr}\right|_{r\to\infty} = 0$. The average incoming beam fluxes are given as $f_{NW,i,\perp} = \frac{f_i\sin(\varphi_i)}{\pi}$ and $f_{sub,i,\perp} = f_i\cos(\varphi_i)$, where $\varphi_i$ is the angle of the incoming beam of group *i* with respect to the substrate normal. $\frac{1}{\pi}$ is the fraction of the NW facets which is exposed to the beam which is perfectly consistent with the transition state approach where transitions are independent of the state they are moving into. Solutions are then of the form,

$$\rho_{NW,i}(z) = C_1\exp\left(\frac{z}{\lambda_{NW,i}}\right) + C_2\exp\left(-\frac{z}{\lambda_{NW,i}}\right) + \frac{\lambda_{NW,i}^2\left(\frac{f_i\sin(\varphi_i)}{\pi} + \Gamma_{sa,i} + \sum_n \Gamma_{va,i_n}\right)}{D_{NW,i}} \qquad (31)$$

$$\rho_{sub,i}(z) = C_3 K_{0,\frac{r}{\lambda_{NW,i}}} + \frac{\lambda_{sub,i}^2\left(f_i\cos(\varphi_i) + \Gamma_{sa,i} + \sum_n \Gamma_{va,i_n}\right)}{D_{sub,i}} \qquad (32)$$

where $K_{h,x}$ is the modified Bessel function of order $h$ evaluated at $x$. To solve for the constants ( $C_i$ ) we need three boundary conditions;

$$D_{NW,i}\left.\frac{d\rho_{NW,i}}{dz}\right|_{z=L_{NW}} = \Delta\Gamma_{al,i} \quad (33)$$

$$D_{NW,i}\left.\frac{d\rho_{NW,i}}{dz}\right|_{z=0} = -D_{sub,i}\left.\frac{d\rho_{sub,i}}{dr}\right|_{r=\frac{d_{NW}}{2}} \quad (34)$$

$$\rho_{NW,i}(z=0) = \rho_{sub,i}\left(r = \frac{d_{NW}(z=0)}{2}\right) \quad (35)$$

Eq.(33) assumes quasi steady state growth, combining the adatom to adatom state transition flux at $z = L_{NW}$ with the net adatom to liquid state transition flux, which is driven primarily by the thermodynamic driving force. Because the $\Delta\Gamma_{al,i}$ depends on the adatom density, $\rho_{NW,i}(L_{NW},\delta\mu_{a-ERS,i})$, it needs to be isolated in eq.(31) before it is put into eq.(4). Using eq.(7) for $\Gamma_{sa,i}$ and $\Gamma_{va,i}$, with $c_{v,i} = \frac{p_i}{RT}$ (the ideal gas law), $\delta\mu_{a-ERS,i}$ (which depends on the adatom densities) is solved numerically at every time step and before being put back into $\Delta\Gamma_{al,i}$. Eq.(34) combines the adatom fluxes at the NW root, whereas eq.(35) assumes a continuous adatom density function across the substrate – nanowire interface, see Dubrovskii et al.[89] and Johansson et al.[90]. Eq.(35) requires that the transition state barriers across the NW-substrate interface are symmetric.

Solving the coupled adatom diffusion equations for diffusion along the NW facets and on an isotropic substrate with the boundary conditions, eq.(33) -(35), leads to the following expression for the adatom density on the NW sidewall,

$$\rho_{NW,i}(z) = \frac{\lambda_{NW,i}}{D_{NW,i}} \frac{\left(\begin{array}{l} -\cosh\left(\frac{z}{\lambda_{NW,i}}\right)K_{0,\frac{d_{NW}}{2\lambda_{sub,i}}}D_{NW,i}\lambda_{sub,i}\Delta\Gamma_{al,i} + \cosh\left(\frac{z-L_{NW}}{\lambda_{NW,i}}\right)K_{1,\frac{d_{NW}}{2\lambda_{sub,i}}}\left(\Gamma_{sub,i}D_{NW,i}\lambda_{sub,i}^2 - \Gamma_{NW,i}D_{sub,i}\lambda_{NW,i}^2\right) - \\ \sinh\left(\frac{z}{\lambda_{NW,i}}\right)K_{1,\frac{d_{NW}}{2\lambda_{sub,i}}}D_{sub,i}\lambda_{NW,i}\Delta\Gamma_{al,i} + \cosh\left(\frac{L_{NW}}{\lambda_{NW,i}}\right)K_{1,\frac{d_{NW}}{2\lambda_{sub,i}}}\Gamma_{NW,i}D_{sub,i}\lambda_{NW,i}^2 + \sinh\left(\frac{L_{NW}}{\lambda_{NW,i}}\right)K_{0,\frac{d_{NW}}{2\lambda_{sub,i}}}\Gamma_{NW,i}\lambda_{sub,i}\lambda_{NW,i} \end{array}\right)}{\cosh\left(\frac{L_{NW}}{\lambda_{NW,i}}\right)K_{1,\frac{d_{NW}}{2\lambda_{sub,i}}}D_{sub,i}\lambda_{NW,i} + \sinh\left(\frac{L_{NW}}{\lambda_{NW,i}}\right)K_{0,\frac{d_{NW}}{2\lambda_{sub,i}}}D_{NW,i}\lambda_{sub,i}}$$

(36)

where $\Gamma_{j,i} = f_{j,i,\perp} + \Gamma_{va,i} + \Gamma_{sa,i}$ is the generation flux of $i$ adatoms of the $j$'th surface. As $\Delta\Gamma_{al,i}$ is a function of $\rho_{NW,i}(z = L_{NW})$, $\rho_{NW,i}$ is isolated in eq.(36), without isolating $\rho_{NW,i}$ from $\delta\mu_{a-ERS,i}$ we get,

$$\rho_{NW,i}(L_{NW}) = \frac{\begin{pmatrix} \cosh\left(\frac{L_{NW}}{\lambda_{NW,i}}\right) K_{0,\frac{d_{NW}}{2\lambda_{sub,i}}} D_{NW,i}\lambda_{sub,i}\Xi_{al,i} \exp\left(-\frac{\Delta g_{al,i}^{ERS} - \delta\mu_{l-ERS,i}}{k_B T}\right) \frac{\bar{\rho}_i^{ERS}}{x_i^{ERS}} x_i + K_{1,\frac{d_{NW}}{2\lambda_{sub,i}}} \left(\Gamma_{sub,i} D_{NW,i}\lambda_{sub,i}^2 - \Gamma_{NW,i} D_{sub,i}\lambda_{NW,i}^2\right) \\ -\sinh\left(\frac{L_{NW}}{\lambda_{NW,i}}\right) K_{1,\frac{d_{NW}}{2\lambda_{sub,i}}} D_{sub,i}\lambda_{NW,i}\Xi_{al,i} \exp\left(-\frac{\Delta g_{al,i}^{ERS} - \delta\mu_{l-ERS,i}}{k_B T}\right) \frac{\bar{\rho}_i^{ERS}}{x_i^{ERS}} x_i + \cosh\left(\frac{L_{NW}}{\lambda_{NW,i}}\right) K_{1,\frac{d_{NW}}{2\lambda_{sub,i}}} \Gamma_{NW,i} D_{sub,i}\lambda_{NW,i}^2 + \sinh\left(\frac{L_{NW}}{\lambda_{NW,i}}\right) K_{0,\frac{d_{NW}}{2\lambda_{sub,i}}} \Gamma_{NW,i}\lambda_{sub,i}\lambda_{NW,i} \end{pmatrix}}{\begin{pmatrix} \frac{D_{NW,i}}{\lambda_{NW,i}} \left(\cosh\left(\frac{L_{NW}}{\lambda_{NW,i}}\right) K_{1,\frac{d_{NW}}{2\lambda_{sub,i}}} D_{sub,i}\lambda_{NW,i} + \sinh\left(\frac{L_{NW}}{\lambda_{NW,i}}\right) K_{0,\frac{d_{NW}}{2\lambda_{sub,i}}} D_{NW,i}\lambda_{sub,i}\right) + \frac{1}{\rho}\cosh\left(\frac{L_{NW}}{\lambda_{NW,i}}\right) K_{0,\frac{d_{NW}}{2\lambda_{sub,i}}} D_{NW,i}\lambda_{sub,i}\Xi_{al,i} \exp\left(-\frac{\Delta g_{al,i}^{ERS} - \delta\mu_{a-ERS,i}}{k_B T}\right) \\ -\frac{1}{\rho}\sinh\left(\frac{L_{NW}}{\lambda_{NW,i}}\right) K_{1,\frac{d_{NW}}{2\lambda_{sub,i}}} D_{sub,i}\lambda_{NW,i}\Xi_{al,i} \exp\left(-\frac{\Delta g_{al,i}^{ERS} - \delta\mu_{a-ERS,i}}{k_B T}\right) \end{pmatrix}}$$

(37)

Note that $\Xi_{al,i}$ is a triple line flux, i.e. a particle transfer per length per time. If we assume a barrier free $al$ transition, the exponentials vanish in eq.(37) and the only dependence on $\delta\mu_{a_j-ERS,III(V)}$ is through the diffusion lengths. $\delta\mu_{a_j-ERS,III(V)}\left(\rho_{j,III}, \rho_{j,V}, T\right)$ at $z = L_{NW}$ can now be solved numerically at every step time in a double iterative process for both $\bar{\rho}_{j,III}\left(z = L_{NW}\right)$ and $\bar{\rho}_{j,V}\left(z = L_{NW}\right)$ choosing certain initial values, step size and acceptable error values depending on the computation time available and accuracy needed. The principle of a single numerical computation loop in a typical math language (here Mathcad) is shown below,

$$\delta\mu_{a_j-ERS,III(V)}\left(\rho_{NW,i}, T, \delta\mu_{quess,III}, \delta\mu_{quess,V}, step, error\right)_i = \left| \begin{array}{l} \delta\mu_{III(V)} \leftarrow \delta\mu_{quess,III(V)} \\ \delta\mu_{V(III)} \leftarrow \delta\mu_{quess,V(III)} \\ \Delta_{III(V)} \leftarrow 1eV \\ while\ \left|\Delta_{III(V)}\right| > error \\ \quad \left| \begin{array}{l} F_{III} \leftarrow \delta\mu_{a_j-ERS,III}\left(\rho_{j,III}\left(L_{NW}, \delta\mu_{III}\right), \rho_{j,V}\left(L_{NW}, \delta\mu_{V}\right), T\right) \\ F_{V} \leftarrow \delta\mu_{a_j-ERS,V}\left(\rho_{j,III}\left(L_{NW}, \delta\mu_{III}\right), \rho_{j,V}\left(L_{NW}, \delta\mu_{V}\right), T\right) \\ G_{III} \leftarrow \delta\mu_{III} \\ G_{V} \leftarrow \delta\mu_{V} \\ \Delta_{III} \leftarrow F_{III} - G_{III} \\ \Delta_{V} \leftarrow F_{V} - G_{V} \\ \delta\mu_{III(V)} \leftarrow \delta\mu_{III(V)} + step \cdot sign(\Delta_{III(V)})\ \ if\ \ \left|\Delta_{III(V)}\right| > error \wedge \left|\Delta_{V(III)}\right| > error \end{array} \right. \\ \begin{pmatrix} \delta\mu_{III} \\ \delta\mu_{V} \end{pmatrix} \end{array} \right.$$

(38)

where $\delta\mu_{a_j-ERS,III(V)}\left(\rho_{j,III}\left(L_{NW},\delta\mu_{III}\right),\rho_{j,V}\left(L_{NW},\delta\mu_{V}\right),T\right)$ is given by eq.(4) with $\bar{\rho}_{j,III(V)}\left(L_{NW},\delta\mu_{III(V)}\right)$ being the value from eq.(37). The calculated value of $\delta\mu_{a_{NW}-ERS,III(V)}$ is a 1x2 matrix with $\delta\mu_{a_{NW}-ERS,III}$ and $\delta\mu_{a_{NW}-ERS,V}$ on each position. Note that much computation time is saved by choosing the simplest version $\delta\mu_{a_{NW}-ERS,i}\left(\rho_{NW,i},T\right)\cong k_B T\ln\left(\frac{\bar{\rho}_{NW,i}\left(z=L_{NW}\right)}{\bar{\rho}_{NW,i}^{ERS}}\right)$, which only requires one iteration loop for each element at each time step, which may be a rough but fairly reasonable simplification at low total fluxes and if only looking at axial growth. After this step, $\delta\mu_{a_{NW}-ERS,i}\left(\rho_{NW,i},T\right)$, is finally put into eq.(37) which is again put into eq.(14) and (36).

Solving for the adatom density on the isotropic substrate (which is a reasonable approximation on (111) surfaces and amorphous oxide layers), leads to the following solution,

$$\rho_{sub,i}(r)=\frac{\lambda_{sub,i}}{D_{sub,i}}\frac{\left(\cosh\left(\frac{L_{NW}}{\lambda_{NW,i}}\right)K_{1,\frac{d_{NW}}{2\lambda_{sub,i}}}\Gamma_{sub,i}D_{sub,i}\lambda_{sub,i}\lambda_{NW,i}+\sinh\left(\frac{L_{NW}}{\lambda_{NW,i}}\right)K_{0,\frac{d_{NW}}{2\lambda_{sub,i}}}\Gamma_{sub,i}D_{NW,i}\lambda_{sub,i}^2+\left(\left(\Gamma_{NW,i}\lambda_{NW,i}^2-\Gamma_{sub,i}D_{NW,i}\lambda_{sub,i}^2\right)\cosh\left(\frac{L_{NW}}{\lambda_{NW,i}}\right)+\Delta\Gamma_{al,i}\lambda_{NW,i}D_{sub,i}\right)K_{0,\frac{r}{\lambda_{sub,i}}}\right)}{\cosh\left(\frac{L_{NW}}{\lambda_{NW,i}}\right)K_{1,\frac{d_{NW}}{2\lambda_{sub,i}}}D_{sub,i}\lambda_{NW,i}+\sinh\left(\frac{L_{NW}}{\lambda_{NW,i}}\right)K_{0,\frac{d_{NW}}{2\lambda_{sub,i}}}D_{NW,i}\lambda_{sub,i}}$$

(39)

## 5.3 TEMPERATURE INDEPENDENT PARAMETERS USED FOR GAAS NW GROWTH MODELING

Values without references are fitting parameters or estimated values.

| Parameters | Values | Ref. |
|---|---|---|
| $\Xi'_{al,Ga}$ | $1 \cdot 10^4 nm^{-1} s^{-1}$ | - |
| $\Xi'_{al,As}$ | $1 nm^{-1} s^{-1}$ | - |
| $\Xi_{ls,III-V} \exp\left(-\frac{\delta g^{TS,ERS}_{ls,III-V}}{k_B T}\right)$ | $1 \cdot 10^3 nm^{-2} \cdot s^{-1}$ | $at\ T = 630°C$ |
| $\bar{Z}'_{aa,III}$, $\bar{Z}'_{aa,V}$ | $1 \cdot 10^{-3}$ | - |
| $\bar{Z}'_{as,III}$ $\bar{Z}'_{as,V}$ | $1 \cdot 10^{-15}$ | - |
| $\bar{Z}'_{av,III}$, $\bar{Z}'_{av,V}$ | $1 \cdot 10^{-2}$ | - |
| $\bar{Z}'_{av,III,sub}$, $\bar{Z}'_{av,V,sub}$ | $1 \cdot 10^{-3}$ | - |
| $\Delta h^{ERS}_{aa,\{1\bar{1}0\},Ga}$ | $0.3 eV$ | [91] |
| $\Delta h^{ERS}_{aa,\{1\bar{1}0\},As}$ | $0.65 eV$ | [91] |
| $\Delta h^{ERS}_{aa,\{111\},Ga}$ | $0.3 eV$ | [91] |
| $\Delta h^{ERS}_{av,\{1\bar{1}0\},Ga}$ | $2.3 eV$ | [91] |
| $\Delta h^{ERS}_{av,\{1\bar{1}0\},As_2}$ | $2 eV$ | [91] |
| $\Delta h^{ERS}_{av,SiO_x,As} // \Delta h^{ERS}_{av,SiO_x,Ga}$ | $1 eV // 1.5 eV$ | - |
| $\Delta h^{ERS}_{as,SiO_x,i}$ | $0 eV$ | - |
| $\gamma_{vs,ZB\{1\bar{1}0\}}$ | $4.98 \frac{eV}{nm^2}$ | [92] |
| $\gamma_{vs,WZ\{\bar{1}2\bar{1}0\}}$ | $4.42 \frac{eV}{nm^2}$ | [92] |
| $\gamma_{vs,ZB\{311\}} / \gamma_{vs,WZ\{\bar{1}2\bar{1}2\}}$ | $7\dfrac{eV}{nm^2}$ | - |
| $\gamma_{vs,ZB\{201\}} // \gamma_{vs,WZ\{\bar{1}2\bar{1}1\}}$ | $6\dfrac{eV}{nm^2}$ | - |
| $\gamma_{vs,ZB\{312\}} / \gamma_{vs,WZ\{\bar{1}2\bar{1}3\}}$ | $7\dfrac{eV}{nm^2}$ | - |
| $\gamma_{vs,ZB\{111B\}} / \gamma_{vs,WZ\{1000B\}}$ | $5\dfrac{eV}{nm^2}$ | - |
| $\gamma_{vl}(liquid\ Ga)$ | $4.2\dfrac{eV}{nm^2}$ | [93] |

Facet orientations with the highest symmetry, used in the illustrations in Figure 6.

| ZB facets | | | WZ facets | |
|---|---|---|---|---|
| $\omega = \{-30°, 90°..\}$ | $\omega = \{0°, 60°..\}$ | $\omega = \{30°, 150°..\}$ | $\omega = \{-30°, 30°..\}$ | $\omega = \{0°, 60°..\}$ |
| $\{2\bar{1}\bar{1}\}A$ <br> $\theta_T = 0°$ | $\{10\bar{1}\}$ <br> $\theta_T = 0°$ | $\{11\bar{2}\}B$ <br> $\theta_T = 0°$ | $\{0\bar{1}10\}$ <br> $\theta_T = 0°$ | $\{\bar{1}2\bar{1}0\}$ <br> $\theta_T = 0°$ |
| $\{3\bar{1}\bar{1}\}$ <br> $\theta_T = 10°$ | $\{31\bar{1}\}$ <br> $\theta_T = 31.5°$ | $\{11\bar{1}\}$ <br> $\theta_T = 19.5°$ | $\{0\bar{2}21\}$ <br> $\theta_T = 15.0°$ | $\{\bar{1}2\bar{1}1\}$ <br> $\theta_T = 17.1°$ |
| $\{100\}$ <br> $\theta_T = 35.3°$ | $\{210\}$ <br> $\theta_T = 50.8°$ | $\{22\bar{1}\}$ <br> $\theta_T = 35.3°$ | $\{0\bar{1}11\}$ <br> $\theta_T = 28.1°$ | $\{\bar{1}2\bar{1}2\}$ <br> $\theta_T = 31.7°$ |
| $\{211\}$ <br> $\theta_T = 70.5°$ | $\{321\}$ <br> $\theta_T = 67.8°$ | $\{110\}$ <br> $\theta_T = 54.7°$ | $\{0\bar{1}12\}$ <br> $\theta_T = 46.9°$ | $\{\bar{1}2\bar{1}3\}$ <br> $\theta_T = 42.8°$ |

**Table 1. Facets for ZB and WZ structure for the upper hemisphere with the lowest predicted surface energies are described with a set of angles $(\omega, \theta)$ as shown in Figure 6. Here $90° - \theta = \theta_T = 90°$ is defined to be the growth axis.**

| Liquid | $g_{Ga_{1-x_V}As_{x_V}}$ $\left[\frac{J}{mole}\right]$ | $g_{In_{1-x_V}As_{x_V}}$ $\left[\frac{J}{mole}\right]$ |
|---|---|---|
| $g_{l,III}{}'(T)$ | $-1389.2+114.049T-26.069299T\ln(T)$ $+1.0506\cdot10^{-4}T^2-4.0173\cdot10^{-8}T^3-118332T^{-1}$ | $-3479.81+116.8358T-27.4562T\ln(T)$ $+5.4607\cdot10^{-4}T^2-8.367\cdot10^{-8}T^3-211708T^{-1}$ |
| $g_{l,V}{}'(T)$ | $1.717245\cdot10^4+99.78639T-23.3144T\ln(T)$ $-0.00271613T^2+11600T^{-1}$ | $1.717245\cdot10^4+99.78639T-23.3144T\ln(T)$ $-0.00271613T^2+11600T^{-1}$ |
| $L_0(T)$ | $-25503.6-4.3109\cdot T$ | $-15851-11.27053\cdot T$ |
| $L_1(T)$ | $-5174.7$ | $-1219.5$ |
| $x_V^{ERS}(T)$ | $6.752\cdot10^{-7}\exp(0.0141\cdot T)/100$ | $(9.9\cdot10^{-4}\exp(0.00972\cdot T)-0.3)/100$ |

**Table 2. The coefficients of the free energy expressions of the pure elements in the case of InAs and GaAs are taken from the SGTE database[94], and are relative to the to the enthalpy of the standard element reference (HSER). The interaction parameters are taken from Ansara et. al.[75]. $T$ is the corresponding Kelvin temperature and all values are in Joule per mole. The equilibrium As mole fraction $x_{V,eq}$ is found from fitting liquidus values from ref.[95], in the range $T=400-800^{\circ}C$ for ZB GaAs and $T=350-550^{\circ}C$ for ZB InAs. All equilibrium data are found from experimental measurements and are relying on thermodynamical parameters which therefore should coexist in kinetic equilibrium.**

| Gas | $\Psi_i{}'^{n}(T)$ $\left[\frac{J}{mole}\right]$ |
|---|---|
| Ga | $263612.519 + 33.4871429T - 30.75007T\ln(T)$ $+ 0.00537745T^2 - 5.46534\cdot10^{-7}T^3 -150942.65T^{-1}$ |
| In | $237868.024 - 110.524313T - 8.405227\,T\ln T$ $- 0.0156847T^2 + 2.21196333\cdot10^{-6}T^3 -110674.05T^{-1}$ |
| As | $272027.85 - 32.2533338T - 21.21551T\ln T$ $+ 4.3891495\cdot10^{-4}T^2 - 7.393995\cdot10^{-8}T^3 + 9666.555\,T^{-1}$ |
| $As_2$ | $179351.548 + 10.5519715T - 37.35966\,T\ln T$ $- 5.61806\cdot10^{-5}T^2 - 2.13098\cdot10^{-8}T^3 + 104881.15T^{-1}$ |
| $As_4$ | $129731.745 + 230.754352T - 83.04465\,T\ln T$ $- 2.5148475\cdot10^{-5}T^2 + 1.0444733\cdot10^{-9}T^3 + 252728.45T^{-1}$ |

**Table 3. Thermodynamic data taking from Ansara et al.[75], where $g_{v,i_n}^{pure}{}'^{n}(T)=\Psi_i{}'^{n}(T)+RT\ln(P)$, with $P$ being the total vapor pressure in units of 0.1 MPa. $T$ is the corresponding Kelvin temperature and all values are in Joule per mole.**

## 5.4 TRIGONOMETRIC RELATIONS FOR THE SINGLE SLICE MODELING

Following the single slice construction shown in Fig.1 (b), the associated trigonometric quantities are given by:

$$\vartheta\left(\Delta z_+, \Delta z_-, d_{NW}\right) = arctan\left(\frac{\Delta z_+ - \Delta z_-}{d_{NW}}\right)$$

$$d_{eff}\left(\Delta z_+, \Delta z_-, d_{NW}\right) = \frac{d_{NW}}{\cos\left(\vartheta\right)}$$

$$d_{top}\left(\Delta z_+, \Delta z_-, \theta_T^+, \theta_T^-, d_{NW}\right) = d_{NW} - \Delta z_- \tan(\theta_{T-}) - \Delta z_+ \tan(\theta_{T+})$$

$$R_l\left(\Delta z_+, \Delta z_-, d_{NW}, \xi\right) = \frac{d_{NW}}{2\cos\left(\vartheta\right)\sin(\xi)}$$

$$H_l^{'}\left(\Delta z_+, \Delta z_-, d_{NW}, \xi\right) = R_l + \Xi(\xi)\sqrt{R_l^{\,2} - \left(\frac{d_{eff}}{2}\right)^2}$$

$$\Xi(\xi) = \begin{matrix} 1 & for & \xi > \pi/2 \\ -1 & for & \xi < \pi/2 \end{matrix}$$

The volumes of the slice shown in Fig. 1(c) are given by:

$$\Delta V_{s,low\pm}\left(\Delta z_\pm, d_{NW}\right) = \frac{d_{NW}{}^2 d\omega^2}{8}(\Delta z_{ref} - \Delta z_\pm)$$

$$\Delta V_{s,top\pm}\left(\Delta z_\pm, \theta_{T\pm}, d_{NW}\right) = \frac{d\omega}{6}\Delta z_\pm\left[3\left(\frac{d_{NW}}{2}\right)^2 + \left(\Delta z_\pm \tan(\theta_{T\pm})\right)^2 - \frac{3}{2}d_{NW}\Delta z_\pm \tan(\theta_{T\pm})\right]$$

$$\Delta V_{s,T\pm}\left(\Delta z_\pm, \theta_{T\pm}, d_{NW}\right) = \frac{1}{12}d\omega\left(\Delta z_\pm\right)^2 \tan\left(\theta_{T\pm}\right)\left(3d_{NW} - 2\Delta z_\pm \tan\left(\theta_{T\pm}\right)\right)$$

and the total liquid and solid volumes are therefore given by

$$\Delta V_l\left(\Delta z_+, \Delta z_-, \theta_{T+}, \theta_{T-}, d_{NW}, \xi\right) = \frac{2d\omega}{2\pi}\frac{1}{3}\pi H_l^{'2}\left(3R_l - H_l^{'}\right) - \left(\Delta V_{s,top-} + \Delta V_{s,top+}\right)$$

$$\Delta V_S\left(\Delta z_+, \Delta z_-, \theta_{T+}, \theta_{T-}, d_{NW}\right) = \Delta V_{s,low-} + \Delta V_{s,top-} + \Delta V_{s,low+} + \Delta V_{s,top+}$$

respectively. The corresponding number of atoms in the respective phases are given by, $\Delta N_{l(s)} = \frac{\Delta V_{l(s)}}{\Omega_{l(s)}}$, with $\Omega_{l(s)}$ being the atomic volumes. The areas of the side-, truncation-, and top-facet (see Fig. 2(a)) are given by:

$$\Delta A_{vl}\left(\Delta z_{+},\Delta z_{-},d_{NW},\xi\right)=\frac{2d\omega}{2\pi}2\pi R_{l}H_{l}^{'}$$

$$\Delta A_{ls111}\left(\Delta z_{+},\Delta z_{-},\theta_{T+},\theta_{T-},d_{NW}\right)=\frac{1}{2}d\omega\left[\left(d_{NW}-\Delta z_{+}\tan(\theta_{T+})\right)^{2}+\left(d_{NW}-\Delta z_{-}\tan(\theta_{T-})\right)^{2}\right]$$

$$\Delta A_{ls_T\pm}\left(\Delta z_{\pm},\theta_{T\pm},d_{NW}\right)=\frac{1}{2}d\omega\frac{\Delta z_{\pm}}{\cos(\theta_{T\pm})}\left(d_{NW}-\tan(\theta_{T\pm})\Delta z_{\pm}\right)$$

$$\Delta A_{vs}\left(\Delta z_{+},\Delta z_{-},d_{NW}\right)=\left(2\Delta z_{ref}-\Delta z_{-}-\Delta z_{+}\right)\frac{d_{NW}}{2}d\omega$$

## 5.5 SIMULATIONS OF THE TRUNCATION DYNAMICS

Parameters for the simulations of the truncation dynamics are shown in Table 4.

| Simulation no. | $\theta_{T_,0}=\theta_{T+,0}$ | $\xi_0$ | $w_{hkl}$ | $\sigma$ | $\gamma_{vs0}$ | $c_{hkl}$ |
|---|---|---|---|---|---|---|
| S1 | 50.8° | 120° | 10° | 0.35 | $10\frac{eV}{nm^2}$ | 1 |
| S2 | 50.8° | 110° | 8° | 0.4 | $9\frac{eV}{nm^2}$ | 1 |
| S3 | 50.8° | 120° | 15° | 0.4 | $9\frac{eV}{nm^2}$ | 0.71 |
| S4 | 31.5° | – | 3° | 0.4 | $8\frac{eV}{nm^2}$ | 1 |
| S5 | 50.8° | – | 15° | 0.4 | $9\frac{eV}{nm^2}$ | 0.71 |
| S6 | 50.8° | 130° | 15° | 0.5 | $9\frac{eV}{nm^2}$ | 0.71 |

**Table 4. Six types of simulations of 15 sec of Ga catalyzed GaAs growth in the single slice construction. Initial conditions marked with a subscript $_0$. 15 sec of growth was in all these cases enough to go into a quasi steady state growth mode. Basic growth conditions are in all cases: $\frac{f_V}{f_{III}}=10$, $GR_{planar}=0.3\frac{\mu m}{hr}$ and $T=630°C$. The time**

steps are $\Delta t = 0.001s$ and at $t = 0$ the liquid composition is $x_{As} = 0.01$ and the truncation height is $\Delta z_{-,0} = \Delta z_{+,0} = -1nm$. All simulations are modeling the formation of ZB structure at $\omega = 0°$ where the structure is symmetric around the growth axis in the single slice construction. Parameters not given here are given in Appendix 5.3.

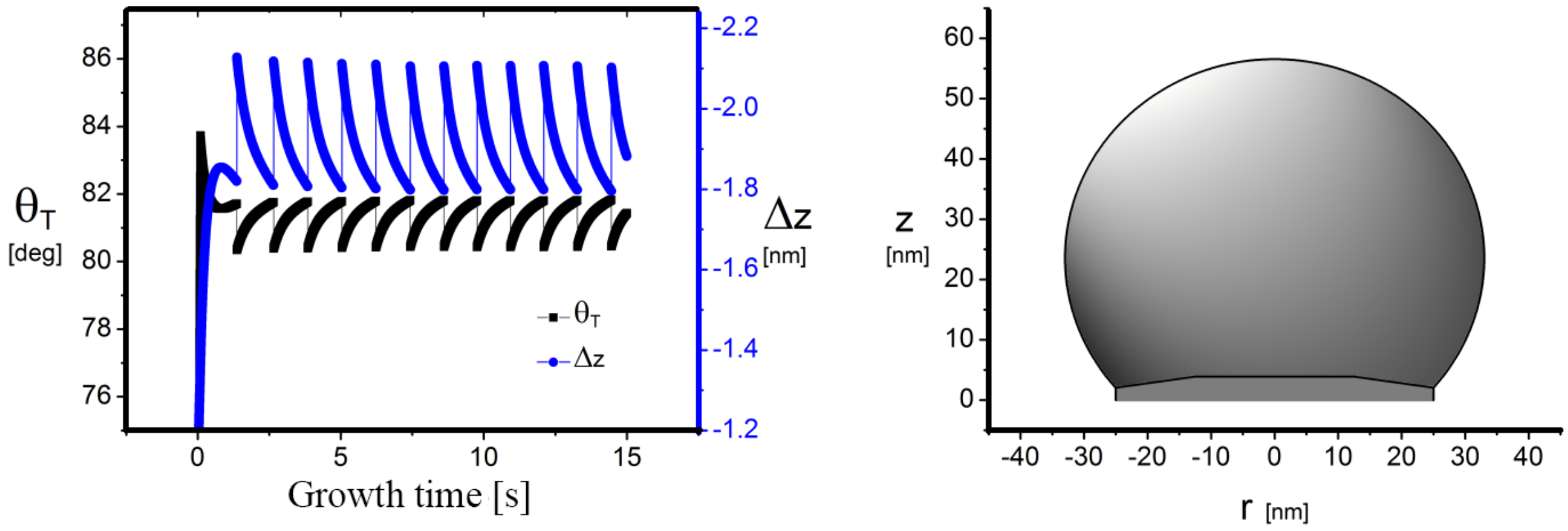

Figure 19. 15 sec growth simulation (using simulation parameters S6 in Table 4) of self-catalyzed GaAs NW growth in the single slice construction. On the left, the truncation height and truncation angle are oscillating in coherent manner with periods of the formation of a ML at the topfacet. In this simulation the oscillations only fill up approximately a single ML at the truncation facets between each nucleation event at the topfacet. This means that the oscillations would be difficult to detect even in in-situ TEM experiments, however in this single slice construction it should be seen as kind of an average of the whole growth region. For the real 3D system, the oscialltions might dominate on certain facets[40,41,42]. The NW morphology after 15 sec of growth is shown on the right.

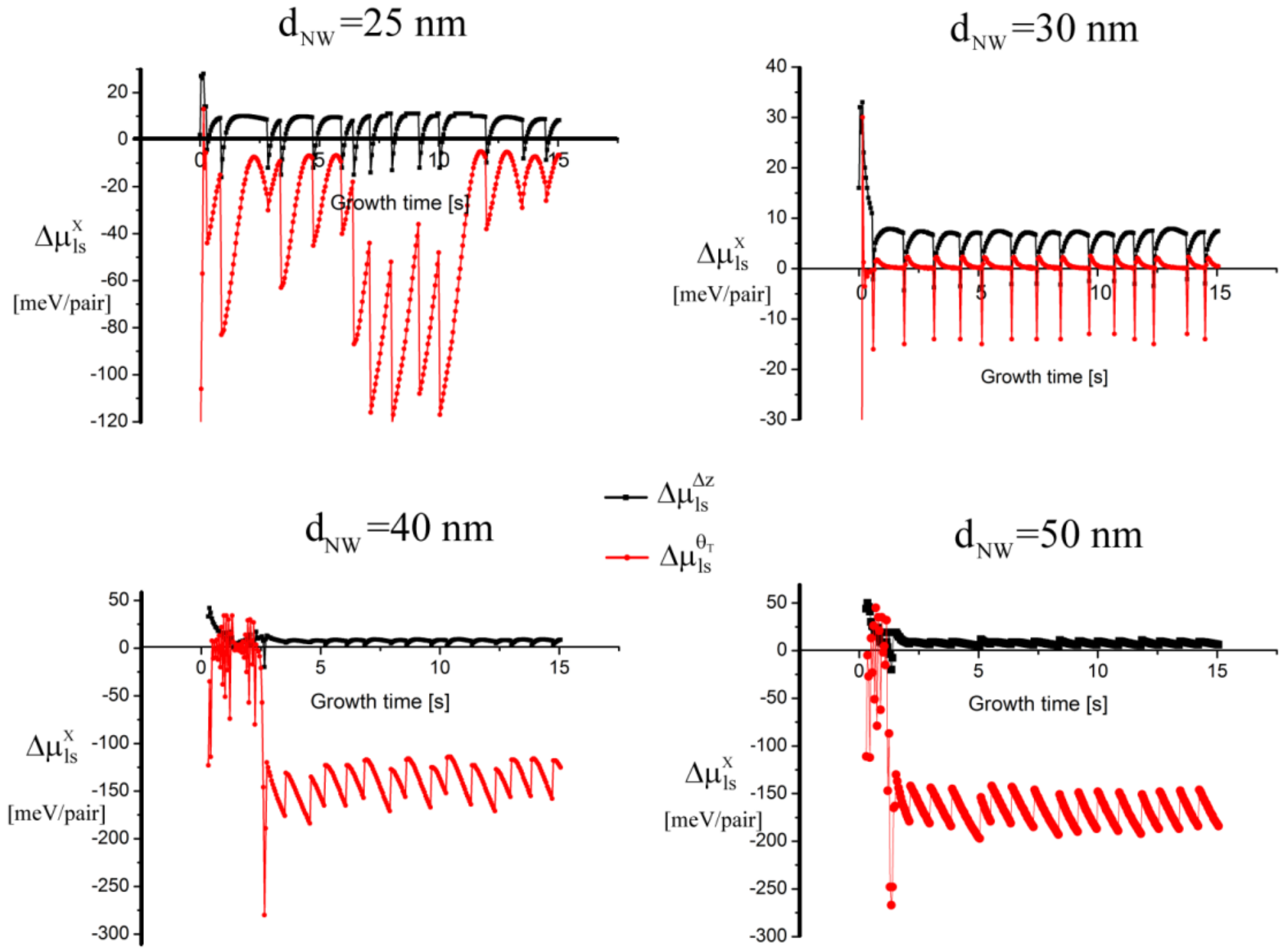

Figure 20. For a given set of initial conditions the *ls* driving forces forming the truncation facets are plotted as a function of 15 seconds of growth time (Growth simulations S1). At smaller diameters the growth system seems highly unstable which is due to not only the initial conditions but also to the solid anisotropy. For $d_{NW} = 30nm$ the system

**moves into a steady state regime not far from equilibrium, this means that the initial conditions are fitting well with the energies chosen for the system. At larger diameters the truncation facets moves into a steady state regime far from local equilibrium.**

## 5.6 INCLUDING THE EFFECT OF A TRIPLE LINE TENSION

An additional contribution to the free energy of the system may come from an in-balance of capillary forces meeting at the TL.[45] A change in one of the involved interface orientations implies a change in such a TL energy, and in order to reach mechanical equilibrium, an increase in strain per unit TL length in the solid and/or by a local change in the $vl$ curvature on the cost of more $vl$ interface is induced. . Both effects alter the chemical potentials and can therefore have an influence on the growth dynamics. The effect of TL forces on the growth of NWs was introduced by Schwarz and Tersoff[45], who used the tangential component of the TL force on a locally smooth solid surface to describe the TL motion, and the normal component altering the solid chemical potential at the TL. We will here take a slightly different approach and let the TL equilibration allow to take part in the total free energy minimization process in all dimensions. Because changes in the liquid volume induce changes in the TL excess, we assign the TL excess to the liquid phase for convenience and add an extra term to the liquid chemical potential as

$$\delta\mu_{l,i}(x_{III}, x_V, T, \omega) = \mu_{l,i}^{\infty}(x_{III}, x_V, T) + \gamma_{vl}\frac{\partial A_{vl}}{\partial N_{l,i}} + \frac{d\Upsilon(\omega)}{dN_{l,i}} - \mu_i^{ERS}$$

Here $d\Upsilon(\omega)$ is the TL excess free energy per length at $\omega$ and $\Upsilon$ is the total TL excess. The effect of the TL force on crystal growth is difficult to quantify mainly because it has been difficult to measure experimentally. Nevertheless if we as in ref.[45] define an effective width of the TL, $w_{eff}$, the TL force along the $pq$ interfacial component can be written as,

$$f_{pq} = w_{eff}\left(\gamma_{pq} + \gamma_{qw}\cos(\theta_q) + \gamma_{pw}\cos(\theta_p) + \tau\kappa_{pq}\right)$$

where $pqw$ is any cyclic permutation of $vls$. $\tau$ is the line excess free energy depending on $f_{pq}$ itself and $\kappa_{pq}$ the line curvature at $\omega$ projected on the $pq$ component. Assuming the TL curvature is negligible, the net force along all interfaces at the vanish at equilibrium, and the surface energies and corresponding contact angles are given by $\gamma_{pq} = \frac{\sin(\theta_{w,eq})}{\sin(\theta_{p,eq})}\gamma_{qw}$ and $\cos(\theta_{p,eq}) = \frac{\gamma_{qw}^2 - \gamma_{pq}^2 - \gamma_{pw}^2}{2\gamma_{pq}\gamma_{pw}}$. Away from equilibrium, we will describe the TL excess per length as,

$$d\Upsilon(\omega) = \frac{d_{NW}(\omega)}{2} d\omega \left| f_{TL}(\theta_l(\omega), \theta_s(\omega)) \right|$$

with

$$\left| f_{TL}(\theta_l, \theta_s) \right| = \sqrt{f_{ls\|}^2 + f_{ls\perp}^2 + \tau^2\kappa^2} = w_{eff}\sqrt{\left(\gamma_{ls} + \gamma_{vs}\cos(\theta_s) + \gamma_{vl}\cos(\theta_l)\right)^2 + \left(\gamma_{vs}\sin(\theta_s) - \gamma_{lv}\sin(\theta_l)\right)^2 + \tau^2\kappa^2}$$

being the net force per length at $\omega$. To see the effect induced by the TL tension around the TL, evolution of the local morphology may be described by a local curvature dependent driving force as in ref[45].

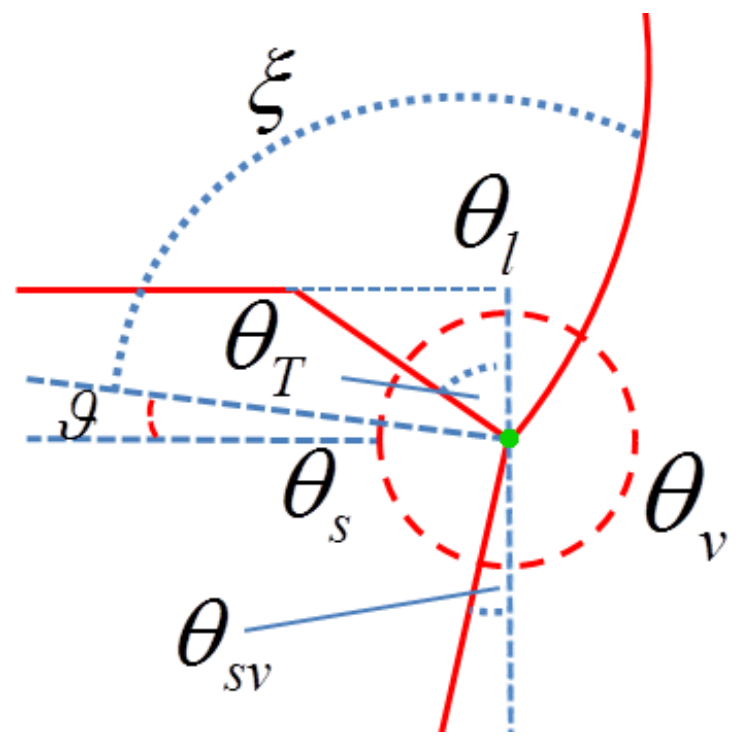

**Figure 21. Cross sectional view on the TL (green dot). The red solid lines illustrate the actual morphology at the TL region at the given $\omega$, and the blue dotted lines illustrates the construction lines.**

## 5.7 GEOMETRICAL ANALYSIS OF A CONSTANT VL CURVATURE CONSTRUCTION AND TOTAL FACETTING, $\eta(\omega) = 0$

To analyze the total liquid-solid dynamics we need to include the $\omega$- dependence on all parameters in the parameter set, $\{X(\omega)\}$. However this is as mentioned a very complex problem and we will here only make some rough simplifications in order to get qualitative ideas and better understanding about the three dimensional system. As compared to the single slice construction above at least one additional parameter, namely $\eta(\omega)$ (see eq.(18)), needs to be added to the parameter set. If the liquid is assumed to have a constant curvature an example of a choice of parameter set could be; $\{X(\omega)\} \in \{d_{NW}, \eta, \Delta z_+, \Delta z_-, \theta_{T+}, \theta_{T-}, \xi\}$. The form of $\eta(\omega)$ which describes the cross sectional shape of the growth interface, is very important for the total configuration but it is also a very complex parameter to include. It has not been possible in the time of writing to find a consistent method to solve this system. In this section we will instead show some implications of typical assumptions used for modeling NW growth, which will serve as instructive and informative insight to the three dimensional anisotropic *ls* system. Such as under which configurations and conditions the NW growth system will move in and out of regime *I* and *II*. The free energy minimization process of the *ls* system during growth is complex mainly due to the interplay between the isotropic liquid and the anisotropic solid. If we imagine that the cusps of the gamma-plot shown in Figure 6 B or C are very sharp and deep, then the system will choose total sidewall facetting even at the TL, and the liquid phase will 'adjust' to this as long as the system is regime *II*. In an ideal regime *I* (a single planar *ls* topfacet) the nucleation statistics can be treated in the framework proposed in ref.[26], which is mainly a relevant regime during changes in growth conditions where the relative size of the liquid is decreasing. If we furthermore assume that the *vl* interface tension is strong, the isotropic (and assumed homogenous) liquid prefers a constant curvature due to a strong Laplace pressure. To describe such a system we will first choose a single slice construction which is oriented in such a way that $\omega = 0^{\circ}$ is in the direction of the liquid-solid displacement, $\Delta r$ (see Figure 22 (A)). Using cylindrical coordinates, $(r, \omega, z)$, we can write two intersections between the wire and liquid as $(r_-, 180^{\circ}, z_-)$ and $(r_+, 0^{\circ}, z_+)$. For a given radius of curvature there exist two solutions, one for $\xi \geq 90^{\circ}$ and one for $\xi < 90^{\circ}$, as seen in Figure 22 (B). Here $z_- = -\frac{\Delta}{2} - \frac{d_{NW0}}{2}\sqrt{\frac{4R_l^2 - \Delta^2 - d_{NW0}^2}{\Delta^2 + d_{NW0}^2}}$, and the two intersections are given by

$$\left.\begin{aligned} (r_-, 180^{\circ}, z_-) &= \left(-\sqrt{R_l^2 - z_-^2}, 180^{\circ}, z_-\right) \\ (r_+, 0^{\circ}, z_+) &= \left(-\sqrt{R_l^2 - z_+^2}, 0^{\circ}, \Delta + z_-\right) \end{aligned}\right\} \quad for \quad \xi \geq 90$$

$$\left.\begin{aligned} (r_-, 180^{\circ}, z_-)' &= (-r_+, 0^{\circ}, -z_+) \\ (r_+, 0^{\circ}, z_+)' &= (-r_-, 180^{\circ}, -z_-) \end{aligned}\right\} \quad for \quad \xi \leq 90$$

where $\Delta = \Delta z_- - \Delta z_+$ is difference in truncation in the two sides and $d_{NW0}$ is the diameter at $\omega = 0^{\circ}$. The z-coordinate for intersection between wire and liquid as a function of $\omega$ are then given by, $z(\omega) = -\sqrt{R_l^2 - \left(\cos(\omega) d_{NW}(\omega) + \Delta r\right)^2 - \left(\sin(\omega) d_{NW}(\omega)\right)^2}$, where $\Delta r = \sin(\vartheta)\sqrt{R_l^2 - \frac{d_{NW0}^2}{4\cos(\vartheta)^2}}$ is the displacement between of center of the NW crystal and the liquid center at $\omega = 0^{\circ}$. Now analyzing the wetting consequences when assuming total sidefacetting ($\eta(\omega)=0$ in eq.(18)) will give us some qualitative ideas about the real system and under which conditions TL nucleation can take place, as shown in C.

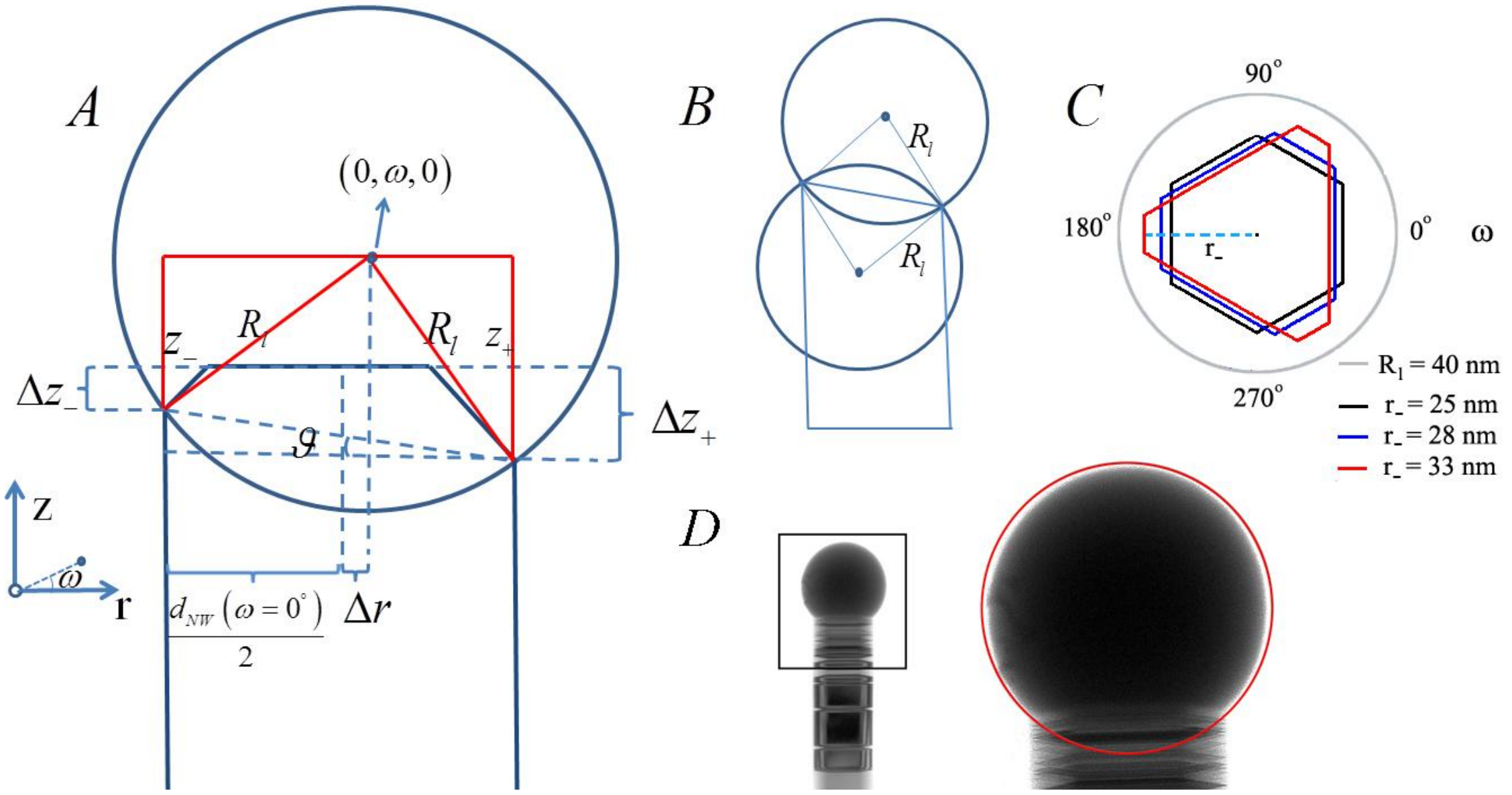

**Figure 22. (A) Illustration of the trigonometry used to derive the $\Delta z$ vs $\omega$ relation for the constant vapor-liquid curvature construction and $\eta(\omega)=0$. (B) For a given radius of curvature there exist two solutions, one for $\xi \geq 90^{\circ}$ and one for $\xi < 90^{\circ}$. (C) If we allow for a three-fold symmetric morphology we need to define two lengths, $r_-$ and $r_+$, to describe the diameter. However for a given crystal volume only $r_-$ is needed, see eq.(19). (D) A TEM image along the $[1\bar{1}0]$ zone axis of a GaAs NW with a Ga droplet on top. The red circle on the enlarged view is a perfect circle, which fits almost perfect to the shape of the Ga droplet.**

In Figure 23 the truncation heights are plotted for different relative sizes of the droplets and for six-fold and three-fold facetting. It is obvious that a relative large droplet will have a smaller probability of inducing positive truncations and it is also obvious that a hexagonal shape is the most convenient shape in relation with a liquid. If we allow for a three-fold facetting we need to define two lengths, $r_-$ and $r_+$, to describe the diameter, see Figure 23(C). It is very likely that for real systems that the truncation is negative all the way around the TL and only becomes positive at a given location when the liquid supersaturation is high and induce either TL nucleation or move the TL into regime *I*. A TEM image along the $[1\bar{1}0]$ zone axis of a GaAs NW with a Ga droplet on top is shown in Figure 20 (D). The red circle on the enlarged view is a perfect circle, which fits almost perfect to the shape of the Ga droplet. Thus if the liquid curvature is constant also as a function of $\omega$, we can say that if the NW crystal completely faceted ($\eta_0 = 0$), the truncation height would vary as a function of $\omega$ as shown in . In the other extreme if the system is completely axi-symmetric $\eta_0 = 1$ the truncation height would be independent of $\omega$. It is important to note that for real liquid-solid growth systems using the $\eta_0$ parameter to describe the system would give a value somewhere in between 0 and 1, and the amplitude of the curves in Figure 23 will be smaller. See the figure text for a discussion of different cases of total facetting and constant liquid curvature.

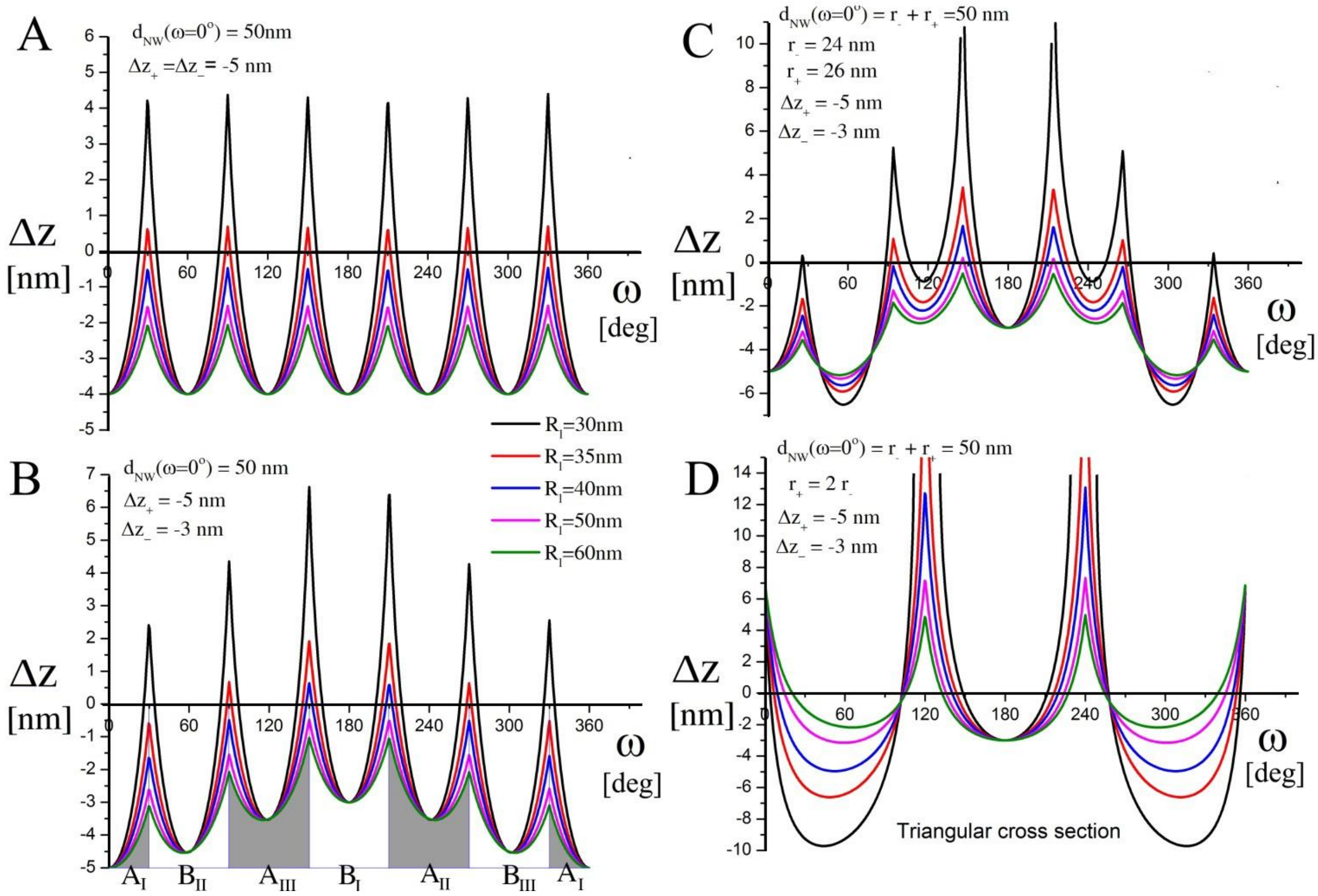

**Figure 23 Geometrical representation of the truncation size, $\Delta z$, as a function of $\omega$ and different liquid sizes, for a NW system with $d_{NW}(\omega = 0^{\circ}) = 50\ nm$ assuming constant vapor-liquid curvature and total facetting at the same time. (A) Assuming equal truncation $\Delta z_- = \Delta z_+$ at both sides at $\omega = 0^{\circ}$, as initial conditions on a complete facetted solid hexagonal cross section. (B) If the crystal has three-fold symmetry but takes on a six-fold morphology it can be favorable to incline the growth system. However, assuming vertical sidewalls it can be shown that the system does not lower the free energy because the areas of A and B type facets are the same in total (indicated by grey and white regions) and the system either chooses to make the A facets smaller and the B facets larger as shown in (C). It should be noted that an inclination angle could be initiated by a non-isotropic incoming vapor flux due to the Marangoni effect[96] but this is out of the scope of this study. In (C) it is seen that if the solid induce even a small derivation from the hexagonal shape, it has a huge impact on the growth system which will most likely also be present in the real system. (D) In the extreme case of a triangular shaped NW and constant *vl* curvature, the system will be in regime *I* for all truncation sizes in the case of $R_l > 37\ nm$ around the edges of the triangle. This is because there is no solution to the sidefacet-liquid intersection problem. In this case the edges will be either rounded or TL has moved in on the topfacet and the facet edges may be completed by surface diffusion.**

In Figure 24 we see the relationship between parameters; $d_{NW}$, $\xi$, $\Delta z$ and $\omega$ under six fold and three fold symmetric sidewall facetting in the case of constant liquid curvature.

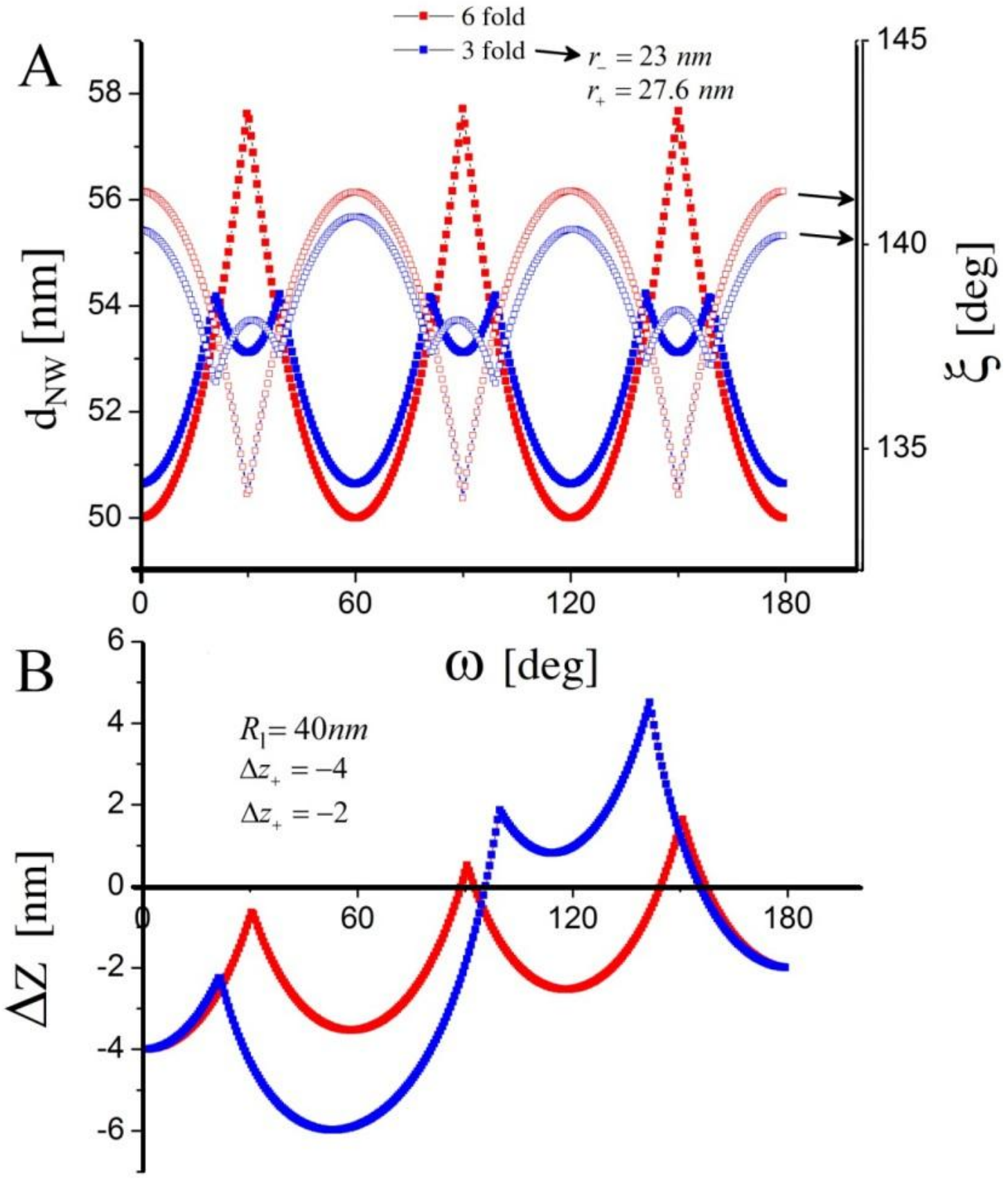

**Figure 24 Comparison between a perfect 6-fold hexagonal shaped NW (red) and a three-fold cross sectional morphology (blue) for a given crystal volume, under the assumption of constant *vl* curvature and total facetting. (A) The solid dots represent the NW diameter and the open dots represent the contact angle. (B) In the case of a strong driving force towards 3-fold sidewall facetting it is likely that the TL will move in on the topfacet, here in the region around $\omega \sim 120°$. For $\omega \in [180°, 360°]$, the curves are mirrored in $\omega = 180°$.**

# 6. List of symbols and abbreviations

*ERS*: Equilibrium reference state

$i$: Refers to the $i$'th element

$j$: Refers to the $j$'th interface (unless other stated)

$T$: Substrate temperature

$f_{i(,\perp)}$: Beam flux in the direction of the beam ($\perp$ refers to the flux perpendicular to the given interface)

$f_i^{ERS}$: Pressure equivalent beam (PEB) flux of element $i$ needed to attain ERS conditions in the absence of a vapor phase.

$p_i$: Vapor pressure

$\rho_{j,i}$: Density of adatoms

$x_i$: Atomic fraction in the liquid phase

$\bar{c}_{p,i}$: General symbol for the normalized density in phase $p$

$G_p$: Global Gibbs free energy of the $p$ phase

$g_p$: Gibbs free energy per atom in phase $p$

$\mu_{p,i}^{(\infty)}$: Chemical potential in state $p$ ( $\infty$ refers to infinitely large phases)

$\delta\mu_{p-ERS,i}$: Chemical potential in phase $p$ with respect to the $ERS$

$\Delta\mu_{pq,i} = \delta\mu_{p-ERS,i} - \delta\mu_{q-ERS,i}$: Change in free energy due to a $p$ to $q$ atomic state transition

$\Delta\varepsilon_s$ : The difference in bulk free energy between the crystal with stacking sequence $s$ and the standard reference ($ERS$)

$\delta g_{pq,i}^{TS}$: The activation free energy per $p$ atom needed to reach to transition state between $p$ and $q$

$\Gamma_{pq,i}$: $p$ to $q$ state transition flux

$\Delta\Gamma_{pq,i}$: The net flux of the $p$ to $q$ state transitions

$S_{b(v),i}$: Sticking coefficient of beam or vapor elements

$A_{pq}$: Area of the $pq$ interface

$\lambda_{j,i}$: The effective adatom diffusion length

$D_{j,i}$: The effective diffusivity coefficient

$\tau_{j,i}$: The mean lifetime in the adatom state

$\Xi_{pq,i}$: Rate constant of the p to q transition

$\bar{Z}_{pq}^{'}$: The effective coordination number of the p to q transition. (‘) includes activation entropy.

$N_{p,i}$: Number of atoms of element $i$ in phase $p$

$n_p^{(*)}$: Total number of III-V pairs in a cluster (* refers to the solid critical nucleus)

$h_{ML}$ : Monolayer height along the growth axis

$\gamma_j$: The tension of the $j$'th interface

$L_{TL}$: Total length of the triple phase line

$\varphi_j$: The wetting angle given by Young's equation

$\theta(\omega)$: The angle between the $lv$ and the $sl$ interface at $\omega$

$\omega$: The angle between the middle of the side facet and the nucleation site, as measured from the center of the top-facet

$\eta(\omega)$: Parameter determining the cross sectional shape at the growth interface, see eq.(18)

$R_l$: Radius of curvature for the liquid-vapor interface

$\mathrm{I}_{hkl}$ : The difference in interface energy between the *hkl* facet and 'off facet' energy

$\Omega_p$ : The average atomic volume in phase *p*.

$I_i, I_{i,des}, I_{i,inc}$ : The liquid sorption current, the liquid desorption current, the incorporation current

$Z$ : The Zeldovich factor

$\sigma$ : The ratio between *vs* and *ls* interfacial energies

$w_{hkl}$ : Parameter specifying the half-width half maximum of the cusp in the gamma function around the $(hkl)$ facet

$c_{hkl}$ : Correction parameter at high $w_{hkl}$ values

$\xi$ : Contact angle of the constant curvature construction, see Figure 14 (B)

$\theta$ : The angle from the top facet to a given orientation

$\theta_T$ : Truncation angle of a given facet defined as $\theta_T = 90 - \theta$

$GR_{planar}$ : Corresponding planar growth rate

$\Delta t$ : Time step in simulation

[54] V. Schmidt, S. Senz and U. Gösele, *Appl. Phys. A,* 80, 445–450 (2005)

[55] The steric factor is here defined as the fraction of the total collision cross section of the $p$ state which effectively has the transition state barrier specified for the $p \rightarrow q$ transition

[56] Even though the prefactors in principle could be derived analytically, it is a complex problem and will eventually involve estimations and simplifications for different entropy contributions and effective coordination numbers and they are therefore more reasonably used as fitting parameters.

[57] F. Glas, Phys. *Status Solidi B* 247, 2, 254–258 (2010)

[58] T. Rieger, S. Heiderich, S. Lenk, M. I. Lepsa, D. Grützmacher, *J. Cryst. Growth* 353, 1, 2012

[59] M. Tchernycheva, L. Travers, G. Patriarche, F. Glas, J.-C. Harmand, G. E. Cirlin and V. G. Dubrovskii, *J. Appl. Phys.* 102, 094313 (2007)

[60] E. D. Jong, R. R. LaPierre and J. Z. Wen, *Nanotechnology* 21, 045602 (2010)

[61] J. G. Connell et al., *Nano Lett.*, 13, 199 (2013)

[62] K. A. Dick, J. Bolinsson, B. M. Borg, and J. Johansson, *Nano Lett.*, 12, 6 (2012)

[63] J. W. Cahn, Acta Metall. 8 (1960)

[64] C.-Y. Yeh, Z. W. Lu, S. Froyen, and A. Zunger, *Phys. Rev. B* 45, 12130 (1992)

[65] C.-Y. Yeh, Z. W. Lu, S. Froyen, and A. Zunger, *Phys. Rev. B* 46, 10086 (1992)

[66] The Wulff shape is the shape of a crystal with the lowest free energy possible for a given crystal volume. See textbooks on crystal growth for a more detailed explanation.

[67] F. Glas, J-C. Harmand and G. Patriarche, *Phys. Rev. Lett.* 104, 135501 (2010)

[68] R. Becker and W. Döring, *Ann. Phys.* 24, 719 (1935)

[69] W. K. Burton, N. Cabrera and F. C. Frank, *Phil. Trans. R. Soc. Lond. A* 243, 299-358 (1951)

[70] I. Markov, Crystal growth for beginners, 2nd edition, World Scientific, Singapore (2003)

[71] Yeh, C-Y.; Lu, Z. W.; Froyen, L. S.; Zunger, A.; *Phys. Rev. Lett.,* 46, 10086 (1992)

[72] V. G. Dubrovskii, G. E. Cirlin, N. V. Sibirev, F. Jabeen, J. C. Harmand, and P. Werner, *Nano Lett.*, 11, 1247–1253 (2011)

[73] E.J. Schwalbach, S.H. Davis, P.W. Voorhees, D. Wheeler and J.A. Warren, *J. Mat. Res.*, 26, 17 (2011)

[74] O. Redlich and A. Kister, *Ind. Eng. Chem.*, 40, 345-348 (1948)

[75] I. Ansara et al., *Calphad*, 18, 2 (1994)

[76] E. Russo-Averchi, M. Heiss, L. Michelet, P. Krogstrup, J. Nygard, C. Magen, J.R. Morante, E. Uccelli, J. Arbiol, A Fontcuberta i Morral, *Nanoscale,* 4, 5 (2012)

[77] E. Uccelli, J. Arbiol, C. Magen, P. Krogstrup, E. Russo-Averchi, M. Heiss, G. Mugny, F. Morier-Genoud, J. Nygard, J.R. Morante, A Fontcuberta i Morral, *NanoLetters,* 11, 9 (2011)

[78] S. Ambrosini, M. Fanetti, V. Grillo, A. Franciosi and S. Rubini, *AIP Advances* 1, 042142 (2011)

[79] C. Colombo, D. Spirkoska, M. Frimmer, G. Abstreiter and A. Fontcuberta i Morral, *Phys. Rev. B* 77, 155326 (2008)

[80] K. A. Dick, P. Caroff, J. Bolinsson, M. E. Messing, J. Johansson, K. Deppert, L. R. Wallenberg and L. Samuelson, *Semicond. Sci. Technol*. 25**,** 024009 (2010)

[81] G. E. Cirlin, V. G. Dubrovskii, Yu. B. Samsonenko, A. D. Bouravleuv, K. Durose, Y. Y. Proskuryakov, B. Mendes, L. Bowen, M. A. Kaliteevski, R. A. Abram and D. Zeze, *Phys. Rev. B* 82, 035302 (2010)

[82] F. Jabeen, V. Grillo, R. Rubini, F. Martelli, *Nanotechnology* 9, 275711 (2008)

[83] D. Spirkoska et al. *Phys. Rev. B*, 80, 245325 (2009)

[84] B. Bauer, A. Rudolph, M. Soda, A. Fontcuberta i Morral, J. Zweck, D. Schuh and E. Reiger, *Nanotechnology* 21, 43 (2010)

[85] S Plissard, G Larrieu, X Wallart and P. Caroff, *Nanotechnology*, 22, 27 (2011)

[86] G Wulff, *Zeitschrift fur Krystallographie und Mineralogie*, 34, 449 (1901)

[87] M. Tchernycheva, L. Travers, G. Patriarche, F. Glas, J.-C. Harmand, G. E. Cirlin and V. G. Dubrovskii, *J. Appl. Phys*. 102, 094313 (2007)

[88] In this continuum approach, these activation free energies should be seen as effective values depending on the surface reconstruction and the roughness of the facet, and is in principle also temperature dependent. However for a given surface reconstruction, surface roughness and therefore $\Delta g_{sa,i}$ is a slowly varying function of $T$ and can reasonably be ignored at typical growth temperatures of III-V NWs.

[89] V. Dubrovskii et al. *Physical Rev B*, 79, 205316 (2009)

[90] J. Johansson, C. P. T. Svensson, T. Mårtensson, L. Samuelson, W. Seifert, *J. Phys. Chem. B*., 2005, 109 (2005)

[91] Activation enthalpies are based on DFT calculations are taken from the works; V. Pankoke, S. Sakong and P. Kratzer, *Phys. Rev. B* 86, 085425 (2012); J. N. Shapiro, A. Lin and D. L. Huffaker, *Phys. Rev. B*, 84, 085322 (2011)

[92] V. Pankoke, P. Kratzer and S. Sakong, *Phys. Rev. B* 84, 075455 (2011)

[93] G. E. Cirlin et al. *Phys. Rev B* 82, 035302 (2010)

[94] A. T. Dinsdale, *Calphad*, 15,4 317-425 (1991)

[95] M. Tmar, A. Gabriel, C. Chatillon and I. Ansara, *J. Cryst. Growth* 69, 421-441, (1984)

[96] S. Abbasoglu and I Sezai, *Engineering with Computers* 23, 2 (2007)